\documentclass{emulateapj}
\usepackage{apjfonts}

\def\degrees{^\circ}
\def\kms{{\ }{\rm km}\,{\rm s}^{-1}}
\def\LCDM{$\Lambda$CDM}

\begin{document}
\submitted{The Astrophysical Journal, submitted}

\shortauthors{KAZANTZIDIS ET AL.}

\shorttitle{On the Efficiency of the Tidal Stirring Mechanism for the
  Origin of dSphs}

\title{On the Efficiency of the Tidal Stirring Mechanism for the
  Origin of Dwarf Spheroidals:\\ Dependence on the Orbital and
  Structural Parameters of the Progenitor Disky Dwarfs}

\author{Stelios Kazantzidis\altaffilmark{1},  
        Ewa L. {\L}okas\altaffilmark{2},
        Simone Callegari\altaffilmark{3},\\
        Lucio Mayer\altaffilmark{3,4}, and
        Leonidas A. Moustakas\altaffilmark{5}}

\altaffiltext{1}{Center for Cosmology and Astro-Particle Physics; and
  Department of Physics; and Department of Astronomy, The Ohio State 
  University, Columbus, OH 43210, USA; {\tt stelios@mps.ohio-state.edu}}
\altaffiltext{2}{Nicolaus Copernicus Astronomical Center, 00-716 Warsaw,
  Poland; {\tt lokas@camk.edu.pl}}
\altaffiltext{3}{Institute for Theoretical Physics, University of Z\"urich, 
  CH-8057 Z\"urich, Switzerland; {\tt callegar@physik.uzh.ch}} 
\altaffiltext{4}{Institute of Astronomy, Department of Physics, ETH Z\"urich,
  CH-8093 Z\"urich, Switzerland; {\tt lucio@phys.ethz.ch}}
 \altaffiltext{5}{Jet Propulsion Laboratory, California Institute of Technology, 
   Pasadena, CA 91109, USA; {\tt leonidas@jpl.nasa.gov}}
 
\begin{abstract}
  
  The tidal stirring model posits the formation of dwarf spheroidal
  galaxies (dSphs) via the tidal interactions between late-type,
  rotationally-supported dwarfs and Milky Way-sized host galaxies.
  Using a comprehensive set of collisionless $N$-body simulations, we
  investigate the efficiency of the tidal stirring mechanism for the
  origin of dSphs. In particular, we examine the degree to which the
  tidal field of the primary galaxy affects the sizes, masses, shapes,
  and kinematics of the disky dwarfs for a range of dwarf orbital and
  structural parameters. Our study is the first to employ
  self-consistent, equilibrium models for the progenitor dwarf
  galaxies constructed from a composite distribution function and
  consisting of exponential stellar disks embedded in massive,
  cosmologically-motivated dark matter halos. Exploring a wide variety
  of dwarf orbital configurations and initial structures, we
  demonstrate that in the majority of cases the disky dwarfs
  experience significant mass loss and their stellar distributions
  undergo a dramatic morphological, as well as dynamical,
  transformation. Specifically, the stellar components evolve from
  disks to bars and finally to pressure-supported, spheroidal systems
  with kinematic and structural properties akin to those of the
  classic dSphs in the Local Group (LG) and similar environments. The
  self-consistency of the adopted dwarf models is crucial for
  confirming this complex transformation process via tidally-induced
  dynamical instabilities and impulsive tidal heating of the stellar
  distribution. Our results suggest that such tidal transformations
  should be common occurrences within the currently favored
  cosmological paradigm and highlight the key factor responsible for
  an effective metamorphosis to be the strength of the tidal shocks at
  the pericenters of the orbit. We also demonstrate that the
  combination of short orbital times and small pericentric distances,
  characteristic of dwarfs being accreted by their hosts at high
  redshift, induces the strongest and most complete transformations.
  Our models also indicate that the efficiency of the transformation
  via tidal stirring is affected significantly by the structure of the
  progenitor disky dwarfs.  While the mass-to-light ratios, $M/L$, of
  the dwarf galaxies typically decrease monotonically with time as the
  extended dark matter halos are efficiently tidally stripped, we
  identify a few cases where this trend is reversed later in the
  evolution when stellar mass loss becomes more effective.  We also
  find that the dwarf remnants satisfy the relation $V_{\rm max} =
  \sqrt{3}\,\sigma_{\ast}$, where $\sigma_{\ast}$ is the
  one-dimensional, central stellar velocity dispersion and $V_{\rm
    max}$ is the maximum halo circular velocity, which has intriguing
  implications for the missing satellites problem. Assuming that the
  distant dSphs in the LG, such as Leo I, Tucana, and Cetus are the
  products of tidal stirring, our findings suggest that these galaxies
  should have only been partially stirred by the tidal field of their
  hosts. We thus predict that these remote dwarfs should exhibit
  higher values of $V_{\rm rot}/\sigma_{\ast}$, where $V_{\rm rot}$ is
  the stellar rotational velocity, compared to those of dSphs located
  closer to the primary galaxies. Overall, we conclude that the action
  of tidal forces from the hosts constitutes a crucial evolutionary
  mechanism for shaping the nature of dwarf galaxies in environments
  such as that of the LG.  Environmental processes of this type should
  thus be included as ingredients in models of dwarf galaxy formation
  and evolution.
  
\end{abstract}
 

\section{Introduction}
\label{sec:introduction}

The currently favored cold dark matter (CDM) paradigm of hierarchical
structure formation \citep[e.g.,][]{White_Rees78,Blumenthal_etal84}
generically predicts that dwarf galaxies comprise the primary building
blocks of more massive systems. Observational evidence has recently
confirmed this expectation with the discovery of streams and complex
stellar structures associated with accreted and tidally disrupted
dwarfs in the Milky Way (MW)
\citep[e.g.,][]{Ibata_etal94,Yanny_etal00,Ibata_etal01a,
  Newberg_etal02,Majewski_etal03,Martin_etal04,
  Martinez-Delgado_etal05,Grillmair_Dionatos06,Belokurov_etal06}, the
Andromeda galaxy (M31)
\citep{Ibata_etal01b,Ferguson_etal02,Ferguson_etal05,
  Kalirai_etal06,Ibata_etal07}, and beyond the Local Group (LG)
\citep[e.g.,][]{Malin_Hadley97,Shang_etal98,Peng_etal02,Forbes_etal03,Pohlen_etal04}.
Understanding the formation and evolution of dwarf galaxies is crucial
for testing the predictions of the CDM theory and gaining insight into
the physical processes of structure formation.

In this context, the dwarf spheroidal galaxies (dSphs) of the LG (see
\citealt{Mateo98} and \citealt{Tolstoy_etal09} for comprehensive
reviews) constitute excellent candidates for constraining the CDM
model, as they are believed to be highly dark matter (DM) dominated,
with mass-to-light ratios of $M/L_V \sim 10^{1-3}$
\citep[e.g.,][]{Mateo98,Gilmore_etal07,Simon_Geha07,Walker_etal09}
(for alternative explanations of their extreme DM content see, e.g.,
\citealt{Kuhn_Miller89,Milgrom95,Kroupa97,Lokas01}). dSphs are the
faintest galaxies known and their stellar components are supported by
random motions, with a ratio of rotational velocity to line-of-sight,
central velocity dispersion of $V_{\rm rot}/\sigma_{\ast} \lesssim 1$
\citep{Mateo98}. Among the dwarf galaxies in the LG, dSphs are also
the most numerous and they tend to be clustered around the massive
spirals MW and M31 (though some outliers exist including the distant
dwarf Leo I and the isolated dSphs Cetus and Tucana that lie on the
outskirts of the LG). This tendency is referred to as the
morphology-density relation where rotationally-supported dwarf
irregular galaxies (dIrrs) are found in the periphery of the LG
\citep[e.g.,][]{Mateo98,Grebel99}.  Moreover, dSphs are gas poor or
completely devoid of gas \citep[e.g.,][]{Mateo98,Grcevich_Putman09}
and they exhibit a wide diversity in their star formation histories
\citep[e.g.,][]{Grebel00,Orban_etal08}.  Owing to their proximity,
dSphs have been most thoroughly studied in the LG.  However, dwarfs
with the properties of dSphs have been recently identified in nearby
groups and clusters of galaxies \citep{Chiboucas_etal09,Penny_etal09}
suggesting that these intriguing galaxies are not unique to the LG.

While our understanding of dSphs has grown impressively in the past
decade, a definitive model for their origin still remains elusive (see
\citealt{Mayer10} and \citealt{Kravtsov10} for recent reviews).  Two
main classes of models have been proposed so far to explain their
formation and present-day properties.  In the first, dSphs are the
result of the interplay between cosmic reonization and stellar
feedback suppressing gas accretion and star formation in low-mass
galaxies
\citep[e.g.,][]{Dekel_Silk86,Bullock_etal00,Susa_Umemura04,Ricotti_Gnedin05,Tassis_etal08,Sawala_etal10,Maccio_etal10}.
In the second, the origin of dSphs is intimately linked to various
environmental mechanisms, including tidal and ram pressure stripping
\citep[e.g.,][]{Einasto_etal74,Faber_Lin83,Mayer_etal01a,Mayer_etal01b,Kravtsov_etal04,
  Mayer_etal06,Mayer_etal07,Klimentowski_etal07,Penarrubia_etal08,Klimentowski_etal09a}
and resonant stripping \citep{D'Onghia_etal09}.

Using controlled simulations of individual disky dwarf galaxies
orbiting inside a MW-sized host, \citet{Mayer_etal01a,Mayer_etal01b}
demonstrated for the first time that the repeated action of tidal
forces from the primary galaxy can transform the dwarfs into
pressure-supported stellar systems with the structural and kinematic
properties of dSphs \citep[see
also][]{Klimentowski_etal07,Klimentowski_etal09a}.  This
transformation mechanism, termed ``tidal stirring'', involves a
combination of tidally-induced dynamical instabilities in stellar
disks (e.g., the bar and buckling instabilities) and impulsive tidal
heating of the stellar distribution. In the context of CDM, tidal
stirring should be particularly effective since satellites of massive
galaxies are affected by strong tidal forces due to their highly
eccentric orbits \citep[e.g.,][]{Ghigna_etal98,Diemand_etal07}.

Although the tidal stirring model naturally explains the tendency of
dSphs to be concentrated near the dominant spiral galaxies, it is only
applicable to classic dSphs, namely those that were known before the
discovery of the ultra-faint dwarfs by the Sloan Digital Sky Survey
(SDSS) \citep[e.g.,][]{Simon_Geha07}. When combined with ram pressure
stripping and the effect of radiation fields at high redshift, such as
the cosmic ionizing ultraviolet background, the tidal stirring
mechanism can also account for both the low gas fraction and the
extremely high DM content in some of the classic dSphs such as Draco
and Ursa Minor \citep{Mayer_etal06,Mayer_etal07}. Recently, the
structure and kinematics of the stellar core of the nearby Sagittarius
dwarf galaxy have also been successfully modeled within the framework
of tidal stirring \citep{Lokas_etal10a}.

In addition to controlled numerical experiments, significant
theoretical effort has been devoted to performing fully cosmological,
hydrodynamical simulations of dwarf galaxy formation
\citep[e.g.,][]{Ricotti_Gnedin05,Read_etal06a,Tassis_etal08,
  Governato_etal10,Sawala_etal10}. While pressure-supported objects
with the properties of LG dSphs have been produced in some cases,
\citep[e.g.,][]{Ricotti_Gnedin05,Sawala_etal10}, other studies have
demonstrated the formation of systems with properties similar to those
of the likely progenitors of dSphs according to the tidal stirring
model \citep{Governato_etal10}.  Successes notwithstanding and despite
the continuing increase in dynamic range, limited resolution prohibits
current cosmological simulations with hydrodynamics to address in
detail the dynamical and structural evolution of dwarf {\it satellite}
galaxies. In addition, due to numerical loss of angular momentum and
overcooling, satellites in cosmological simulations of galaxy
formation are too bright and too dense
\citep[e.g.,][]{Governato_etal07}.

Given these facts, we are motivated to explore the tidal evolution of
individual, rotationally-supported dwarf galaxies inside MW-sized
hosts using a large ensemble of controlled $N$-body simulations. While
our work is informed by many past numerical investigations aimed to
elucidate the tidal stirring scenario for the origin of dSphs
\citep{Mayer_etal01a,Mayer_etal01b,Mayer_etal02,Klimentowski_etal07,Klimentowski_etal09a},
our numerical experiments extend those of earlier studies in several
important respects. For example, unlike previous work, we employ
self-consistent, equilibrium numerical models of disky dwarfs for our
experiments. These models are derived from three-integral composite
distribution functions (DFs) and are thus superior to those
constructed from simpler approximate prescriptions that produce models
that are initially out of equilibrium \citep[e.g.,][]{Hernquist93}.
The transformation via tidal stirring is an intricate process with
several distinct stages that depends sensitively on the development of
tidally-induced dynamical instabilities (e.g., bars) and the detailed
response of the stellar distribution to tidal shocks.  For this
reason, the self-consistency of the adopted models is crucial.

Moreover, we explore a wide variety of orbital configurations and
structural parameters for the progenitor dwarf galaxies, conducting a
simulation campaign that allows the investigation of a much larger
parameter space than before. Our ultimate goal is to determine the
degree to which the evolution of rotationally-supported dwarfs is
affected by the strong tidal field of their host galaxies under a
broad range of initial conditions, and through this to establish the
generic efficiency of the tidal stirring mechanism for the origin of
dSphs.  Lastly, our numerical experiments are characterized by much
higher numerical resolution than that of previous related studies.
This fact, in conjunction with the increasing accuracy of current
observational data \citep[e.g.,][]{Walker_etal09}, offers unique
opportunities for a systematic and quantitative comparison with
observations, and we undertake such a task in a companion paper
({\L}okas et al. 2010, in preparation).

In this study, we focus on the evolution of the intrinsic, global
parameters of accreted disky dwarfs orbiting inside a MW-sized primary
galaxy. Our results firmly establish that tidal interactions between
late-type dwarf galaxies and their hosts can produce objects with
kinematic and structural properties akin to those of the {\it classic}
dSphs in the LG and similar environments under a wide variety of
initial conditions. We conclude that such tidal transformations should
be common occurrences within the currently favored cosmological
paradigm. Environmental mechanisms of the type highlighted in the
present work should thus constitute important ingredients in models of
dwarf galaxy formation and evolution.

The outline of this paper is as follows. In
\S~\ref{sec:methods_simulations}, we introduce the dwarf and host
galaxy models. In this section, we also describe the methods adopted
and the setup of the numerical experiments performed in the present
study. In \S~\ref{sec:orbital_parameters} and
\S~\ref{sec:struct_properties}, we investigate the degree to which the
orbital parameters and initial structure of the progenitor disky
dwarfs can influence their morphological and dynamical transformation
into dSphs. Implications and extensions of our findings are presented
in \S~\ref{sec:discussion}, which also includes a discussion of the
caveats of the current study and a number of promising directions for
future work. Lastly, in \S~\ref{sec:summary} we summarize our main
results and conclusions.

\section{Simulations and Methods}
\label{sec:methods_simulations}

\subsection{Models of Dwarf Galaxies}
\label{sub:dwarf_models}

The present investigation utilizes fully self-consistent, equilibrium
models of dwarf galaxies for tidal stirring experiments. This aspect
of the modeling constitutes one of the major improvements we introduce
in this study. As we discussed in the previous section, the nature of
the transformation mechanism is such that the self-consistency of the
adopted dwarf models is essential. Specifically, we employ the method
of \citet{Widrow_Dubinski05} to construct numerical realizations of
dwarf galaxies consisting of exponential stellar disks embedded in
cuspy, cosmologically-motivated \citet[][hereafter
NFW]{Navarro_etal96} DM halos. The \citet{Widrow_Dubinski05} models
are specified by a large number of parameters. They are derived from
three-integral, composite DFs and thus represent self-consistent,
equilibrium solutions to the coupled Poisson and collisionless
Boltzmann equations. The \citet{Widrow_Dubinski05} method has been
recently used in a variety of numerical studies associated with
instabilities in disk galaxies, including the dynamics of warps and
bars \citep{Dubinski_Chakrabarty09,Dubinski_etal09} and the heating of
galactic disks by halo substructure
\citep{Gauthier_etal06,Kazantzidis_etal08,Purcell_etal09,Kazantzidis_etal09}.
We refer the reader to \citet{Widrow_Dubinski05} for an overview of
all relevant parameters and a detailed description of this technique.

The density distributions of the dwarf DM halos are given by
\begin{equation}
   \rho_{\rm NFW}(r) = 
      \frac{\rho_s} {\left (r/r_s\right) 
        \left (1 + r/r_s\right )^2} \ ,
\end{equation}
where $\rho_s$ is a characteristic inner density, and $r_s$ denotes
the scale radius of the density profile defined as the distance from
the center where the logarithmic slope, $\rm d \ln \rho(r)/\rm d \ln
r$, is equal to $-2$. The NFW density profile is formally infinite in
extent with a cumulative mass that diverges as $r \rightarrow \infty$.
In order to keep the total mass finite, \citet{Widrow_Dubinski05}
impose a tidal radius in the DM halo, $R_h$, which represents the
outer edge of the system. In our modeling, $R_h$ is chosen in such a
way that it becomes roughly equal to the cosmologically-motivated
virial radius, $R_h \approx r_{\rm vir}$. As a result of this choice,
the total mass of the halo within $R_h$, which we denote $M_h$, would
be equivalent to the virial mass. We control the shape of the halo
density profile via the concentration parameter $c\equiv r_{\rm
  vir}/r_s \approx R_h/r_s$.  Higher values of concentration
correspond to a larger fraction of the mass contained in the inner
regions of the halo.  Lastly, all DM halos were constructed with no
net angular momentum.

The surface density profiles of the dwarf disks follow an exponential
distribution in cylindrical radius $R$, while the vertical structure
is modeled by constant-thickness, self-gravitating isothermal sheets
\citep{Spitzer42}
\begin{equation}
   \rho_d(R,z) \propto \exp\left(-\frac{R}{R_d}\right) 
       {\rm sech}^2\left(\frac{z}{z_d}\right) \ ,
   \label{disk_density}
\end{equation}
where $R_d$ and $z_d$ denote the radial scale length and the
(sech$^2$) vertical scale height of the disk, respectively. The
phase-space DF of the disk is fully determined once the disk velocity
ellipsoid is specified.  The radial velocity dispersion,
$\sigma_R(R)$, is assumed to be exponential with $\sigma_R^{2}(R) =
\sigma_{R0}^{2}\exp\left(-R/R_d\right)$, where $\sigma_{R0}$ denotes
the central radial velocity dispersion. The disk azimuthal dispersion,
$\sigma_\phi(R)$, is related to $\sigma_R(R)$ via the epicycle
approximation \citep{Binney_Tremaine08}, while the vertical velocity
dispersion, $\sigma_z(R)$, is set by the requirement that the adopted
value of the scale height $z_d$ is maintained in the total potential
of the galaxy model. Here we parametrize the disk thickness as
$z_d/R_d$ and the disk mass as a given fraction, $m_d$, of the halo
mass, $M_h$. For simplicity, we also assume a constant value of
$\sigma_{R0}=10\kms$ for all our dwarf models. This choice results in
a decreasing ratio of $V_{\rm rot}/\sigma_{\ast}$ with decreasing
mass, where $V_{\rm rot}$ and $\sigma_{\ast}$ denote the stellar
rotational velocity and line-of-sight central velocity dispersion, a
trend that is indeed observed in dIrrs of the LG \citep{Mateo98}.


\begin{table*}
\caption{Initial Structural Parameters of the Dwarf Galaxy Models}
\begin{center}
  \vspace*{-12pt}
\begin{tabular}{lcccccccccccccccc}
\hline
\hline 
\\
\multicolumn{1}{c}{}              &
\multicolumn{1}{c}{}              &
\multicolumn{1}{c}{$m_d$}         & 
\multicolumn{1}{c}{}              &
\multicolumn{1}{c}{}              &
\multicolumn{1}{c}{$M_h$} &
\multicolumn{1}{c}{$r_s$} &
\multicolumn{1}{c}{$R_d$}         &
\multicolumn{1}{c}{} &                   
\multicolumn{1}{c}{$V_{\rm max}$} &
\multicolumn{1}{c}{$r_{\rm max}$} &
\multicolumn{1}{c}{$M/L$}         &
\multicolumn{1}{c}{}              &
\multicolumn{1}{c}{}              &
\multicolumn{1}{c}{}              &
\multicolumn{1}{c}{}              &
\multicolumn{1}{c}{}              
\\
\multicolumn{1}{c}{Model}&
\multicolumn{1}{c}{$z_d/R_d$} &
\multicolumn{1}{c}{($M_h$)}       & 
\multicolumn{1}{c}{$\lambda$}    & 
\multicolumn{1}{c}{$c$} & 
\multicolumn{1}{c}{($10^9 M_{\odot}$)} &
\multicolumn{1}{c}{(kpc)} &  
\multicolumn{1}{c}{(kpc)}             &
\multicolumn{1}{c}{$Q$}             &
\multicolumn{1}{c}{(km s$^{-1}$)}        & 
\multicolumn{1}{c}{(kpc)}      & 
\multicolumn{1}{c}{$(M_{\odot}/L_{\odot})$}& 
\multicolumn{1}{c}{$V_{\rm rot}/\sigma_{\ast}$} & 
\multicolumn{1}{c}{$\beta$} & 
\multicolumn{1}{c}{$A_2$} &
\multicolumn{1}{c}{$b/a$} & 
\multicolumn{1}{c}{$c/a$} 
\\
\multicolumn{1}{c}{(1)}&
\multicolumn{1}{c}{(2)}&
\multicolumn{1}{c}{(3)}&
\multicolumn{1}{c}{(4)}&
\multicolumn{1}{c}{(5)}&
\multicolumn{1}{c}{(6)}&
\multicolumn{1}{c}{(7)}&
\multicolumn{1}{c}{(8)}&
\multicolumn{1}{c}{(9)}&
\multicolumn{1}{c}{(10)}&
\multicolumn{1}{c}{(11)}&
\multicolumn{1}{c}{(12)}&
\multicolumn{1}{c}{(13)}&
\multicolumn{1}{c}{(14)}&
\multicolumn{1}{c}{(15)}&
\multicolumn{1}{c}{(16)}&
\multicolumn{1}{c}{(17)}
\\
\\
\hline
\\
D1  &  0.2     &  0.02     &  0.040     &   20    &   1      &   1.29   & 0.41 & 3.93 & 19.8 & 2.07 & 29.5 & 2.93  & 0.15  &  0 & 1.00  & 0.12  \\
D2  &{\bf 0.1} &  0.02     &  0.040     &   20    &   1      &   1.29   & 0.41 & 3.90 & 19.9 & 2.12 & 30.4 & 3.04  & 0.29  &  0 & 1.00  & 0.06  \\
D3  &{\bf 0.3} &  0.02     &  0.040     &   20    &   1      &   1.29   & 0.41 & 3.91 & 19.8 & 2.19 & 31.1 & 2.85  & 0.02  &  0 & 1.00  & 0.18  \\
D4  &  0.2     &{\bf 0.01} &  0.040     &   20    &   1      &   1.29   & 0.41 & 7.60 & 19.4 & 2.44 & 64.3 & 2.85  & 0.21  &  0 & 1.00  & 0.12  \\
D5  &  0.2     &{\bf 0.04} &  0.040     &   20    &   1      &   1.29   & 0.41 & 2.05 & 21.0 & 1.69 & 14.2 & 3.06  & 0.03  &  0 & 1.00  & 0.13  \\
D6  &  0.2     &  0.02     &{\bf 0.024} &   20    &   1      &   1.29   & 0.25 & 2.25 & 19.8 & 1.99 & 27.2 & 2.71  & 0.05  &  0 & 1.00  & 0.11  \\
D7  &  0.2     &  0.02     &{\bf 0.066} &   20    &   1      &   1.29   & 0.66 & 6.32 & 19.8 & 2.32 & 35.8 & 3.06  & 0.27  &  0 & 1.00  & 0.12  \\
D8  &  0.2     &  0.02     &  0.040     &{\bf 10} &   1      &   2.58   & 0.41 & 2.97 & 16.5 & 4.54 & 42.4 & 2.07  & 0.27  &  0 & 1.00  & 0.10  \\ 
D9  &  0.2     &  0.02     &  0.040     &{\bf 40} &   1      &   0.65   & 0.41 & 5.00 & 24.8 & 1.19 & 32.6 & 3.78  & 0.13  &  0 & 1.00  & 0.13  \\
D10 &  0.2     &  0.02     &  0.040     &  20     &{\bf 0.2} &   0.76   & 0.24 & 7.95 & 11.6 & 1.29 & 31.2 & 1.38  & 0.58  &  0 & 1.00  & 0.12  \\
D11 &  0.2     &  0.02     &  0.040     &  20     &{\bf 5}   &   2.21   & 0.70 & 2.14 & 32.0 & 3.52 & 25.9 & 4.38  & -0.51 &  0 & 1.00  & 0.12  \\
\hline
\end{tabular}
\end{center}
{\sc Notes.}---The quantities in each column are as follows.
Column 1: Dwarf galaxy model.
Column 2: Scale height of the disk of the dwarf in units of the disk radial scale length, $R_d$.
Column 3: Mass of the disk of the dwarf in units of the mass of the dwarf halo, $M_h$. 
Column 4: Spin parameter of the DM halo of the dwarf used to determine the scale length 
of the dwarf disk (see text for details). The corresponding disk scale lengths are listed in column  8.
Column 5: Concentration parameter of the DM halo of the dwarf.
Column 6: Mass of the DM halo of the dwarf in units of $10^9 M_{\odot}$. This parameter is equivalent to 
the cosmologically-motivated virial mass.
Column 7: Scale radius of the DM halo of the dwarf in kpc. 
Column 8: Radial scale length of the disk of the dwarf in kpc.
Column 9: Toomre stability parameter of the dwarf disk at $R=2.5 R_d$.
Column 10: Maximum circular velocity of the dwarf in $\kms$.
Column 11: Radius at which the maximum circular velocity occurs in kpc.
Column 12: Mass-to-light ratio of the dwarf in units of $M_{\odot}/L_{\odot}$.
Column 13: Ratio of stellar rotational velocity to one-dimensional stellar velocity dispersion of the dwarf.
Column 14: Anisotropy parameter of the stellar distribution of the dwarf.
Column 15: Amplitude of the $m = 2$ Fourier component of the surface density distribution of the dwarf stars.
Column 16: Axis ratio $b/a$ of the stellar component of the dwarf.
Column 17: Axis ratio $c/a$ of the stellar component of the dwarf.
See \S~\ref{subsec:parameters} for details on how the parameters listed in columns 12-17 are calculated.
Note that the entries in these columns are computed within $r_{\rm max}$. 
\label{table:init_param}
\end{table*}


Our goal is to assess the degree to which the structure of late-type
disky dwarfs can influence their tidal evolution inside their host
galaxies. For this reason, we generated a number of dwarf galaxy
models that differed in a number of important structural parameters,
including the disk thickness, mass, and scale length, and the halo
mass and concentration parameter. To investigate the impact of all
these quantities on the tidal evolution of the dwarfs, we first
constructed a ``reference'' or ``default'' dwarf galaxy model and
subsequently initialized additional models by varying (increasing and
decreasing) all relevant parameters in a systematic way, modifying
only one at a time.  Throughout the paper, we compare the effect of
changing a single parameter in three distinct dwarf models, namely the
reference one and those with the largest and smallest values of a
given parameter. We also stress that we do not attempt any explicit
rescaling of our dwarf galaxy models with redshift to account for the
cosmic epoch at which the dwarfs were accreted by their hosts, as was
done in previous studies \citep{Mayer_etal01a,Mayer_etal01b}.  On the
one hand, this allows us to avoid any uncertainties regarding the
applicability of such scalings to dwarf galaxies. On the other hand,
our systematic parameter survey aims to address different choices for
the basic structural parameters of the progenitors of present-day
dSphs. As such, it should account to at least a certain extent for the
wide diversity of dwarf galaxy structures expected at various
cosmological epochs.

In total, we constructed $11$ high-resolution numerical models of
rotationally-supported dwarfs, which we denote D1-D11.  The initial
structural parameters of these models are listed in
Table~\ref{table:init_param} (see \S~\ref{subsec:parameters} for
details on how the parameters corresponding to columns (12)-(17) of
this table are calculated). Particular emphasis should be placed on
column 4 of this table which lists the dimensionless spin parameters
of the dwarf DM halos, $\lambda$. This parameter is a measure of the
total energy content of a DM halo stored in rotation and it is defined
as $\lambda \equiv J |E|^{1/2}/ G M^{5/2}$ \citep{Peebles69}, where
$J$ is the total angular momentum, $E$ is the binding energy, and $M$
is the mass of the halo. Although our halos are non-rotating, we still
employ the parameter $\lambda$ to assign scale lengths to the dwarf
disks. This is because the disk scale length is not a free parameter
in our modeling, but rather is derived via the semi-analytic model of
\citet{Mo_etal98} for the structure of disk galaxies in the {\LCDM}
paradigm\footnote{In reality, the \citet{Mo_etal98} formalism and its
  assumptions for the formation of galactic disks may be inappropriate
  at the scales of dwarf galaxies.  This is due to the greater
  importance of thermal over rotational support in such low-mass
  systems which modify the simple analytic scalings of
  \citet{Mo_etal98}.  Nonetheless, given that our purpose is not to
  derive exact scale lengths for our dwarf galaxies but simply to
  determine the degree to which the tidal evolution of disky dwarfs is
  affected by the sizes of their disks, following \citet{Mo_etal98} is
  reasonable and does not bias our results in any way.}. According to
this model, the baryons settle into a rotationally supported structure
whose scale length is determined by the mass, spin parameter and
concentration of the DM halo, and the fraction of mass and angular
momentum in baryons relative to that of the halo.  Assuming that the
specific angular momentum of baryons is conserved during their infall
and that the halo and baryons start with the same specific angular
momentum, $R_d$ is uniquely determined in our models by $M_h$,
$\lambda$, $c$, and $m_d$.  We note that the \citet{Widrow_Dubinski05}
method for building disk galaxies does not explicitly take into
account the adiabatic contraction of the halo in response to the slow
accumulation of the baryons \citep[e.g.,][]{Blumenthal_etal86}.
Therefore, in order to be consistent, we derive the values for the
disk scale lengths without considering this effect. In summary, the
values of $\lambda$ that we discuss throughout this study do not
reflect the angular momentum content of the DM halos, but rather serve
the practical purpose of enabling us to derive the values of $R_d$ in
the dwarf disks according to \citet{Mo_etal98}.

Our reference dwarf model D1 is characterized by the following values
for the adopted set of parameters: $z_d/R_d = 0.2$, $m_d=0.02$,
$\lambda=0.04$, $c=20$, and $M_h = 10^9 M_{\odot}$. The details of how
we varied these values in our simulation campaign are described in
each relevant subsection of \S~\ref{sec:struct_properties}. In what
follows, we motivate the choices for the default values of the various
parameters in dwarf model D1. Our choices are mainly guided by studies
of the properties of cosmological halos.

Using a constrained simulation of the LG, \citet{Klimentowski_etal10}
studied the distribution of subhalo masses at the time of infall onto
the primary as a function of redshift. These authors found that a
significant fraction of satellites that were accreted by their hosts
since $z \lesssim 2$ and survived until the present time had masses of
$\approx 10^9 M_{\odot}$ (see Figure~8 of
\citealt{Klimentowski_etal10}). Assuming a concordance {\LCDM} model,
the median concentration value for a $z=0$ cosmological halo at this
mass scale is $c \approx 20$
\citep[e.g.,][]{Bullock_etal01a,Maccio_etal07}.  In addition, both
observational and theoretical evidence suggests that dwarf galaxies
are not formed as thin disks, but rather are born as thick, puffy
systems \citep[e.g.,][]{Dalcanton_etal04,Kaufmann_etal07}. This is a
consequence of the greater importance of feedback processes and
turbulent motions in dwarf galaxies. We take this expectation into
account by conservatively adopting $z_d/R_d = 0.2$, instead of the
typical value $z_d/R_d = 0.1$ employed throughout the literature which
would be more appropriate for massive disk galaxies
\citep[e.g.,][]{Kregel_etal02}.

As has been established by a large number of studies, the distribution
of halo spin parameters in $N$-body simulations is well described by a
log-normal distribution, with median values of $\lambda_{\rm med}
\approx 0.04$ and dispersions of $\sigma_{\lambda} \sim 0.5$
\citep[e.g.,][]{Bullock_etal01b,Shaw_etal06,Maccio_etal07,Bett_etal07,Maccio_etal08}.
According to \citet{Maccio_etal07} (see also \citealt{Bett_etal07} and
\citealt{Maccio_etal08}), the distribution of halo spins shows no
dependence on halo mass. We stress that this conclusion does hold for
halos with masses of the order of $\sim 10^9 M_{\odot}$ that we
consider here (A.  Macci\`{o} 2010, private communication). Lastly, we
choose the default value for the disk mass fraction of our dwarfs
equal to $m_d =0.02$. This value is much lower than the universal
baryon fraction but quite typical for present-day low surface
brightness (LSB) or dIrr galaxies
\citep[e.g.,][]{Jimenez_etal03,Geha_etal06,Oh_etal08}, and is also in
agreement with results of hydrodynamical simulations of dwarf galaxy
formation \citep[e.g.,][]{Tassis_etal03,Governato_etal10}.

For each dwarf galaxy model, we generated an $N$-body realization
containing a total of $2.2$ million particles ($N_h = 10^6$ DM
particles and $N_d = 1.2 \times 10^6$ disk particles). The
gravitational softening was set to $\epsilon_h=60$~pc and
$\epsilon_d=15$~pc for the particles in the two components,
respectively. The process of transforming a disky dwarf into a dSph
via tidal stirring is fairly complex and depends on a number of subtle
dynamical effects \citep[e.g.,][]{Mayer_etal01a}. It is thus important
to establish both the quality and the sufficient resolution of our
dwarf galaxy models.  For this reason, we evolved all dwarf galaxies
in isolation for a period of $10$~Gyr.  These test simulations
revealed that the dwarfs retained their equilibrium configuration
within the adopted force resolution over the timescales of the
experiments. Therefore, our models should be largely unaffected by
both two-body relaxation and artificial numerical heating of the dwarf
disk through interactions with the massive particles of the dwarf
halo.  The same simulations also confirmed the stability of all dwarf
galaxy models against bar formation.  Thus, any significant bar growth
identified in the dwarfs during the course of the numerical
experiments should be the result of the tidal field of the host
galaxy, rather than a consequence of noise present in the initial
conditions. All of these precautions imply that our dwarf galaxy
models should indeed be adequate to resolve the generic tidal
evolution of late-type disky dwarfs and to elucidate their
transformation via tidally-induced dynamical instabilities and
impulsive tidal heating.

\subsection{Primary Galaxy Model}
\label{sub:galaxy_model}

Another improvement we introduce in the present study is the fact that
we employ self-gravitating primary galaxies as opposed to static host
potentials adopted in the majority of earlier related investigations.
The motivation behind this choice is twofold.  First and most
importantly, live primary galaxies can trigger instabilities in the
disks of the dwarfs, that otherwise may not develop, and which could
influence the dynamical and morphological evolution of the dwarf
galaxies themselves \citep{Weinberg_Blitz06}. Second, by representing
the host galaxies as a distribution of interacting particles, we
enable the dwarf galaxies to suffer dynamical friction. As a result of
a progressively decaying orbit, the mass loss and tidal stripping
experienced by the dwarfs will be enhanced with obvious consequences
for their dynamical evolution inside the host.

For simplicity, we assume a single primary galaxy with the present-day
structural properties of the MW. In particular, we employ model MWb of
\citet{Widrow_Dubinski05}, which satisfies a broad range of
observational constraints for the MW galaxy. Specifically, the
exponential stellar disk has a mass of $M_D = 3.53 \times 10^{10}
M_{\odot}$, a radial scale length of $R_D=2.82$~kpc, and a sech$^2$
scale height of $z_D=400$~pc. The bulge which follows the
\citet{Hernquist90} density profile has a mass and a scale radius of
$M_B=1.18 \times 10^{10} M_{\odot}$ and $a_B=0.88$~kpc, respectively.
The DM halo has an NFW profile with a tidal radius of $R_H=244.5$~kpc,
a mass of $M_H=7.35 \times 10^{11}M_{\odot}$, and a scale radius of
$r_H=8.82$~kpc.

The simulations reported here use $N_D=10^{6}$ particles in the disk,
$N_B=5\times10^{5}$ in the bulge, and $N_H=2\times10^{6}$ in the DM
halo of the host galaxy, and employ a gravitational softening of
$\epsilon_D=50$~pc, $\epsilon_B=50$~pc, and $\epsilon_H=2$~kpc,
respectively. The choice for the fairly large softening in the DM
particles of the primary galaxy was motivated by our desire to
minimize discreteness noise in the host potential. Such noise may lead
to spurious two-body heating between the excessively massive halo
particles of the primary galaxy and those of the dwarf disk,
potentially interfering with the interpretation of our results.  In
order to confirm the adequacy of our choice for $\epsilon_H$, we
placed the reference dwarf model D1 at rest at various distances
inside the host galaxy and monitored the changes of its basic
properties (e.g., surface density, velocity dispersions, thickness) as
a function of time. The evolution of the structural parameters within
radii of interest (see \S~\ref{subsec:parameters} below) was found to
be fairly small, specifically $\lesssim 20\%$, during a period of
several Gyr. This indicates that the chosen numerical parameters are
indeed appropriate to suppress the effect of two-body heating between
the halo particles of the primary galaxy and those of the dwarf disk.


\begin{table*}
\caption{Summary of Simulations}
\begin{center}
  \vspace*{-12pt}
\begin{tabular}{lcccccccccc}
\hline
\hline
\\
\multicolumn{1}{c}{}              &
\multicolumn{1}{c}{}              &
\multicolumn{1}{c}{$r_{\rm apo}$}   &
\multicolumn{1}{c}{$r_{\rm peri}$}   &
\multicolumn{1}{c}{}             &
\multicolumn{1}{c}{$i$}      &
\multicolumn{1}{c}{}              &
\multicolumn{1}{c}{$m_d$}   &
\multicolumn{1}{c}{}              &
\multicolumn{1}{c}{}              &
\multicolumn{1}{c}{$M_h$}  
\\
\multicolumn{1}{c}{Simulation}    &
\multicolumn{1}{c}{Dwarf Model}    &
\multicolumn{1}{c}{(kpc)}         & 
\multicolumn{1}{c}{(kpc)}         & 
\multicolumn{1}{c}{$r_{\rm apo}/r_{\rm peri}$}             & 
\multicolumn{1}{c}{(deg)}         & 
\multicolumn{1}{c}{$z_d/R_d$}             & 
\multicolumn{1}{c}{($M_h$)}       & 
\multicolumn{1}{c}{$\lambda$} & 
\multicolumn{1}{c}{$c$} & 
\multicolumn{1}{c}{($10^9 M_{\odot}$)} 
\\
\multicolumn{1}{c}{(1)}&
\multicolumn{1}{c}{(2)}&
\multicolumn{1}{c}{(3)}&
\multicolumn{1}{c}{(4)}&
\multicolumn{1}{c}{(5)}&
\multicolumn{1}{c}{(6)}&
\multicolumn{1}{c}{(7)}&
\multicolumn{1}{c}{(8)}&
\multicolumn{1}{c}{(9)}&
\multicolumn{1}{c}{(10)}&
\multicolumn{1}{c}{(11)}
\\
\\
\hline
\\
R1  & D1  &    125   &    25      &     5     &    45   &    0.2    &     0.02   &   0.040    &   20    &    1     \\
R2  & D1  &{\bf 85}  & {\bf 17}   &     5     &    45   &    0.2    &     0.02   &   0.040    &   20    &    1     \\
R3  & D1  &{\bf 250} & {\bf 50}   &     5     &    45   &    0.2    &     0.02   &   0.040    &   20    &    1     \\
R4  & D1  &    125   & {\bf 12.5} & {\bf 10}  &    45   &    0.2    &     0.02   &   0.040    &   20    &    1     \\
R5  & D1  &    125   & {\bf 50}   & {\bf 2.5} &    45   &    0.2    &     0.02   &   0.040    &   20    &    1     \\
R6  & D1  &    125   &    25      &     5     & {\bf 0} &    0.2    &     0.02   &   0.040    &   20    &    1     \\
R7  & D1  &    125   &    25      &     5     &{\bf 90} &    0.2    &     0.02   &   0.040    &   20    &    1     \\
R8  & D2  &    125   &    25      &     5     &    45   &{\bf 0.1}  &     0.02   &   0.040    &   20    &    1     \\
R9  & D3  &    125   &    25      &     5     &    45   &{\bf 0.3}  &     0.02   &   0.040    &   20    &    1     \\
R10 & D4  &    125   &    25      &     5     &    45   &    0.2    &{\bf 0.01}  &   0.040    &   20    &    1     \\
R11 & D5  &    125   &    25      &     5     &    45   &    0.2    &{\bf 0.04}  &   0.040    &   20    &    1     \\
R12 & D6  &    125   &    25      &     5     &    45   &    0.2    &     0.02   &{\bf 0.024} &   20    &    1     \\
R13 & D7  &    125   &    25      &     5     &    45   &    0.2    &     0.02   &{\bf 0.066} &   20    &    1     \\
R14 & D8  &    125   &    25      &     5     &    45   &    0.2    &     0.02   &   0.040    &{\bf 10} &    1     \\
R15 & D9  &    125   &    25      &     5     &    45   &    0.2    &     0.02   &   0.040    &{\bf 40} &    1     \\
R16 & D10 &    125   &    25      &     5     &    45   &    0.2    &     0.02   &   0.040    &   20    &{\bf 0.2} \\
R17 & D11 &    125   &    25      &     5     &    45   &    0.2    &     0.02   &   0.040    &   20    & {\bf 5}  \\
\hline
\end{tabular}
\end{center}
{\sc Notes.}---Columns 3-5 refer to the initial orbital parameters of the disky dwarfs. Columns 7-11 list 
the initial structural parameters of the dwarf galaxy models. The quantities in each column are as follows.
Column 1: Abbreviation for the tidal stirring simulations.
Column 2: Dwarf galaxy model.
Column 3: Apocentric distance of the orbit of the dwarf in kpc.
Column 4: Pericentric distance of the orbit of the dwarf in kpc.
Column 5: Eccentricity of the orbit of the dwarf.
Column 6: Inclination of the disk of the dwarf with respect to the orbital plane in degrees.
Column 7: Scale height of the disk of the dwarf in units of the disk radial scale length, $R_d$.
Column 8: Mass of the disk of the dwarf in units of the mass of the dwarf halo, $M_h$. 
Column 9: Spin parameter of the DM halo of the dwarf used to determine the scale length 
of the dwarf disk (see text for details).The corresponding disk scale lengths are listed in column 8 
of Table~\ref{table:init_param}.
Column 10: Concentration parameter of the DM halo of the dwarf.
Column 11:  Mass of the DM halo of the dwarf in units of $10^9 M_{\odot}$. This parameter is equivalent to 
the cosmologically-motivated virial mass.
\label{table:simulations}
\end{table*}


\subsection{Description of Tidal Stirring Simulations}
\label{subsec:sims}

Our simulation campaign comprised $17$ numerical experiments, which we
denote R1-R17, of the tidal interactions between
rotationally-supported dwarf galaxies and their hosts.
Table~\ref{table:simulations} provides a summary of all simulations we
performed in this study. The various dwarf galaxy models are placed on
bound orbits inside the primary galaxy. Except for varying the initial
structure of the disky dwarf galaxies, we also investigated the degree
to which their tidal evolution is affected by the orbital parameters
and, in particular, on the sizes and the eccentricities of the dwarf
orbits.

The size of the orbit is expected to be crucial for the efficiency of
tidal stirring, since the tidal force exerted by the primary galaxy
depends strongly on the distance from its center.  To explore this
effect, we conducted a set of simulations where the default dwarf
galaxy model D1 is placed on three orbits with different sizes.
Although the orbits of LG dwarfs are currently poorly constrained
observationally, their current distances, which give an indication of
the apocenters of their orbits, coupled with studies of the orbital
properties of cosmological halos, can be used to inform our choices.
Indeed, using a cosmological simulation of the LG,
\citet{Klimentowski_etal10} studied the orbital distribution of
present-day satellites that were identified inside the virial radius
of their hosts and had completed at least one orbit around them.
These authors found that a significant fraction of this satellite
population had apocentric distances between $r_{\rm apo} \approx 250$
and $\approx 85$~kpc, with typical values of $r_{\rm apo} \approx
125$~kpc (see Figure~6 of \citealt{Klimentowski_etal10}). Adopting
$R_{\rm vir} \approx 250$~kpc as the virial radius of the MW halo at
$z=0$ \citep[e.g.,][]{Klypin_etal02}, the previous numbers would
roughly correspond to $R_{\rm vir}$, $R_{\rm vir}/3$, and $R_{\rm
  vir}/2$.  The value $r_{\rm apo} = 125$~kpc also approximately
matches the virial radius of the MW halo at $z\sim 1$
\citep[e.g.,][]{Wechsler_etal02}. A critical reader may note that the
characteristic orbits of the $z=0$ subhalo population that we have
considered may not be representative of those at the time of satellite
infall onto the primary. This is especially true for systems that are
accreted at high redshift ($z \gtrsim 1$). Given the complexity of
halo formation in a cosmological context, this is a valid concern.
While addressing this issue is certainly beyond the scope of the
present paper, we note that at least dynamical friction should not
have a major effect, if any, in altering the orbits of satellites with
masses of the order of $\sim 10^9 M_{\odot}$ that we have adopted here
\citep{Colpi_etal99}.

Motivated by the previous discussion, we adopt $r_{\rm apo} = 125$,
$85$, and $250$~kpc as the apocentric distances of the dwarfs in the
experiments with different sizes of orbits. We refer to these
simulations as ``R1'', ``R2'', and ``R3'', respectively, and discuss
them in \S~\ref{subsec:orbit_size}. Fixing the eccentricity of the
orbits will determine the corresponding pericentric distances.
Specifically, we adopt an eccentricity of $r_{\rm apo}/r_{\rm
  peri}=5$, close to the median ratio of apocentric to pericentric
radii found in cosmological $N$-body simulations
\citep[e.g.,][]{Ghigna_etal98,Diemand_etal07}.  Interestingly, the
pericentric distance of the tightest orbit with $r_{\rm peri} =
17$~kpc roughly corresponds to that inferred for the Sagittarius dwarf
galaxy \citep[e.g.,][]{Law_etal05}. In addition, the pericentric
distance of the most extended orbit with $r_{\rm peri} = 50$~kpc is
similar to that of the Large Magellanic Cloud (LMC), provided that the
LMC has just crossed the pericenter of its orbit around the MW
\citep{Besla_etal07}.  Given the typical parameters associated with
experiment R1, throughout this work we refer to it as the
``reference'' or ``default'' simulation that we use as the basis for
the comparison with other experiments.

Apart from the size of the orbit, we also investigate the impact of
the orbital eccentricity on the tidal transformation of the dwarfs.
Cosmological simulations
\citep[e.g.,][]{Ghigna_etal98,Diemand_etal07,Klimentowski_etal10}.
indicate that satellite orbits range from nearly circular ($r_{\rm
  apo}/r_{\rm peri} \approx 1$) to highly eccentric ($r_{\rm
  apo}/r_{\rm peri} \gtrsim 10$). On high-eccentricity orbits, the
effective duration of the tidal shock becomes so short that the
response of the system is prone to be impulsive rather than adiabatic
and, as a result, tidal heating is particularly efficient
\citep[e.g.,][]{Gnedin_etal99}. To this end, we considered two
additional simulations in which we placed the dwarf galaxy model D1 on
a highly and a mildly eccentric orbit, respectively. In particular, we
considered orbits with eccentricities that were by a factor of $2$
larger ($r_{\rm apo}/r_{\rm peri} = 10$) and smaller ($r_{\rm
  apo}/r_{\rm peri} = 2.5$) compared to that of R1. We refer to these
experiments as ``R4'' and ``R5'' (see Table~\ref{table:simulations})
and discuss them in \S~\ref{subsec:orbit_eccentricity}.  In order to
meaningfully compare the results of this set of simulations, we kept
the apocentric distance of the reference experiment R1 constant
($r_{\rm apo} = 125$~kpc). The desired value of eccentricity is then
achieved by simply varying the pericentric distance, $r_{\rm peri}$.
This modeling results in a pericentric distance of $r_{\rm peri} =
12.5$~kpc in simulation R4, which is the smallest in all experiments
we performed in the present study.

Initial conditions for the tidal stirring experiments were generated
by building models of dwarf galaxies and placing them at the
apocenters of their orbits.  In the coordinate system chosen to
describe the simulations, the orbital plane is the $xy$ plane, and the
center of mass of the combined system of dwarf and primary coincides
with the coordinate origin. We stress that we did not impose any
truncation in the density distribution of the dwarfs (i.e.,, at the
corresponding Jacobi tidal radius) to reflect their placement in the
host tidal field. In addition, the dwarf galaxies are not grown
adiabatically in their orbits, but rather are introduced in the
simulations directly.  We have explicitly checked that this set of
assumptions does not bias our results.  Indeed, the initial starting
positions of the dwarfs correspond to a rather low density in the host
galaxy. This results in a fairly weak tidal perturbation that does not
affect the very inner regions of the dwarf, which constitute the
primary target of the present study.

According to the LG cosmological simulation of
\citet{Klimentowski_etal10}, most satellites that survived until the
present time and whose masses at the epoch of infall onto the primary
were in the range that we consider here (around $10^9 M_{\odot}$),
entered their hosts at $z \lesssim 2$. Motivated by this finding, we
follow the tidal evolution of the dwarfs inside their host galaxies
for $10$~Gyr.  In the great majority of our experiments, the
alignments of the internal angular momentum of the dwarf, that of the
primary disk and the orbital angular momentum were all mildly prograde
and equal to $45 \degrees$.  We discuss the implications of this
choice in \S~\ref{subsec:caveats_directions} but emphasize at the
outset that our results should not be affected by any strong coupling
of angular momenta.

Lastly, all numerical experiments discussed in this work were
performed with PKDGRAV, a multistepping, parallel, tree $N$-body code
\citep{Stadel01}. In all experiments, we set the base time step to be
equal to $1\%$ of the dynamical time at the half-mass radius of the
dwarf model and allowed the individual particle time steps to be at
most a factor of $2^{20}$ smaller. The time integration was performed
with high enough accuracy such that the total energy was conserved to
better than $0.5\%$ in all cases, which is adequate for the type of
study that we undertake in this paper. As we have already stated, our
main goal is to investigate the dynamical evolution of the dwarfs
which is driven by the response of their inner regions to the tidal
shocks. The total energy contained in the inner parts of our dwarf
models is a few percent of that of the entire system, so the accuracy
of the energy conservation must be at least comparable to that in
order to resolve meaningfully the dynamics of the region of interest.

\subsection{Parameters of Relevance}
\label{subsec:parameters}

In order to illustrate the evolution of the intrinsic, global
properties of the dwarf galaxies as they orbit inside their hosts, we
calculated a set of parameters as a function of time.  Specifically,
we examined the degree to which the mass, size, shape, and kinematics
of the dwarfs is affected subject to the strong tidal field of the
primary galaxy.

For each simulation output, we first constructed the circular velocity
profile of the dwarf galaxy, $V_c (r) = [G M(<r)/r]^{1/2}$, where
$M(<r)$ is the spherically-averaged total mass profile about its
center. As a measure of how the mass of the dwarf is affected at any
stage, we employed the maximum circular velocity, $V_{\rm max}$, which
reflects the mass distribution in the inner regions and is a commonly
adopted quantity throughout the literature. The attractive feature of
$V_{\rm max}$ is that it is well defined and allows us to overcome
difficulties regarding the determination of the tidal radius
\citep[e.g.,][]{Read_etal06b}.

In order to demonstrate how the size of the dwarf galaxy is affected
by tides, we investigated the evolution of the radius at which $V_{\rm
  max}$ occurs, $r_{\rm max}$. Following \citet{Klimentowski_etal09a},
we chose $r_{\rm max}$ as the characteristic scale where we computed
the remaining properties of the dwarfs.  This radius is large enough
to include a significant fraction of the mass of the dwarf at all
times, while at the same time is small enough not to be affected by
the tidal tails
\citep[see][]{Klimentowski_etal07,Klimentowski_etal09b}.  Due to the
fact that tidal stripping truncates the dwarf galaxies at increasingly
smaller physical radii, this characteristic radius decreases with
time, from a few kpc initially down to a fraction of a kpc at the end
of the evolution.

Because of the copious amounts of DM in dSphs, another quantity of
particular interest is the mass-to-light ratio, $M/L$. Since the bulk
of the stellar populations in these galaxies is old ($t \gtrsim
8-10$~Gyr; see, e.g., \citealt{Grebel00}), we chose to calculate the
$M/L$ ratio by conservatively adopting a fixed stellar mass-to-light
ratio of $(M/L)_{\ast} = 3 M_{\odot}/L_{\odot}$
\citep[e.g.,][]{Schulz_etal02}.  We note that while such values for
$(M/L)_{\ast}$ should be appropriate for the present time, they may
not be necessarily representative for the whole evolution, as the
dwarfs might undergo periodic bursts of star formation triggered by
tidal compression at pericentric passages \citep{Mayer_etal01a}.
However, for dwarf galaxies accreted by their hosts as early as
$10$~Gyr ago, as assumed here, most of the gas should be quickly
stripped by ram pressure aided by the cosmic ionizing background
radiation \citep{Mayer_etal07}. As a result, star formation should
cease soon after the dwarfs are accreted by the host galaxy. In
addition, using the stellar population synthesis models of
\citet{Bruzual_Charlot93}, \citet{Mayer_etal01a} showed that even when
a burst of star formation occurs at first pericentric approach, the
final $(M/L)_{\ast}$ ratio would correspond to that of an old stellar
population after $7-8$~Gyr of evolution inside the primary potential.
This is because the effect of the newly-formed population on the
$(M/L)_{\ast}$ ratio and on the color of the dwarf galaxy diminishes quite
rapidly. The above discussion suggests that the assumption of
$(M/L)_{\ast} = 3 M_{\odot}/L_{\odot}$ in the present study is fairly
reasonable.

Obviously, one of the most representative properties of dSphs is their
spheroidal shape. We quantified the shape of the stellar component by
calculating the moments of the inertia tensor for all stars within
$r_{\rm max}$ and deriving principal axis ratios $b/a$ and $c/a$,
where $a$, $b$, and $c$ denote the major, intermediate, and minor axis
of the stellar distribution, respectively. Another interesting
quantity which can illuminate the dynamical and morphological
evolution of the dwarf galaxies inside their hosts, is the amplitude
of the $m = 2$ Fourier component of the stellar surface density
distribution, $A_2$, given by
\begin{equation}
   A_2 = \frac{1}{N} \left| \sum_{j=1}^{N} e^{2 i \phi_j}\right| \ .
   \label{Fourier}
\end{equation}
Here $\phi_j$ denotes the two-dimensional cylindrical polar angle
coordinate of particle $j$ projected onto the $xy$ plane (along the
shortest axis) and the summation is performed over all stars within
$r_{\rm max}$. The motivation behind studying the evolution of $A_2$
is twofold. First, it can be used as an auxiliary measure of the shape
of the stellar component since the triaxiality parameter, $T = (a^2 -
b^2) / (a^2 - c^2)$ \citep{Franx_etal91}, can be very noisy for
systems close to spherical symmetry. We note, however, that $A_2$ can
still be ambiguous as its value is close to zero for both nearly
spherical and disky shapes. Second, as we discuss below, in the
majority of cases we studied, the strong tidal shocks at pericentric
passages trigger the formation of bars in the disks of the dwarfs.
The parameter $A_2$ will allow us to determine when a bar forms and to
quantify the strength of the bar instability.

For the purposes of the present study, we designate the formation of a
bar when the amplitude of the $m = 2$ Fourier component satisfies $A_2
\gtrsim 0.2$ between two consecutive pericentric passages, namely for
one full orbital period of the dwarf around its host. These conditions
ensure that any tidally-induced bars that we detect in our simulations
will be relatively strong (see, e.g., \citealt{Debattista_etal06} for
even weaker bars in isolated disk galaxies). They also guarantee that
any temporary increase of $A_2$, when the dwarfs are strongly
elongated by tidal forces at the pericenters of the orbit, will not be
ascribed to a true bar instability.

dSphs are low-angular momentum systems, as highlighted by their low
ratio of rotational velocity to line-of-sight central velocity
dispersion, $V_{\rm rot}/\sigma_{\ast} \lesssim 1$
\citep[e.g.,][]{Mateo98}.  It is thus important to investigate the
evolution of the kinematics in the tidally-stirred disky dwarfs.  To
this end, after determining the directions of the principal axes of
the stellar component in each output, we introduced a spherical
coordinate system ($r$, $\theta$, $\phi$) so that the $z$-axis is
along the shortest axis of the stellar distribution and the angle
$\phi$ is computed on the $xy$ plane. We measured the kinematics of
the dwarfs within $r_{\rm max}$ by calculating the rotational velocity
around the shortest axis $V_{\rm rot}=V_{\phi}$ and the dispersions
$\sigma_r$, $\sigma_{\theta}$ and $\sigma_{\phi}$ around the mean
values. We combined these dispersions into the one-dimensional stellar
velocity dispersion parameter, $\sigma_{\ast} \equiv [(\sigma_r^2 +
\sigma_{\theta}^2 + \sigma_{\phi}^2)/3]^{1/2}$, which we adopted as a
measure of the amount of random motions in the stars. We note that
throughout this paper we compare the derived values of $\sigma_{\ast}$
with those of observed dSphs. Defining $\sigma_{\ast}$ as an average
over three directions makes such comparisons more meaningful, since
the observed one-dimensional stellar velocity dispersions depend on
the random line-of-sight.

Another illuminating quantity related to stellar kinematics is the
anisotropy parameter, $\beta \equiv 1 - \sigma_t^2/2 \sigma_r^2$,
where $\sigma_r$ and $\sigma_t = (\sigma_\theta^2 +
\sigma_\phi^2)^{1/2}$ denote the radial and tangential velocity
dispersion, respectively.  This quantity describes the degree of
anisotropy of the velocity distribution by indicating the amount of
radial ($\beta=1$) versus circular ($\beta \rightarrow - \infty$)
orbits in the stellar population. We stress that this parameter cannot
be measured directly from observations, but only determined by
dynamical modeling \citep[e.g.,][]{Lokas02}. Given that a well
established range for the velocity anisotropy of observed dSphs does
not yet exist, our simulations can provide useful predictions for the
possible values of $\beta$ in such systems.


\begin{figure*}
\centerline{\epsfxsize=5.5in \epsffile{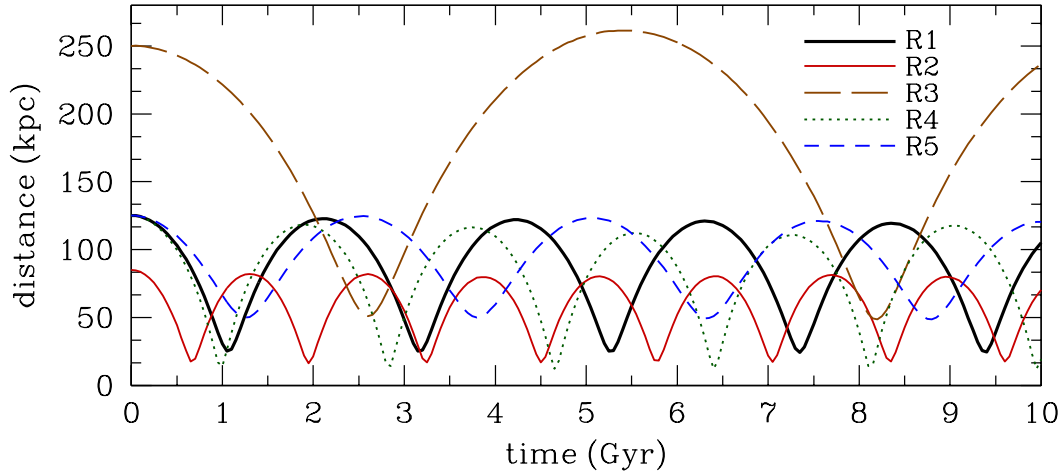}}
\caption{Distances of the orbiting dwarf galaxies from the centers
  of their hosts as a function of time. Results are presented for
  simulations R1-R5, where the dwarfs are characterized by different
  orbital parameters.
\label{fig1}}
\end{figure*}



\begin{figure*}[t]
\begin{center}
  \includegraphics[scale=0.44]{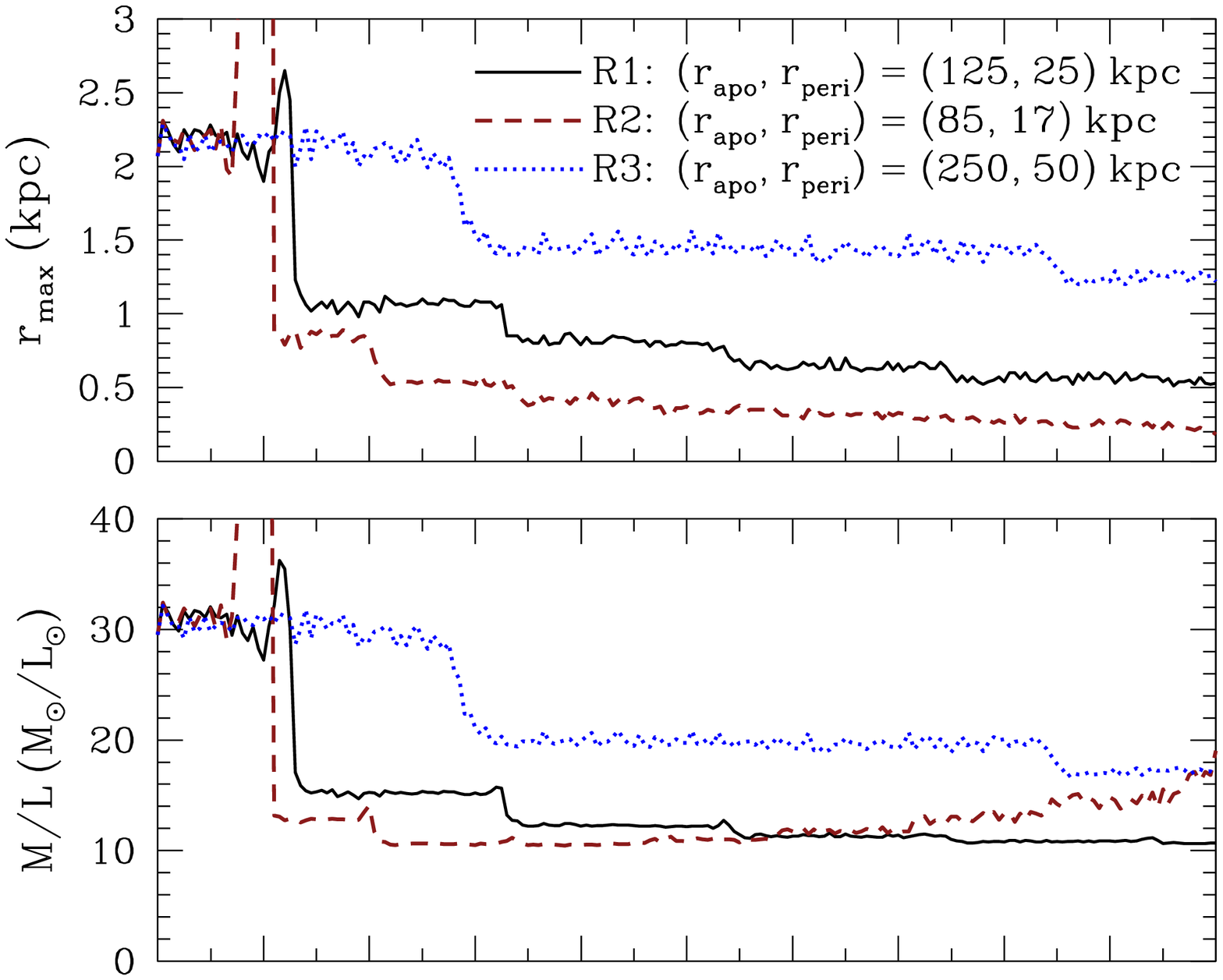}
  \includegraphics[scale=0.44]{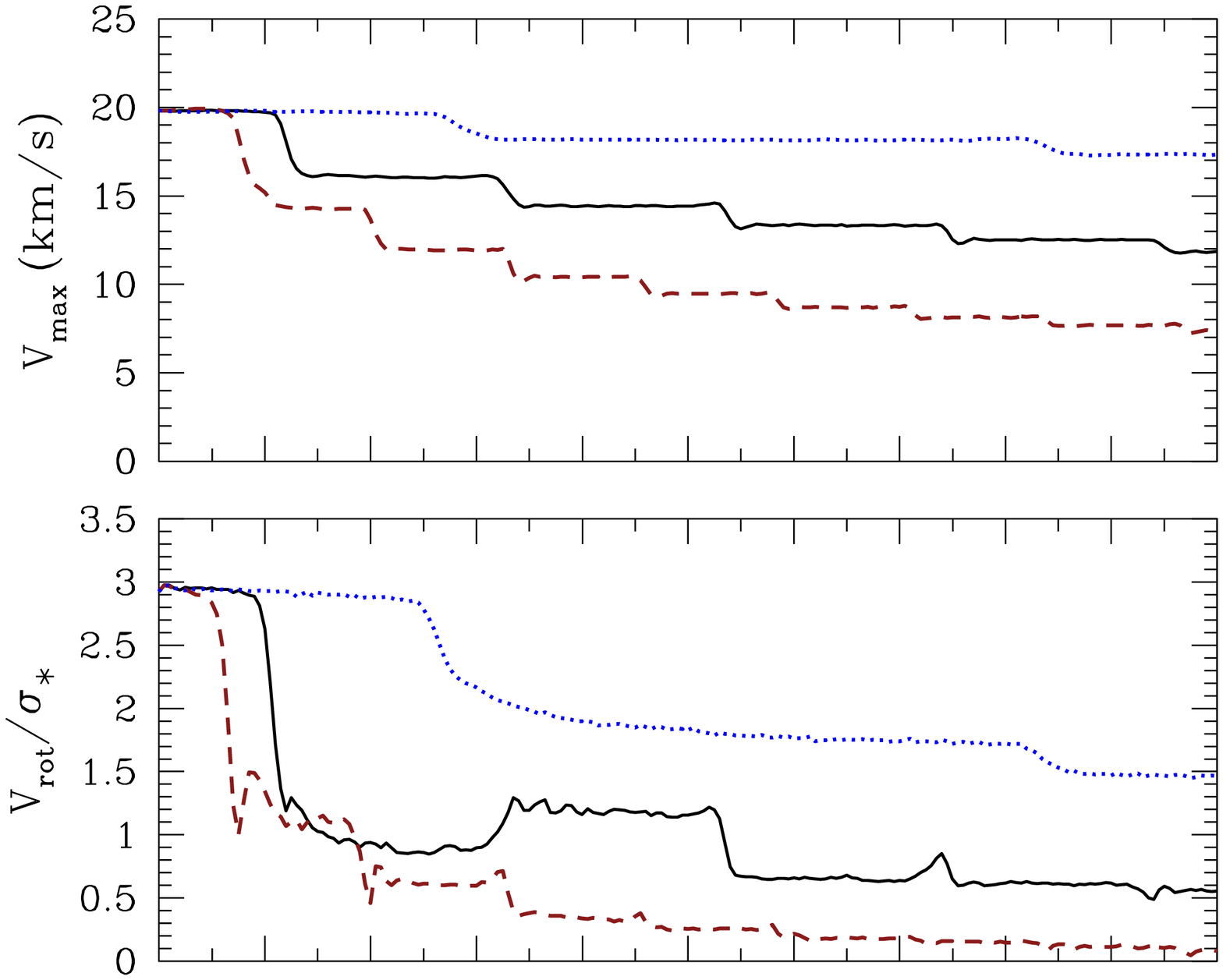}
  \includegraphics[scale=0.44]{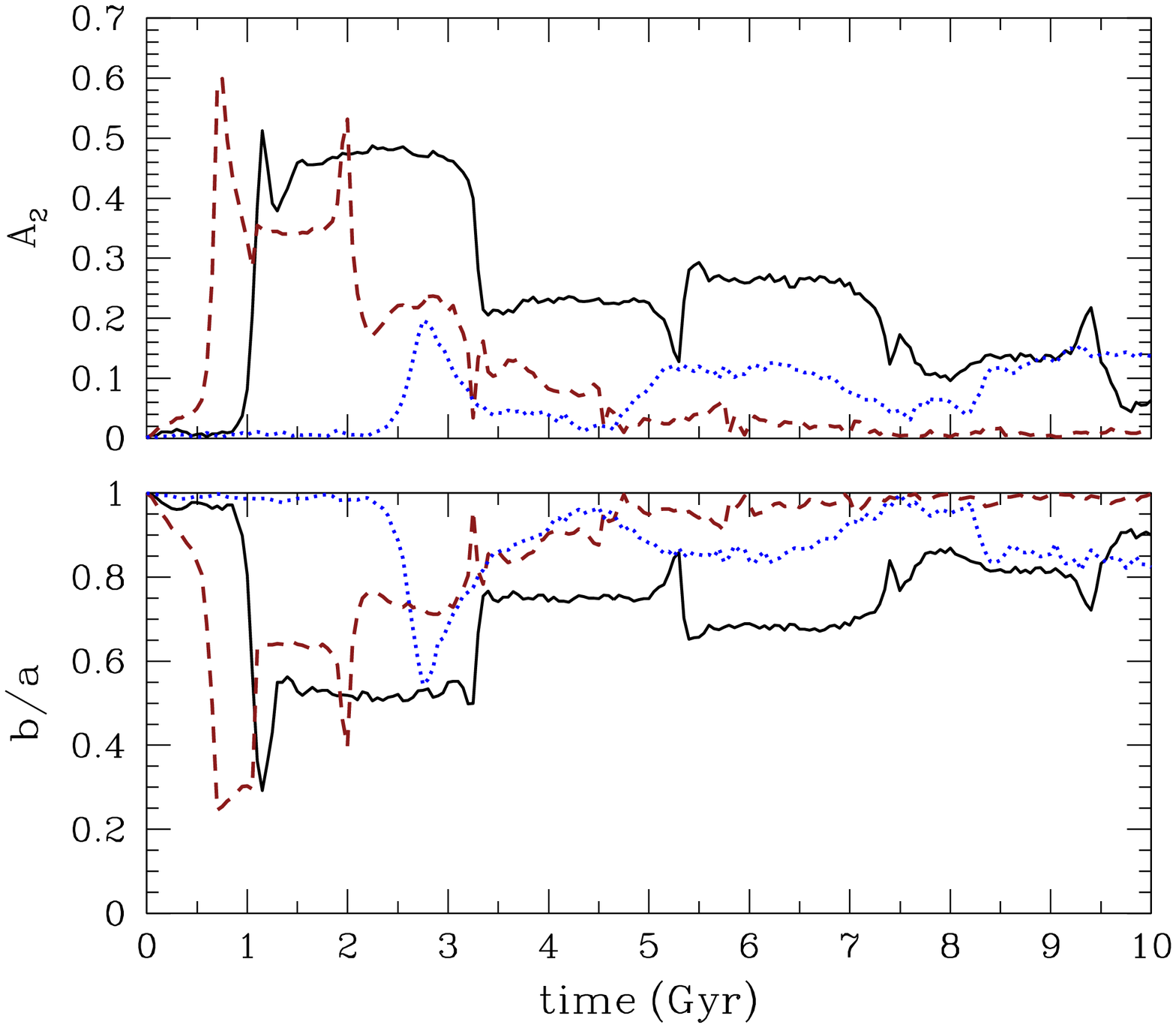}
  \includegraphics[scale=0.44]{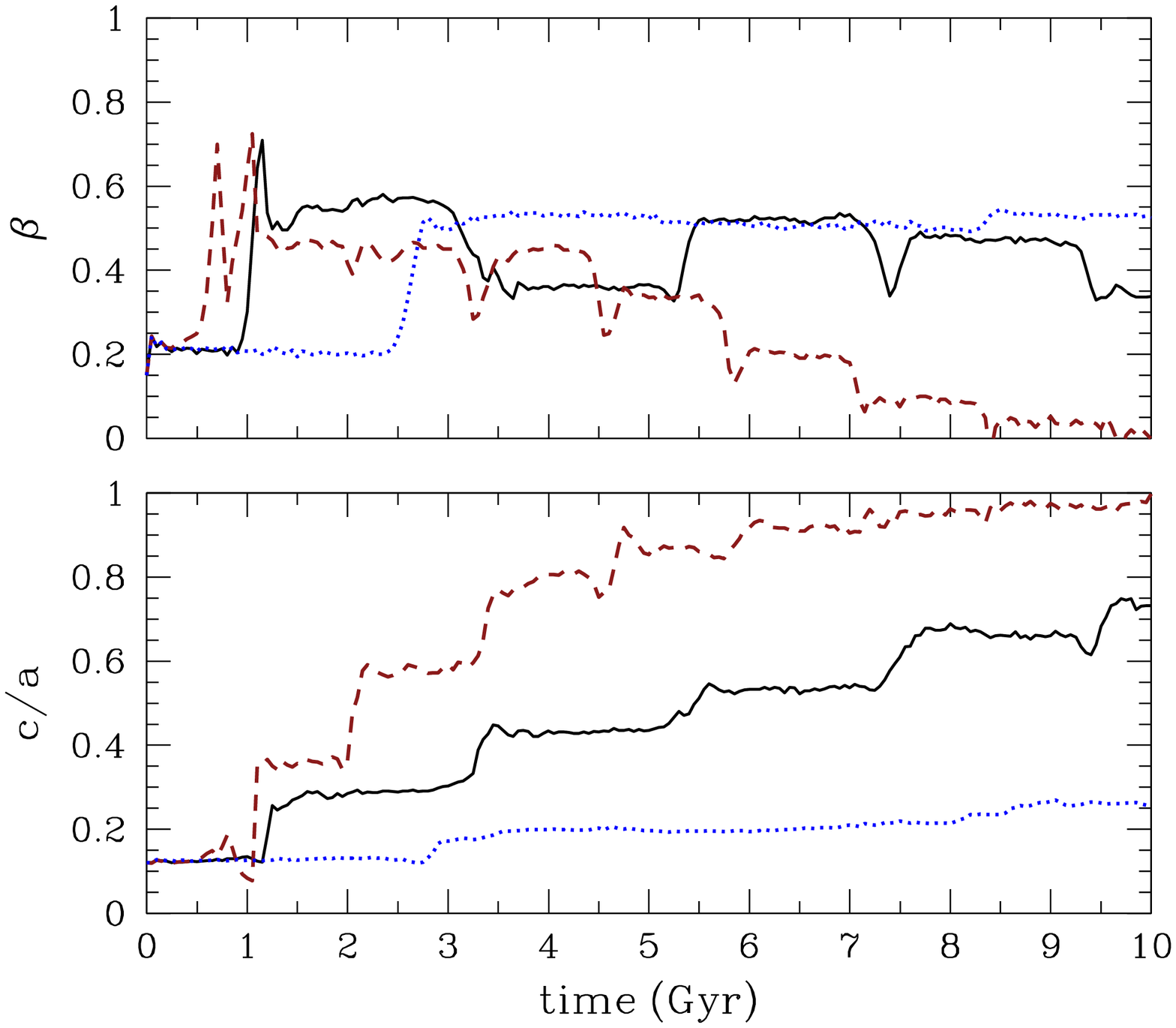}
\end{center}
\caption{Comparison of the evolution of various parameters as a function of
  time illustrating the dependence of the tidal transformation of
  disky dwarf galaxies on the sizes of their orbits. Results are
  presented for the default simulation R1 and for experiments R2 and
  R3. The description of the simulations is presented in
  Table~\ref{table:simulations}. For a fixed eccentricity, $r_{\rm
    apo}/r_{\rm peri}$, rotationally-supported dwarfs on tighter
  orbits with smaller pericentric distances, $r_{\rm peri}$, exhibit
  stronger tidal evolution inside their host galaxies and the
  efficiency of their transformation into dSphs is enhanced
  considerably.
\label{fig2}}
\end{figure*}


\subsection{Classification Criteria}
\label{subsec:criteria}

The goal of the present study is to determine the conditions under
which systems with the properties of dSphs can be produced via the
tidal interactions between disky dwarfs and their host galaxies.
Observed dSphs possess a unique set of kinematic and structural
properties that we can utilize in order to perform comparisons with
simulated dwarfs. In particular, we imposed two criteria for
establishing the formation of a dSph in our simulations.

The first criterion is related to the shape of the stellar component.
Specifically, only simulated dwarf galaxies whose final states are
characterized by $c/a \gtrsim 0.5$ may be regarded as dSphs. The
motivation behind this choice stems from the fact that most classic
dSphs (e.g., Fornax, Draco, Leo I, Tucana, Sextans, Carina) have
projected ellipticities, $\epsilon \equiv 1-b/a$, in the range $0.1
\lesssim \epsilon \lesssim 0.5$
\citep[e.g.,][]{Mateo98,McGaugh_Wolf10}, or equivalently projected
axis ratios $b/a \gtrsim 0.5$, where $b$ and $a$ denote the minor and
major axis of the stellar distribution, respectively. Given that an
elongated shape cannot appear more elongated in projection, the
condition that {\it intrinsic} $c/a \gtrsim 0.5$ in the simulated
dwarfs (n.b. by definition $b/a \geq c/a$) ensures that the {\it
  projected} $b/a$ will also satisfy $b/a \gtrsim 0.5$ in any possible
projection, in accord with the values inferred for observed dSphs.  We
note that the intrinsic axis ratios of the dwarf remnants do not vary
significantly with radius.  Thus, they are appropriate for comparisons
with observed ellipticities that are not computed within $r_{\rm
  max}$, but rather estimated either at the outer parts of the stellar
distribution \citep{Mateo98} or within some observationally-defined
radius (e.g., half-light radius).

The second classification criterion is related to the kinematics of
the stellar component. In particular, we classify as dSphs only those
systems which, after $10$~Gyr of evolution inside the primary galaxy,
exhibit $V_{\rm rot}/\sigma_{\ast} \lesssim 1$ in accordance with the
results of observational studies of classic dSphs
\citep[e.g.,][]{Mateo98}.  We stress that the values we compute for
$V_{\rm rot}$ and $\sigma_{\ast}$ are fairly close to what an observer
would measure using line-of-sight velocities, making the comparison
with observations reasonable.  Indeed, at $r_{\rm max}$ the rotation
is close to the maximum value and can be readily measured if the
line-of-sight is perpendicular to the rotation axis.  Moreover, the
velocity dispersion profiles of our remnants do not strongly vary with
radius (as well as with line-of-sight) suggesting that our derived
values for $\sigma_{\ast}$ should not differ significantly from those
of the central velocity dispersion commonly used by observers.  As a
result of all these facts, the condition $V_{\rm rot}/\sigma_{\ast}
\lesssim 1$ that we impose here guarantees that $V_{\rm
  rot}/\sigma_{\ast} \lesssim 1$ will also be satisfied for a
significant fraction, if not all, of random lines-of-sight.

Overall, for the purposes of the present study, the above
classification scheme is adequate to establish the formation of a dSph
in our simulations.  A more direct comparison between theoretical and
observational measurements for the properties of dSphs will be
performed in a companion paper ({\L}okas et al. 2010, in preparation).

\section{Efficiency of Tidal Stirring and Orbital Parameters 
  of the Progenitor Disky Dwarfs}
\label{sec:orbital_parameters}

In this section, we gauge the dependence of the efficiency of tidal
stirring on the orbital parameters of the progenitor
rotationally-supported dwarfs, and in particular, on the size and
eccentricity of their orbits.

\subsection{Size of the Orbit}
\label{subsec:orbit_size}

We first investigate the degree to which the tidal evolution of a
disky dwarf is affected by the size of its orbit. The size of the
orbit should be crucial for the outcome of tidal stirring since the
tidal force depends strongly on the distance of the dwarf from the
center of the host galaxy. In order to ascertain this, we considered
two additional simulations (R2, R3) in which we placed the dwarf
galaxy model D1 on orbits with sizes that were different from that of
the reference experiment R1. Specifically, in this set of simulations,
we kept the eccentricity constant, but varied the default apocentric
and pericentric distance by the same factor, leaving all other
simulation parameters unchanged (see Table~\ref{table:simulations}).
Figure~\ref{fig1} shows the orbital trajectories of the dwarf galaxies
from the respective center of their host as a function of time for
experiments R1, R2, and R3.  In Figure~\ref{fig2}, we present the time
evolution of the relevant parameters discussed in
\S~\ref{subsec:parameters} ($V_{\rm max}$, $r_{\rm max}$, $b/a$,
$c/a$, $V_{\rm rot}/\sigma_{\ast}$, $\beta$, $M/L$, and $A_2$) for the
same simulations.

In order to highlight some general trends in the evolution of these
quantities with time, we first focus on the reference experiment R1.
As a result of the continuous action of the host galaxy tidal field,
the overall structure of the orbiting dwarf is altered as it
experiences a gradual decrease of its mass and physical size.  This is
reflected in the evolution of $r_{\rm max}$ and $V_{\rm max}$,
respectively, which are both reduced systematically as a function of
time. The preferential stripping of the mass from the outer,
low-density regions of the dwarf galaxy leads to the adjustment of the
circular velocity profile in such a way that both $r_{\rm max}$ and
$V_{\rm max}$ decrease.  The temporary but substantial increase of
$r_{\rm max}$ at the first pericentric approach can be attributed to
the fact that the dwarf is being deformed by strong tidal forces for
the first time and, as a result, its density distribution becomes very
extended. This increase in $r_{\rm max}$ is not observed at subsequent
pericentric passages as the dwarf galaxy becomes progressively more
compact during its evolution and thus the ability of the tidal field
to distort it is reduced. As we show below, in cases where the dwarfs
are initially more concentrated (as a result of either being embedded
in halos of higher concentration or hosting more compact or massive
disks) or are on wider orbits and experience weaker tidal forces, this
increase in $r_{\rm max}$ does not occur.

The evolution of $V_{\rm max}$ shows two interesting features. First,
$V_{\rm max}$ decreases significantly near the pericenters of the
orbit, where the intensity and variation of the time-dependent tidal
force is the strongest, and the tidal shocks occur.  This drop in
$V_{\rm max}$ is associated with mass loss from all regions of the
dwarf galaxy. Because tidal shocks act on a very short timescale, even
the inner regions of the dwarf, which are characterized by the highest
densities and thus the shortest dynamical times, respond impulsively
to the external tidal perturbation and mass is tidally stripped
directly from regions within $r_{\rm max}$.  However, $V_{\rm max}$
also decreases because a fraction of the particles that are removed
from larger radii are, in fact, on eccentric orbits with large
apocentric distances. Although these particles spend most of their
time outside of $r_{\rm max}$, they still contribute to the total mass
in the inner regions. These are preferentially DM particles, since
stars are confined to the central region of the dwarf by construction.
After experiencing the tidal shock, it takes a few hundred million
years, i.e., a multiple of the dynamical time of the system, for the
dwarf to readjust to a new equilibrium state with a new value of
$V_{\rm max}$.


\begin{table*}
\caption{Summary of Results}
\begin{center}
  \vspace*{-12pt}
\begin{tabular}{lccccccccccc}
\hline
\hline
\\
\multicolumn{1}{c}{}              &
\multicolumn{1}{c}{$V_{\rm max}$} &
\multicolumn{1}{c}{$r_{\rm max}$} &
\multicolumn{1}{c}{$M/L$}         &
\multicolumn{1}{c}{}              &
\multicolumn{1}{c}{}              &
\multicolumn{1}{c}{}              &
\multicolumn{1}{c}{}              &
\multicolumn{1}{c}{}              &
\multicolumn{1}{c}{}              &
\multicolumn{1}{c}{}              &
\multicolumn{1}{c}{$T_{\rm orb}$} 
\\
\multicolumn{1}{c}{Simulation} & 
\multicolumn{1}{c}{(km s$^{-1}$)}  &
\multicolumn{1}{c}{(kpc)}      & 
\multicolumn{1}{c}{$(M_{\odot}/L_{\odot})$}& 
\multicolumn{1}{c}{$V_{\rm rot}/\sigma_{\ast}$} & 
\multicolumn{1}{c}{$\beta$} & 
\multicolumn{1}{c}{$A_2$} &
\multicolumn{1}{c}{$b/a$} & 
\multicolumn{1}{c}{$c/a$} &
\multicolumn{1}{c}{Bar Formation} &
\multicolumn{1}{c}{Classification}& 
\multicolumn{1}{c}{(Gyr)}      
\\
\multicolumn{1}{c}{(1)}&
\multicolumn{1}{c}{(2)}&
\multicolumn{1}{c}{(3)}&
\multicolumn{1}{c}{(4)}&
\multicolumn{1}{c}{(5)}&
\multicolumn{1}{c}{(6)}&
\multicolumn{1}{c}{(7)}&
\multicolumn{1}{c}{(8)}&
\multicolumn{1}{c}{(9)}&
\multicolumn{1}{c}{(10)}&
\multicolumn{1}{c}{(11)}&
\multicolumn{1}{c}{(12)}
\\
\\
\hline
\\
R1  & 11.9  &  0.53  & 10.7 &  0.55 &  0.34 & 0.06 & 0.90 & 0.73 & yes & dSph      & 2.09\\
R2  & 7.3   &  0.18  & 19.2 &  0.08 &  0.00 & 0.00 & 1.00 & 1.00 & yes & dSph      & 1.28\\
R3  & 17.3  &  1.22  & 16.9 &  1.46 &  0.52 & 0.14 & 0.82 & 0.26 & no  & non-dSph  & 5.40\\
R4  & 7.2   &  0.24  & 15.3 &  0.03 &  0.03 & 0.03 & 0.95 & 0.94 & yes & dSph      & 1.81\\
R5  & 16.0  &  0.98  & 14.0 &  1.25 &  0.54 & 0.06 & 0.88 & 0.36 & no  & non-dSph  & 2.50\\
R6  & 12.2  &  0.51  & 10.3 &  0.24 &  0.51 & 0.04 & 0.95 & 0.73 & yes & dSph      & 2.09\\
R7  & 12.4  &  0.55  & 10.7 &  0.23 &  0.48 & 0.16 & 0.80 & 0.74 & yes & dSph      & 2.09\\
R8  & 12.7  &  0.50  & 10.3 &  0.74 &  0.42 & 0.10 & 0.86 & 0.66 & yes & dSph      & 2.09\\
R9  & 12.0  &  0.55  & 11.8 &  0.61 &  0.23 & 0.04 & 0.94 & 0.77 & yes & dSph      & 2.09\\
R10 & 10.4  &  0.55  & 18.8 &  0.56 &  0.19 & 0.04 & 0.93 & 0.80 & yes & dSph      & 2.09\\
R11 & 16.0  &  0.49  &  6.7 &  0.40 &  0.48 & 0.28 & 0.67 & 0.58 & yes & dSph      & 2.09\\
R12 & 14.6  &  0.41  &  6.7 &  0.56 &  0.45 & 0.29 & 0.67 & 0.53 & yes & dSph      & 2.09\\
R13 & 10.4  &  0.56  & 22.3 &  0.26 &  0.13 & 0.03 & 0.95 & 0.91 & yes & dSph      & 2.09\\
R14 & 7.3   &  0.40  &  6.5 &  0.31 &  0.15 & 0.01 & 0.98 & 0.89 & yes & dSph      & 2.10\\
R15 & 20.3  &  0.67  & 23.3 &  1.18 &  0.49 & 0.07 & 0.88 & 0.52 & no  & non-dSph  & 2.08\\
R16 & 7.1   &  0.36  & 11.6 &  0.62 &  0.33 & 0.04 & 0.94 & 0.73 & no  & dSph      & 2.14\\
R17 & 19.0  &  0.77  &  9.2 &  0.70 &  0.23 & 0.09 & 0.87 & 0.68 & yes & dSph      & 1.88\\
\hline
\end{tabular}
\end{center}
{\sc Notes.}---Columns 2-9 refer to the final structural parameters of the simulated 
dwarfs. The quantities in each column are as follows.
Column 1: Abbreviation for the tidal stirring simulations.
Column 2: Maximum circular velocity of the dwarf in $\kms$.
Column 3: Radius at which the maximum circular velocity occurs in kpc.
Column 4: Mass-to-light ratio of the dwarf in units of $M_{\odot}/L_{\odot}$.
Column 5: Ratio of stellar rotational velocity to one-dimensional stellar velocity dispersion of the dwarf.
Column 6: Anisotropy parameter of the stellar distribution of the dwarf.
Column 7: Amplitude of the $m = 2$ Fourier component of the surface density distribution of the dwarf stars.
Column 8: Axis ratio $b/a$ of the stellar component of the dwarf.
Column 9: Axis ratio $c/a$ of the stellar component of the dwarf.
Column 10: Formation of a tidally-induced bar during the orbital evolution of the dwarf galaxies according 
to the criteria discussed in \S~\ref{subsec:parameters}.
Column 11: Classification of the final systems according to the criteria imposed in \S~\ref{subsec:criteria}.
Column 12: Orbital time in Gyr defined as the average time elapsed between consecutive apocentric passages.
Note that the entries in columns 4-9 are estimated within $r_{\rm max}$.
\label{table:summary}
\end{table*}



\begin{figure*}[t]
\begin{center}
\begin{tabular}{c}
  \includegraphics[scale=0.35]{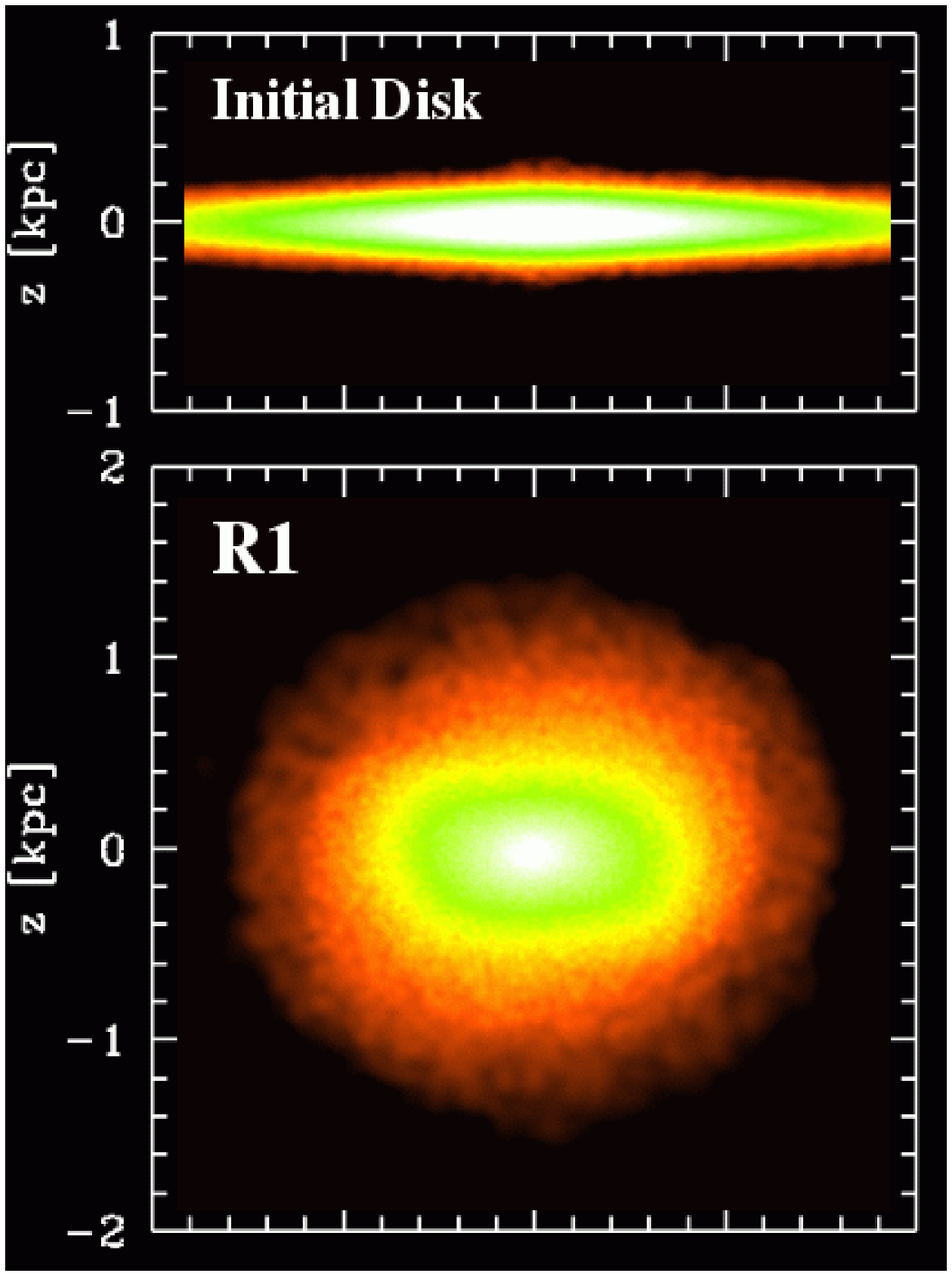}\hspace{-0.1cm}
  \includegraphics[scale=0.35]{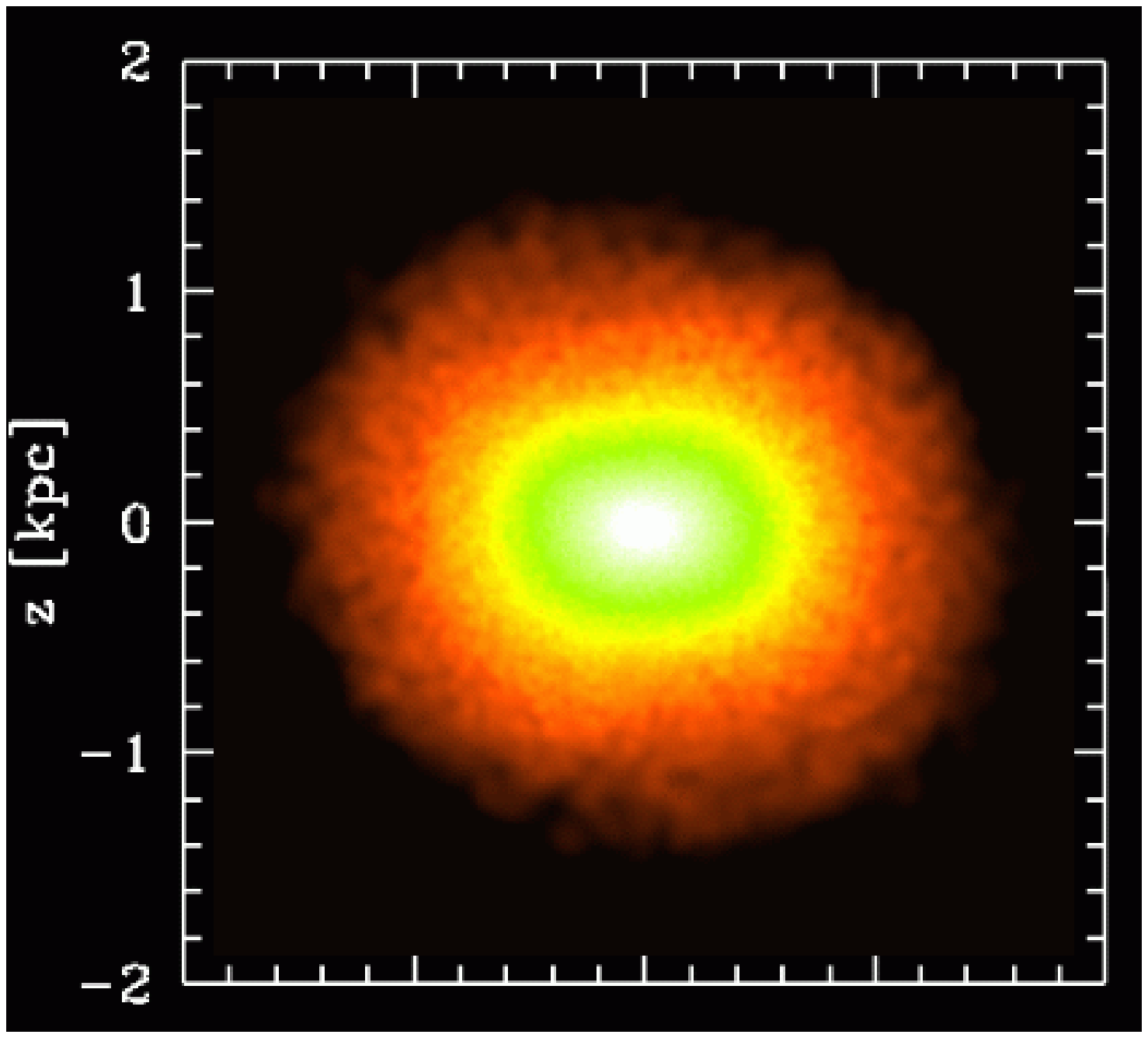}\hspace{-0.1cm}
  \includegraphics[scale=0.35]{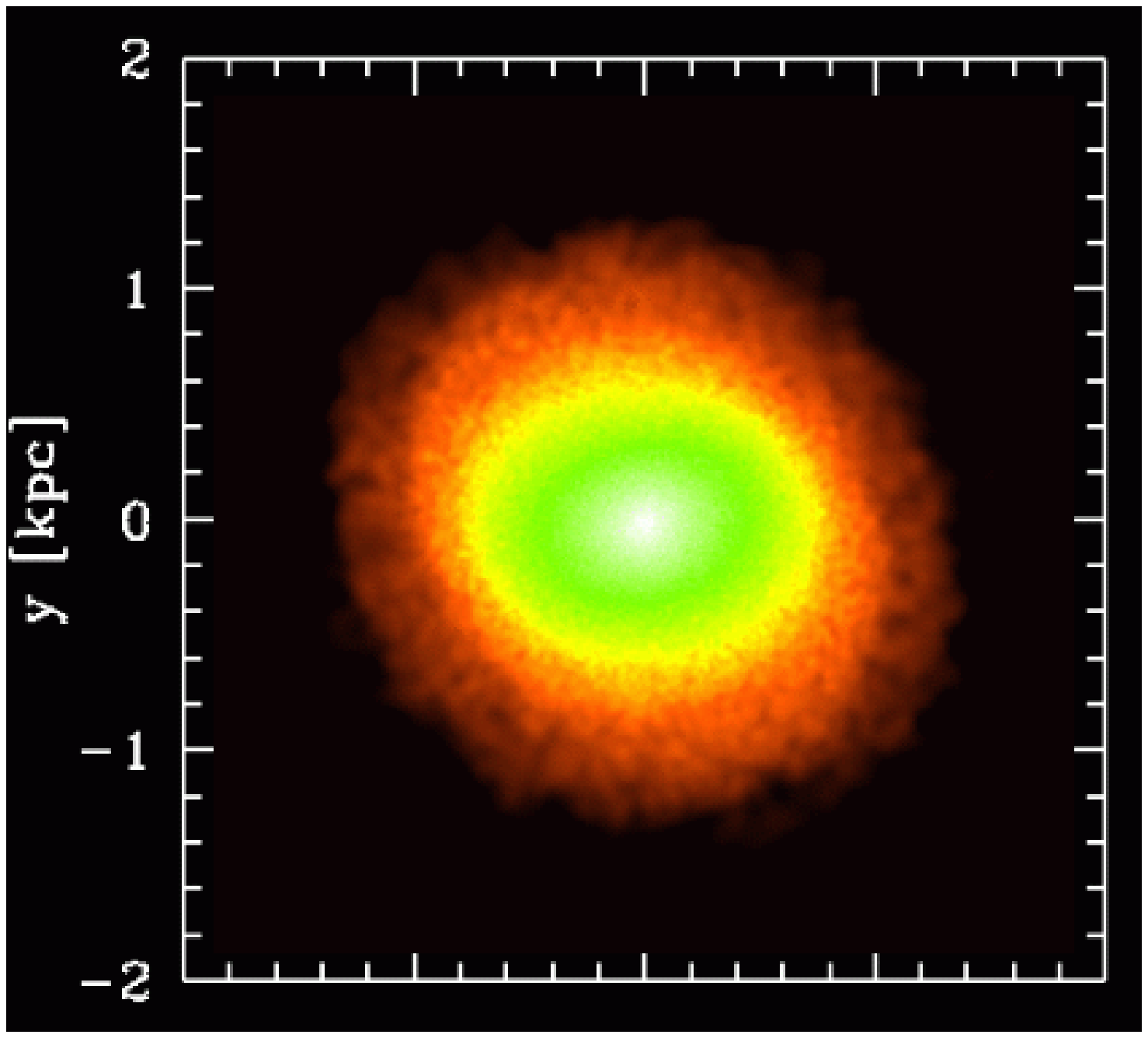}
  \vspace{-0.25cm}
\end{tabular}
\begin{tabular}{c}
  \includegraphics[scale=0.35]{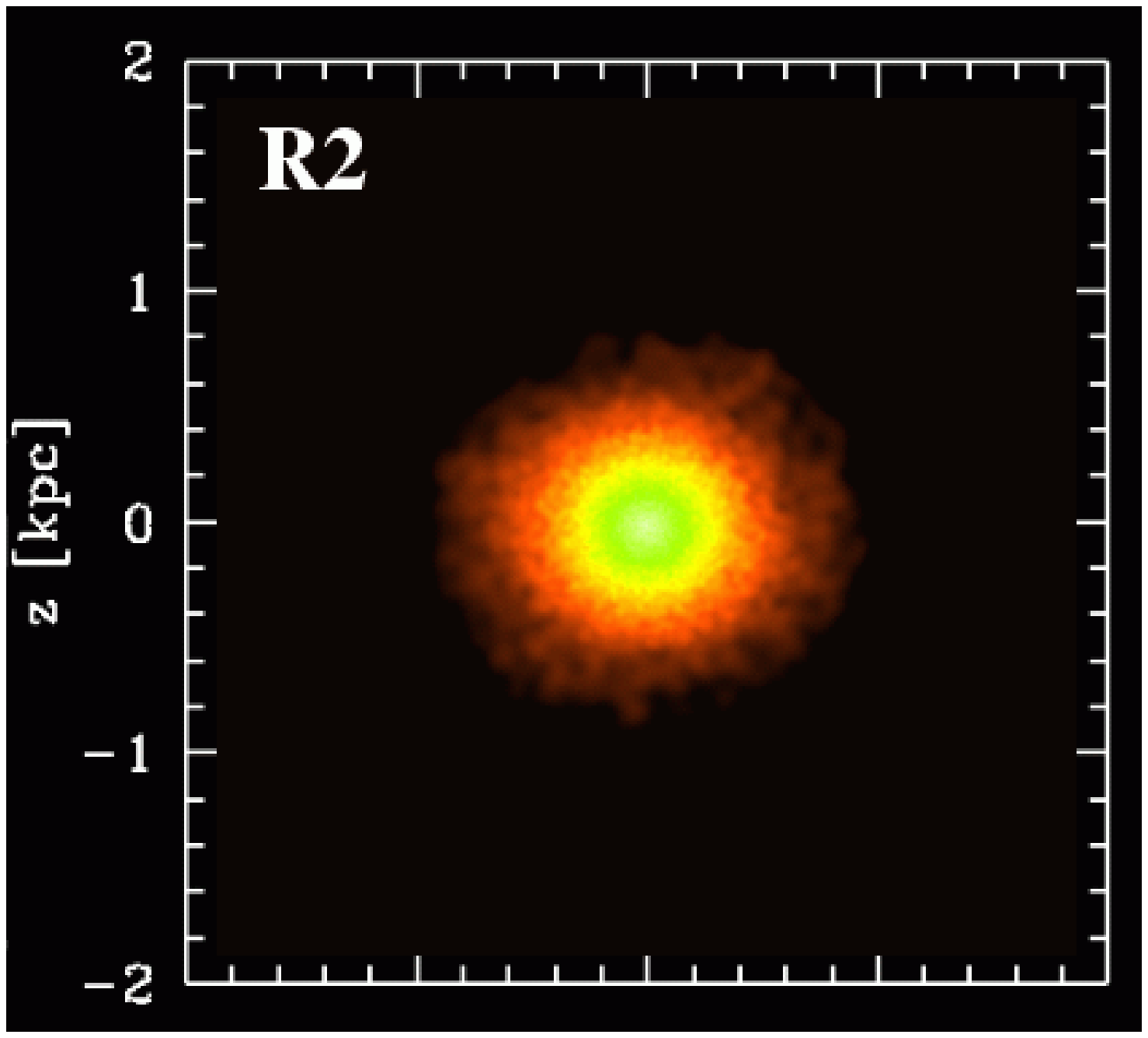}\hspace{-0.1cm}
  \includegraphics[scale=0.35]{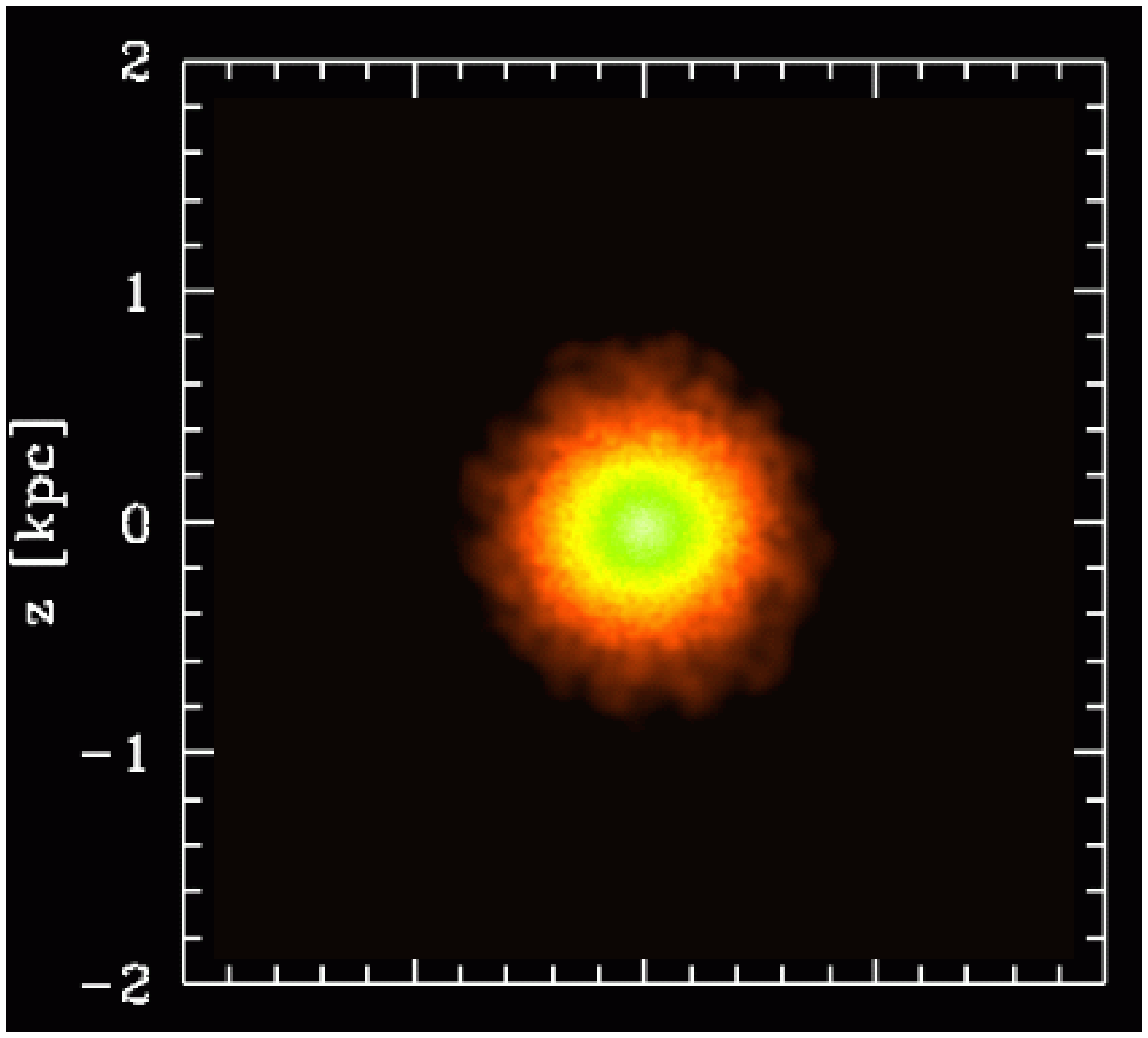}\hspace{-0.1cm}
  \includegraphics[scale=0.35]{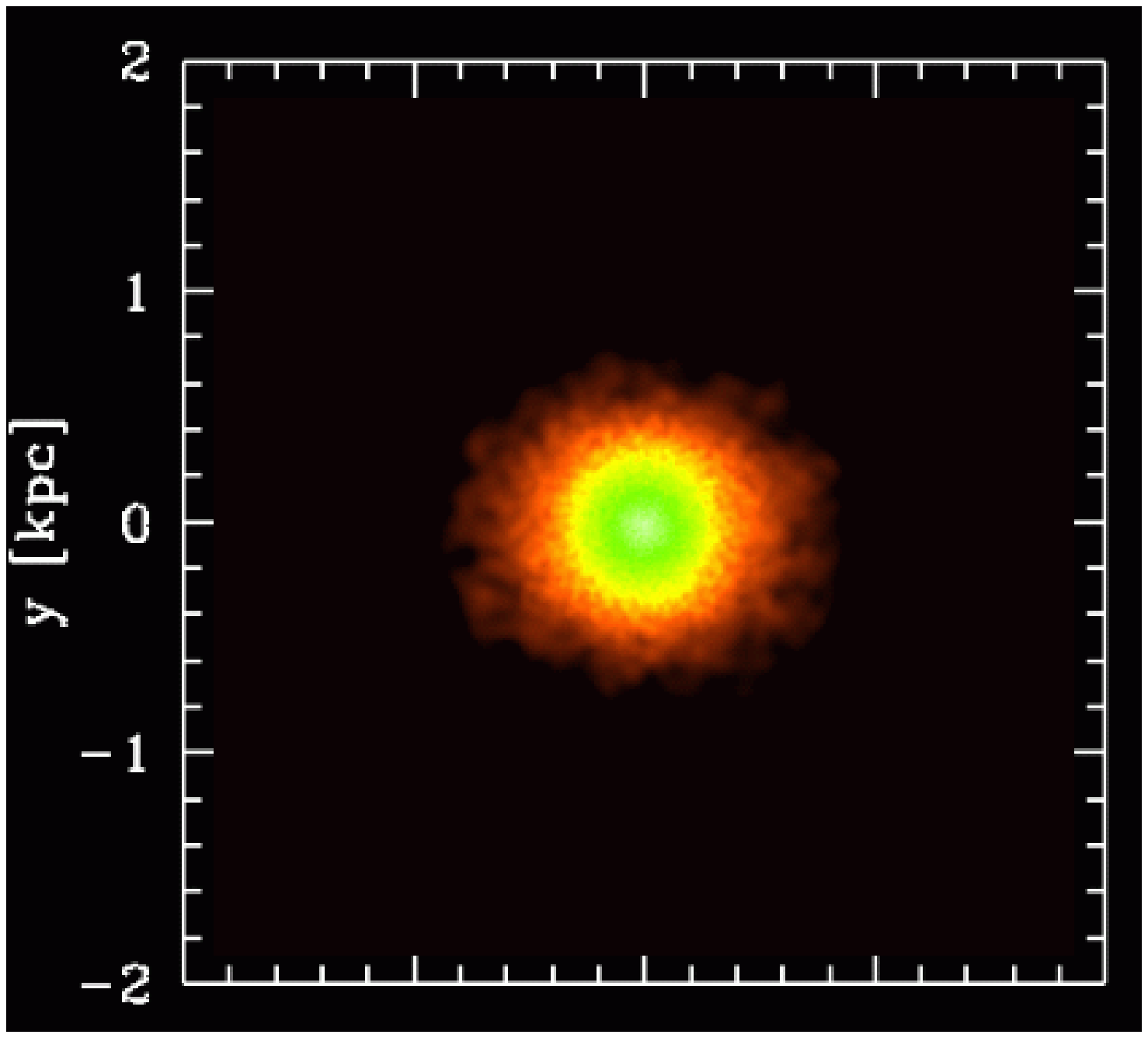}
  \vspace{-0.25cm}
\end{tabular}
\begin{tabular}{c}
  \includegraphics[scale=0.35]{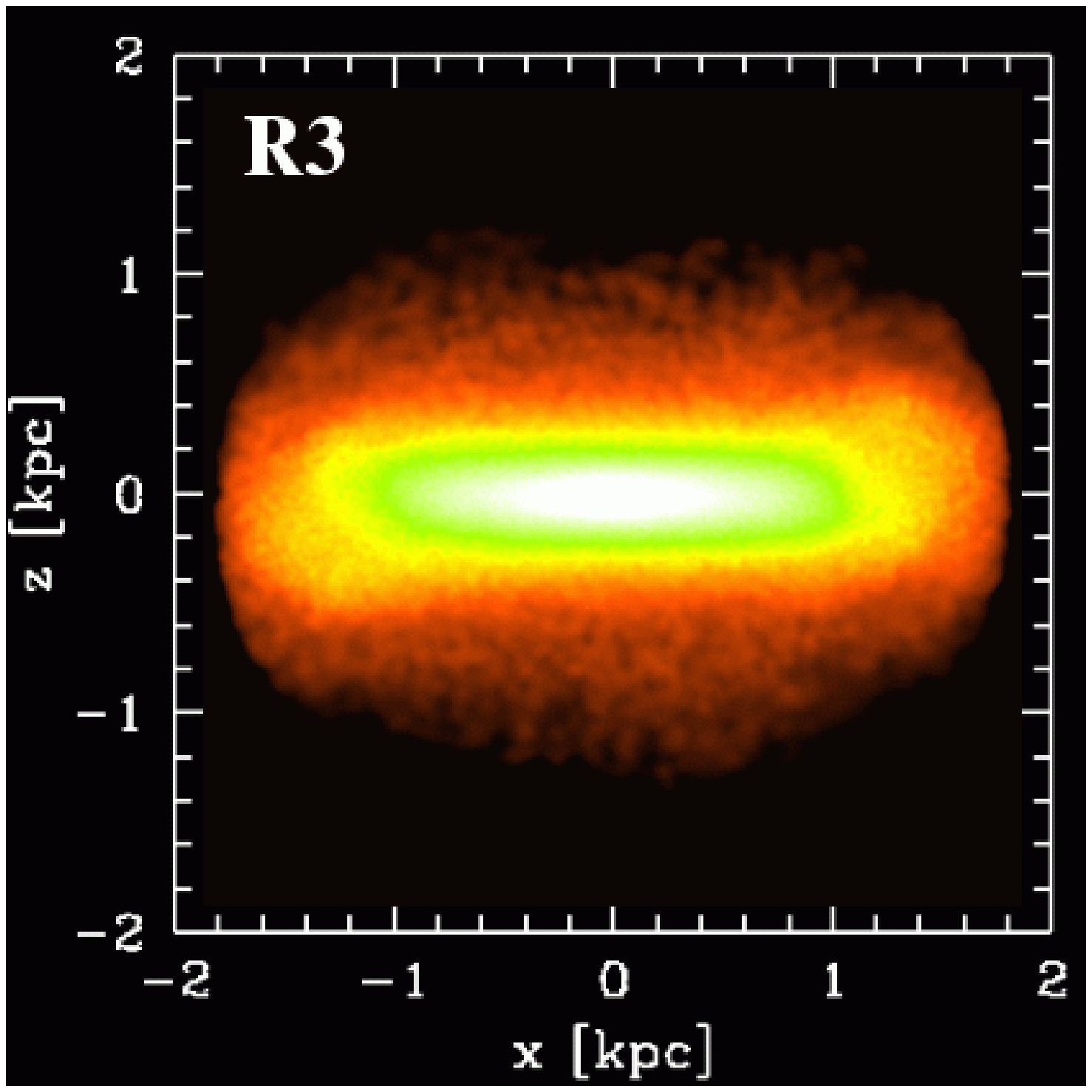}\hspace{-0.1cm}
  \includegraphics[scale=0.35]{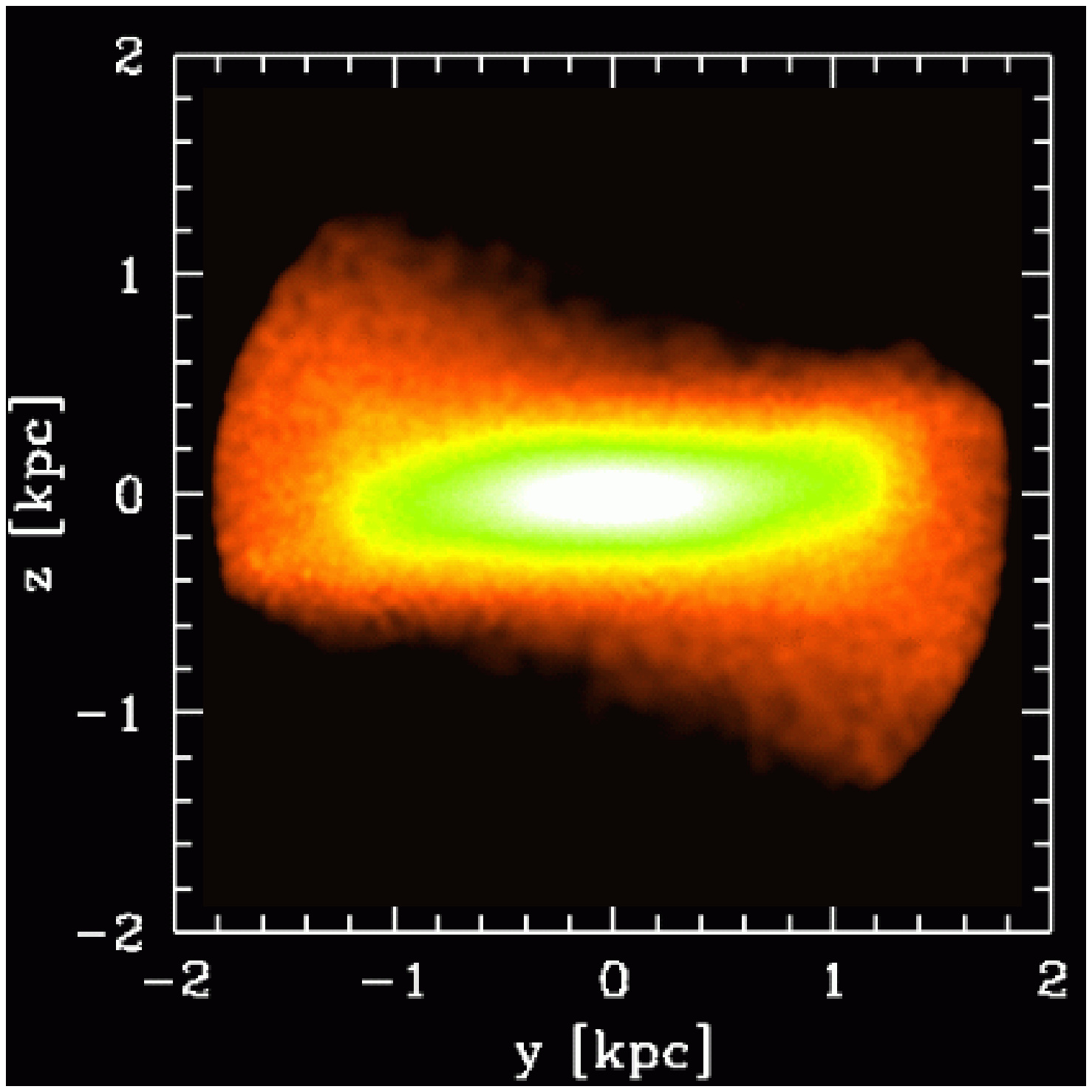}\hspace{-0.1cm}
  \includegraphics[scale=0.35]{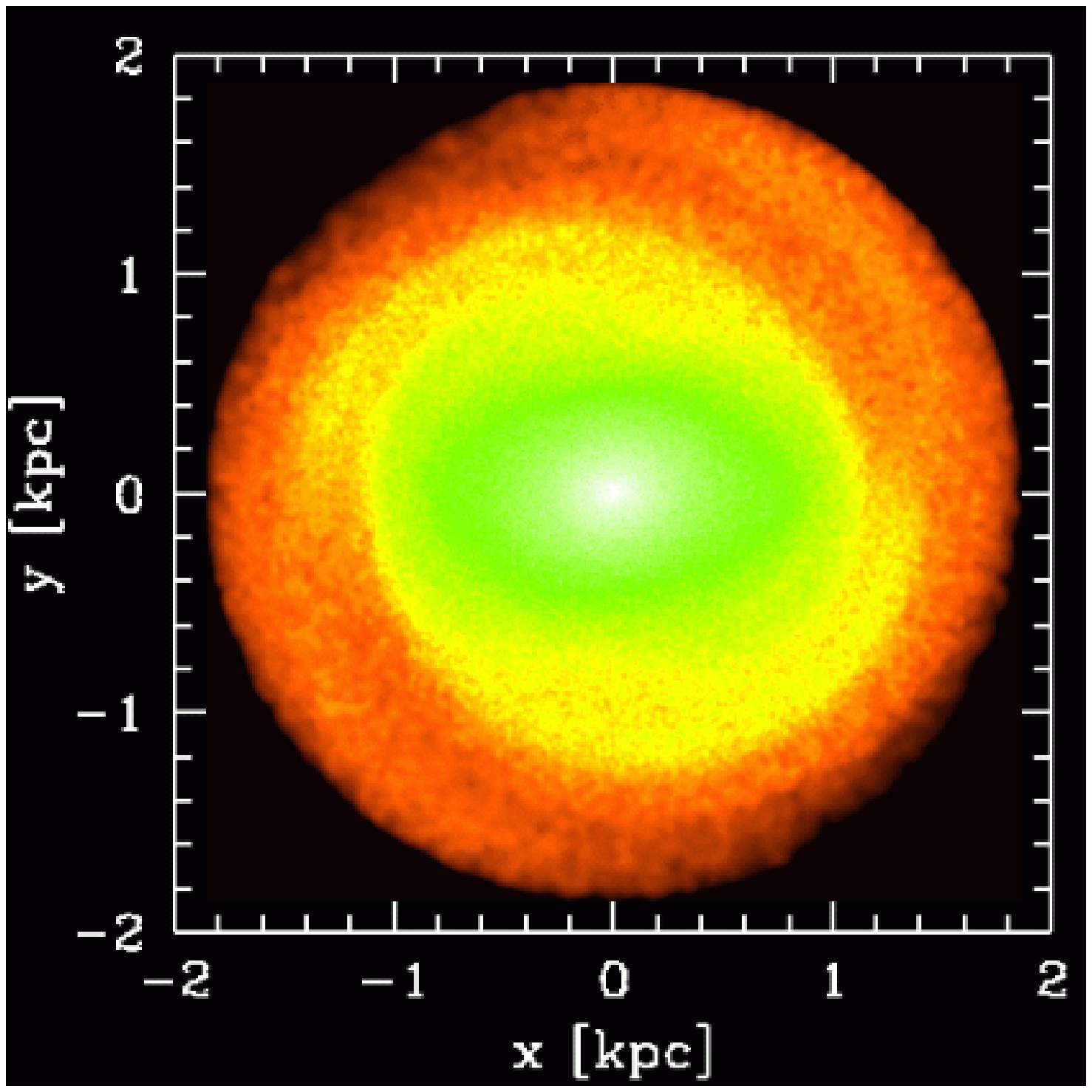}
\vspace{-0.5cm}
\end{tabular}
\end{center}
\vspace{0.1cm}
\caption{Surface density maps of the final stellar distributions of the
  dwarfs in simulations R1 (upper panels), R2 (middle panels), and R3
  (lower panels). In order to aid comparison, the first panel also
  includes the edge-on view of the initial disk. Particles are
  color-coded on a logarithmic scale, with hues ranging from orange to
  white indicating increasing stellar density. Local density is
  calculated using an SPH smoothing kernel of $32$ neighbors. Results
  are presented for projections onto the $xz$ (left columns), $yz$
  (middle columns), and $xy$ planes (right columns), where the $x$,
  $y$, and $z$ axes lie along the major, intermediate, and minor axes
  of the stellar distribution, respectively. R1 is our reference
  simulation.  Experiment R2 with the shortest orbital time and the
  second smallest pericentric distance all our experiments ($T_{\rm
    orb} = 1.28$~Gyr and $r_{\rm peri} = 17$~kpc) yields a
  spherically-symmetric ($b/a \approx c/a \approx 1$) and isotropic
  ($\beta \approx 0$) stellar system with negligible amounts of
  rotation ($V_{\rm rot}/\sigma_{\ast} \lesssim 0.1$) that would be
  classified as a dSph. The dwarf in simulation R3, whose orbit is
  characterized by the longest orbital time and largest pericentric
  distance ($T_{\rm orb} = 5.40$~Gyr and $r_{\rm peri} = 50$~kpc), is
  not transformed into a dSph and remains disky even after $10$~Gyr of
  tidal evolution inside the host galaxy.
\label{fig3}}
\end{figure*}



\begin{figure*}[t]
\begin{center}
  \includegraphics[scale=0.44]{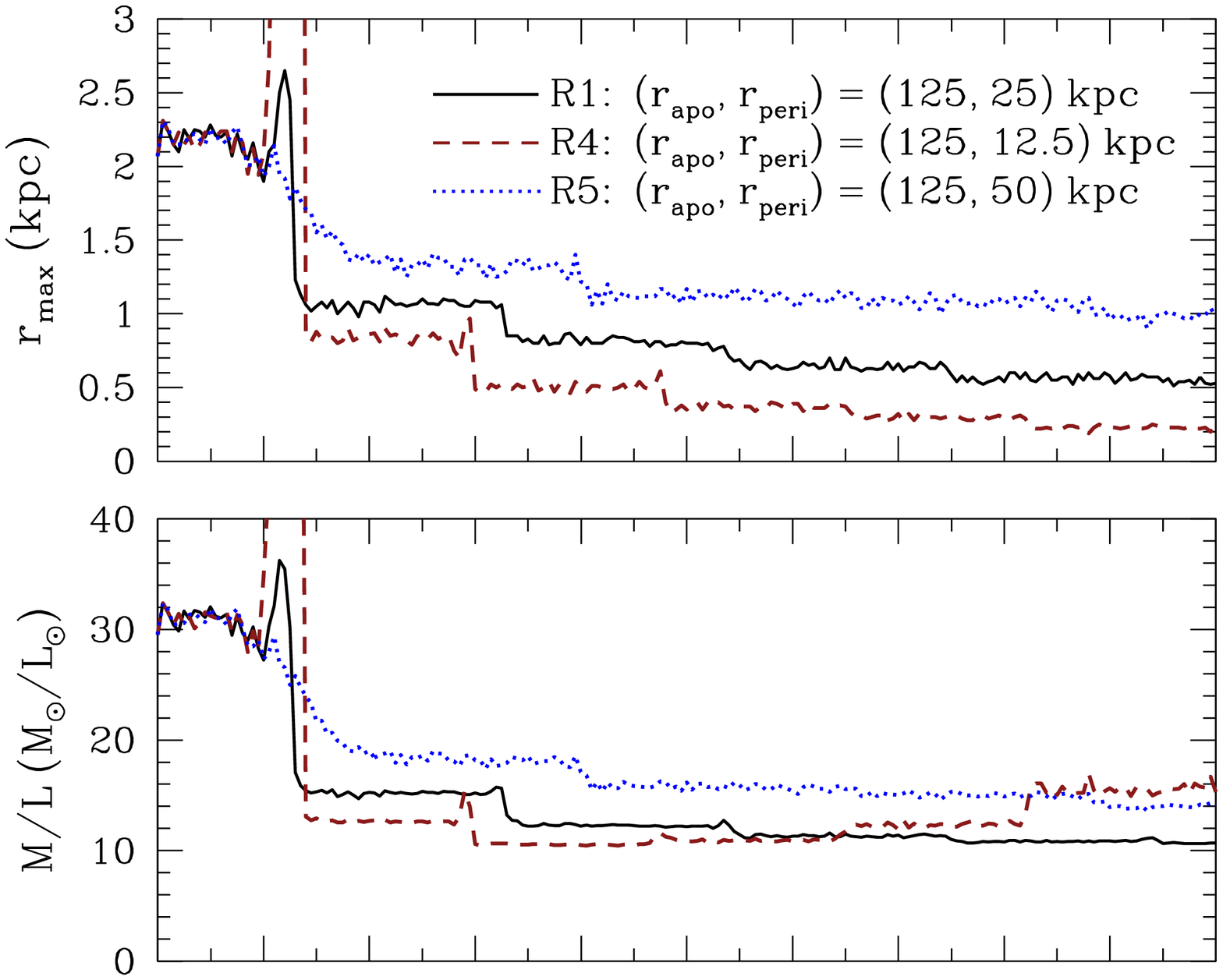}
  \includegraphics[scale=0.44]{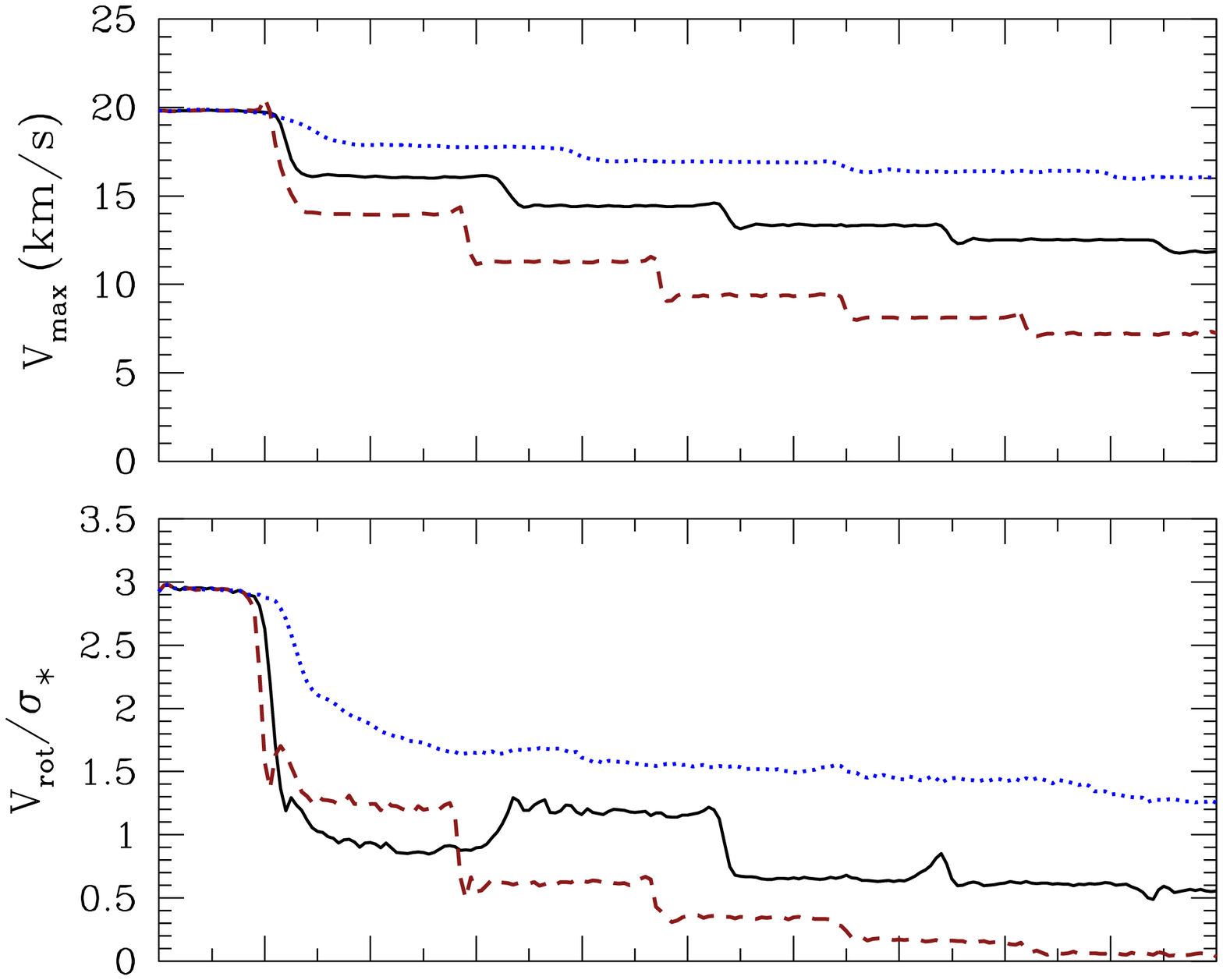}
  \includegraphics[scale=0.44]{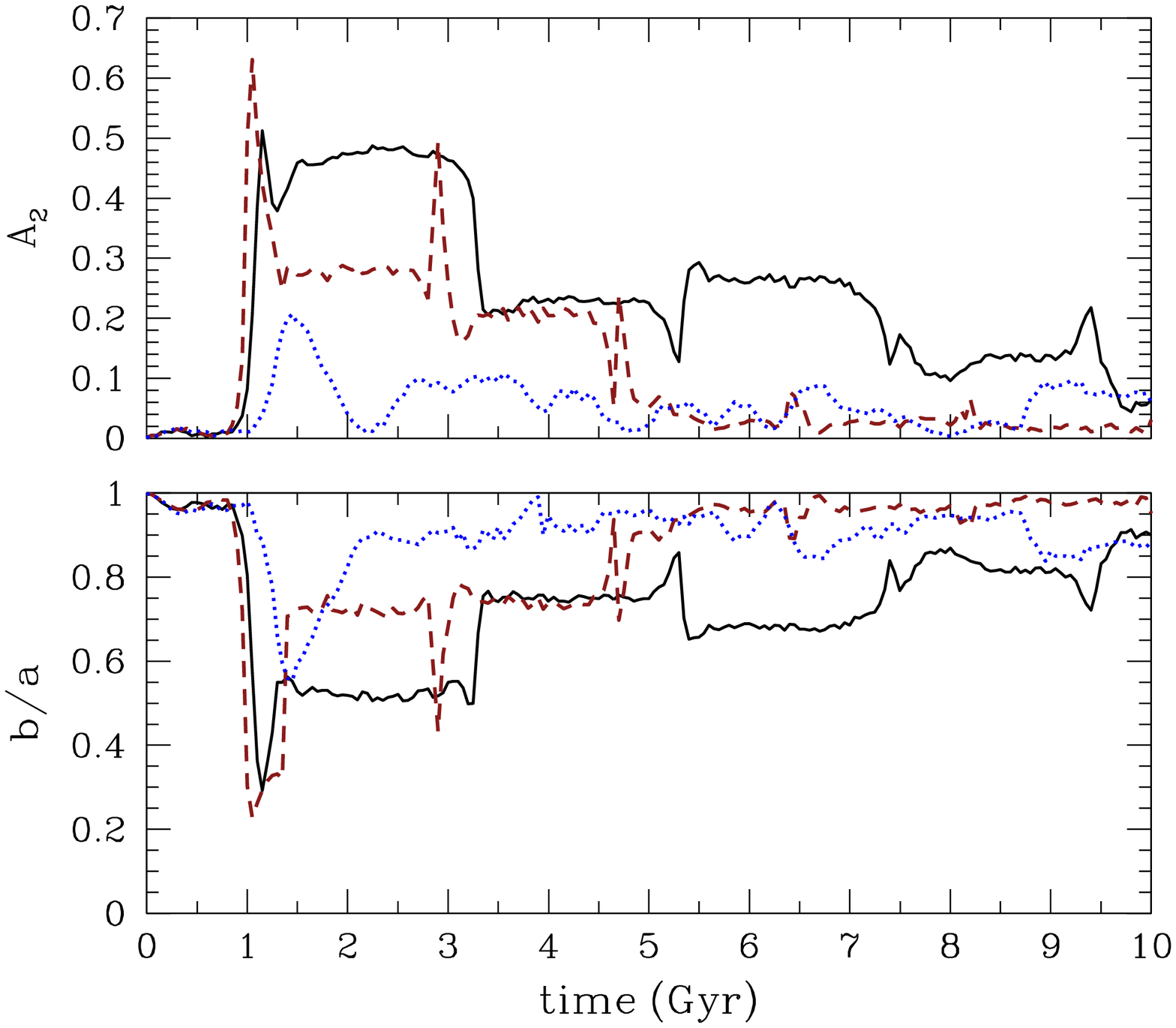}
  \includegraphics[scale=0.44]{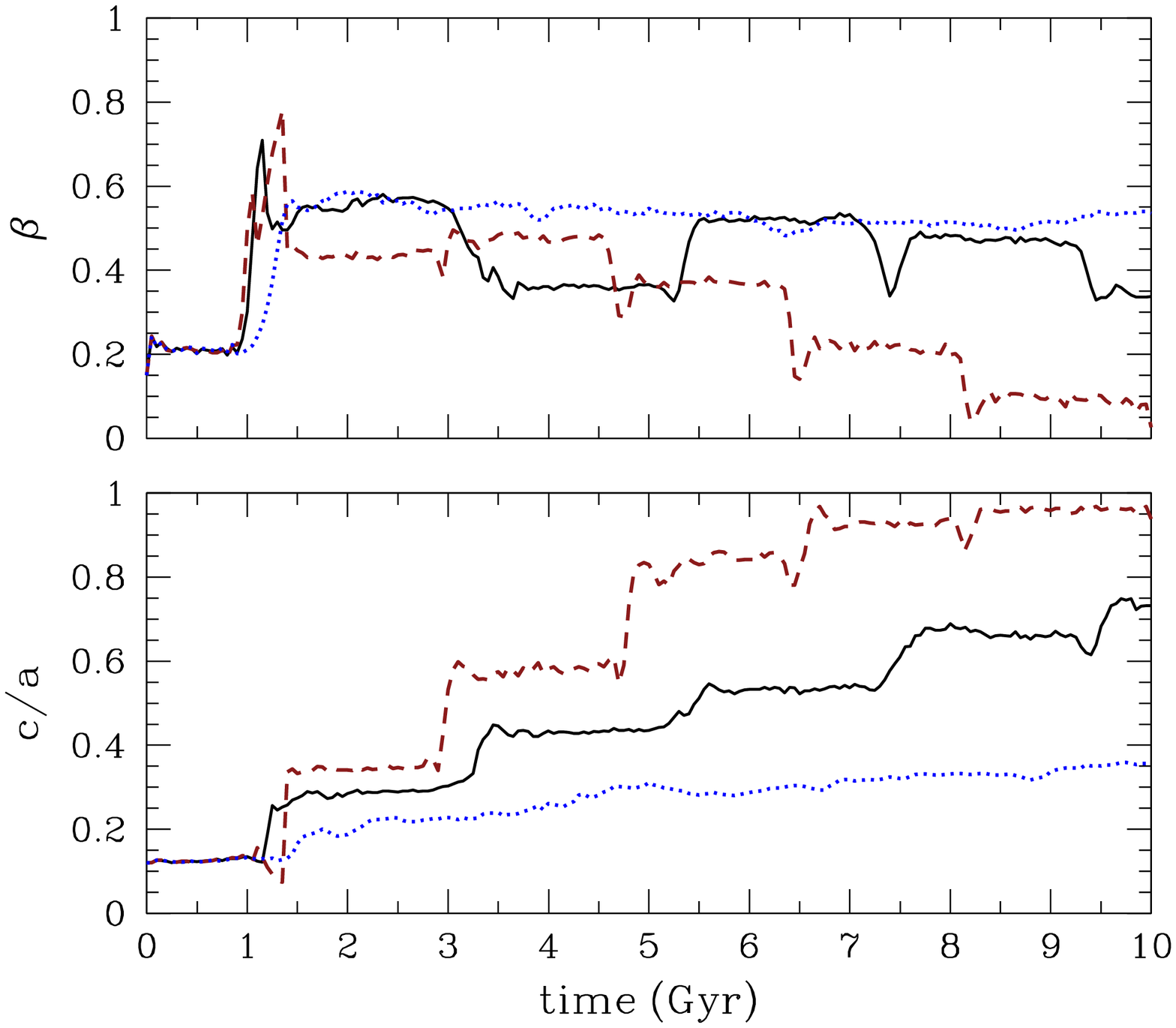}
\end{center}
\caption{Comparison of the evolution of various parameters as a function of
  time illustrating the dependence of the tidal transformation of
  disky dwarf galaxies on the eccentricities of their orbits. Results
  are presented for the default simulation R1 and for experiments R4
  and R5. The description of the simulations is presented in
  Table~\ref{table:simulations}. For a fixed apocentric distance,
  $r_{\rm apo}$, rotationally-supported dwarfs on orbits with higher
  eccentricities, $r_{\rm apo}/r_{\rm peri}$, and thus smaller
  pericentric distances, $r_{\rm peri}$, display stronger tidal
  evolution inside their host galaxies and the efficiency of their
  transformation into dSphs is increased substantially.
\label{fig4}}
\end{figure*}


Second, $V_{\rm max}$ remains remarkably constant between pericentric
approaches as the dwarf responds adiabatically to the weak intensity
and variation of the tidal force. Overall, tides act nearly
impulsively at pericentric passages and have little influence on the
dwarf galaxy between pericentric passages.  This general behavior of
$V_{\rm max}$ and $r_{\rm max}$ is observed in all our simulations,
and is in agreement with results reported in earlier studies
\citep[e.g.,][]{Hayashi_etal03,Kazantzidis_etal04a,Kravtsov_etal04,Penarrubia_etal08}.
We note that, in the reference experiment R1, the dwarf looses $\sim
90\%$ of its initial mass within $r_{\rm max}$ during its orbital
evolution but still survives as a bound entity. For such substantial
mass loss, $V_{\rm max}$ and $r_{\rm max}$ decreased by a factor of
$\sim 1.7$ and $\sim 4$, respectively. This finding has interesting
implications as it suggests that tidal stripping affects the evolution
of the $r_{\rm max}-V_{\rm max}$ relation expected in the {\LCDM}
cosmological model \citep[see also][]{Kravtsov10}.

Focusing on the evolution of the $M/L$ ratio, which we also estimate
within $r_{\rm max}$, Figure~\ref{fig2} shows that it decreases
monotonically as a function of time.  This is a consequence of two
facts. First, the stellar distributions of the dwarfs are naturally
less extended than their DM halos. This implies that, within $r_{\rm
  max}$, the particles with the smallest binding energies and longest
dynamical timescales, namely the particles that are most susceptible
to be unbound by the tidal shocks, belong predominantly to the
extended DM halos. Second, as discussed above, particles that are
stripped from the outer parts of the dwarf but still contribute to the
total mass in the inner regions belong preferentially to the DM
component. For these reasons, DM is stripped more readily than the
stars within $r_{\rm max}$ and this decreases the $M/L$ ratio. In
addition to these two processes, during the orbital evolution of the
dwarfs, $r_{\rm max}$ decreases monotonically and moves towards the
inner, stellar-dominated regions, resulting in a natural increase of
the fraction of mass in stars within $r_{\rm max}$.

The evolution of the Fourier component $A_2$ shows a sudden increase
after the first pericentric approach which signifies the formation of
a tidally-induced bar. The onset of the bar instability coincides with
a drop of $V_{\rm rot}/\sigma_{\ast}$, whose value continuously
decreases thereafter. Loss of angular momentum caused by the bar and
simultaneous increase of the stellar velocity dispersion due to tidal
heating lead to a final value of $V_{\rm rot}/\sigma_{\ast} \approx
0.55$.  As a result of the tidal shocks, the strength of the bar
diminishes as a function of time and the dwarf becomes progressively
more spherical.  After $10$~Gyr of evolution inside the host, the
initially-disky stellar distribution is transformed into a spheroid
with axis ratios of $c/a \approx 0.7$ and $b/a \approx 0.9$.

In what follows, we shall concentrate on comparing the tidal response
of the dwarf in the reference simulation R1 with that in experiments
R2 and R3. Figure~\ref{fig2} demonstrates that the dependence of the
evolution of $r_{\rm max}$, $V_{\rm max}$, and $M/L$ ratio on the size
of the orbit is dramatic.  As expected, the tighter the orbit is, the
stronger and more rapid the decrease in all parameters.  Specifically,
the dwarf galaxy in simulation R2 with the smallest pericentric
distance looses $\sim 99\%$ of its initial mass within $r_{\rm max}$
during its orbital evolution. In this case, $V_{\rm max}$ and $r_{\rm
  max}$ decreased by a factor of $\sim 2.7$ and $\sim 11.5$,
respectively.

In addition, Figure~\ref{fig2} illustrates two interesting features in
the behavior of the dwarfs in experiments R1 and R2. First, the
monotonic decrease of the $M/L$ ratio is reversed for the tightest
orbit in simulation R2 during the late stages of the evolution,
indicating that the stars begin to be stripped more effectively
compared to the DM.  Second, the systematic decrease of $V_{\rm
  rot}/\sigma_{\ast}$ is reversed temporarily for simulation R1
between the second and third pericentric passage. We postpone the
investigation of both issues for \S~\ref{sec:discussion}.

In the case of experiments R1 and R2, the evolution of the anisotropy
parameter $\beta$ reflects the transition to more radial orbits after
the first pericentric approach, as the dwarfs change shapes to prolate
spheroids. The stellar orbits tend to become more isotropic with time,
especially for simulation R2 where the velocity distribution of the
dwarf at the end is completely isotropic ($\beta \approx 0$). The
velocity anisotropy of the remnant in experiment R1 is mildly radial
with $\beta \approx 0.3$.

Interestingly, the most strongly perturbed dwarf in simulation R2 is
almost entirely pressure-supported ($V_{\rm rot}/\sigma_{\ast}
\lesssim 0.1$) and its final shape is also spherically-symmetric ($b/a
\approx c/a \approx 1$). Furthermore, while the dwarf in experiment R2
develops a strong bar, its counterpart in simulation R3 does not. The
least evolved dwarf in experiment R3 remains oblate for the whole time
except for a short period after the first pericentric approach.
According to the criteria described in \S~\ref{subsec:criteria}, the
final dwarfs in simulations R1 and R2 would be classified as dSphs but
the remnant in experiment R3 would not as it has a disky shape and it
is still dominated by rotation ($V_{\rm rot}/\sigma_{\ast} > 1$).  It
is also important to highlight the strong association between bar
formation and the formation of dSphs reported in this set of
simulations.  Such a connection is observed in all but one of our
experiments.

The final structural parameters of the simulated dwarfs after $10$~Gyr
of evolution inside the primary galaxy are listed in
Table~\ref{table:summary}. Special emphasis should be placed on
columns 10 and~11 of this table.  Column 10 refers to whether a
tidally-induced bar was formed in the stellar distribution of the
dwarf galaxy during the course of its evolution inside the host.
Column 11 indicates whether a dSph was produced as a result of the
tidal interaction between the progenitor disky dwarf and the primary
galaxy.  The entries in column 6 demonstrate that the values of
$\beta$ in the dSph remnants range from isotropic to mildly radial ($0
\lesssim \beta \lesssim 0.5$). Given that a well established range for
the velocity anisotropy of observed dSphs does not yet exist, a direct
comparison with observations is not possible.  However, we note that
the magnitude of the radial velocity anisotropy in our dSphs is rather
similar to that observed in the outer parts of CDM halos formed in
cosmological simulations \citep[e.g.,][]{Cole_Lacey96}.  Lastly, the
orbital times listed in the last column of Table~\ref{table:summary}
are defined as the average time elapsed between consecutive apocentric
passages.

Figure~\ref{fig3} shows the surface density maps of the final stellar
distributions of the dwarfs in simulations R1, R2, and R3. Results are
presented for projections onto the $xz$, $yz$, and $xy$ planes and
particles are color-coded on a logarithmic scale, with hues ranging
from orange to white indicating increasing stellar density.  This
figure visually confirms the conclusions advanced above regarding the
degree of morphological transformation experienced by the dwarfs in in
this set of experiments.

\subsection{Eccentricity of the Orbit}
\label{subsec:orbit_eccentricity}

In this section, we explore the extent to which the tidal evolution of
a rotationally-supported dwarf is influenced by the eccentricity of
its orbit. As we discussed earlier, this effect is worth investigating
as the effective duration of the tidal shock, and thus the response of
the system to the tidal perturbation, depends on the orbital
eccentricity \citep[e.g.,][]{Gnedin_etal99}. To this end, we conducted
two additional simulations (R4, R5) in which we placed the dwarf
galaxy model D1 on orbits with eccentricities that are larger ($r_{\rm
  apo}/r_{\rm peri} = 10$) and smaller ($r_{\rm apo}/r_{\rm peri} =
2.5$) by a factor of $2$ from that of the reference experiment R1 (see
Table~\ref{table:simulations}). Simulations R4 and R5 should be viewed
as corresponding to a highly and a mildly eccentric orbit,
respectively. Figure~\ref{fig1} shows the orbital trajectories of the
dwarf galaxies from the centers of their hosts in these experiments.

Figure~\ref{fig4} shows the main results related to this set of
simulations.  As expected, this figure demonstrates that for a fixed
apocentric distance, higher eccentricity orbits induce a much stronger
transformation in the orbiting dwarfs.  The evolution of the various
parameters in this set of experiments proves to be analogous to the
one presented in Figure~\ref{fig2}. The dwarf galaxies in simulations
R4 and R5 evolve similarly to those in R2 and R3, respectively, and
the same strong link between bar formation and the formation of dSphs
is reported.  Indeed, the disky dwarf in experiment R4 develops a bar
and the final product is a nearly spherical ($b/a \approx c/a \gtrsim
0.9$), isotropic ($\beta \approx 0.03$), and non-rotating ($V_{\rm
  rot}/\sigma_{\ast} \approx 0.03$) dSph, with final properties akin
to those of the remnant in simulation R2.  The evolution of $V_{\rm
  max}$, $r_{\rm max}$, and mass within $r_{\rm max}$ is also similar
in the two cases. This is due to the fact that although the orbital
time of the dwarf in experiment R4 is significantly larger, is has a
smaller pericentric distance compared to that of R2.  On the other
hand, the dwarf in experiment R5 does not form a bar and remains
oblate, radially anisotropic and rotating, analogous to the one in
simulation R3, despite the shorter orbital time and larger number of
pericentric passages (see Figure~\ref{fig1}).

A last interesting thing to note is that the bar instability that
develops in the most strongly perturbed dwarf (R4) is weaker compared
to that of the reference simulation, despite the fact that the
pericentric distance is much smaller in the former case. The same
behavior, although to a lesser degree, is observed in
Figure~\ref{fig2}. We speculate that this is a consequence of the very
strong tidal forces in simulations R2 and R4. The tidal shocks may
heat the newly formed bar so substantially that they actually cause
the instability to rapidly dissipate.

To summarize, the results of this and the previous section highlight
that the effectiveness of the transformation into a dSph via tidal
stirring depends crucially on the orbital parameters of the progenitor
disky dwarfs. The relative importance of the orbital time and the
pericentric distance in this process will be addressed in
\S~\ref{sec:discussion}.


\begin{figure*}[t]
\begin{center}
  \includegraphics[scale=0.44]{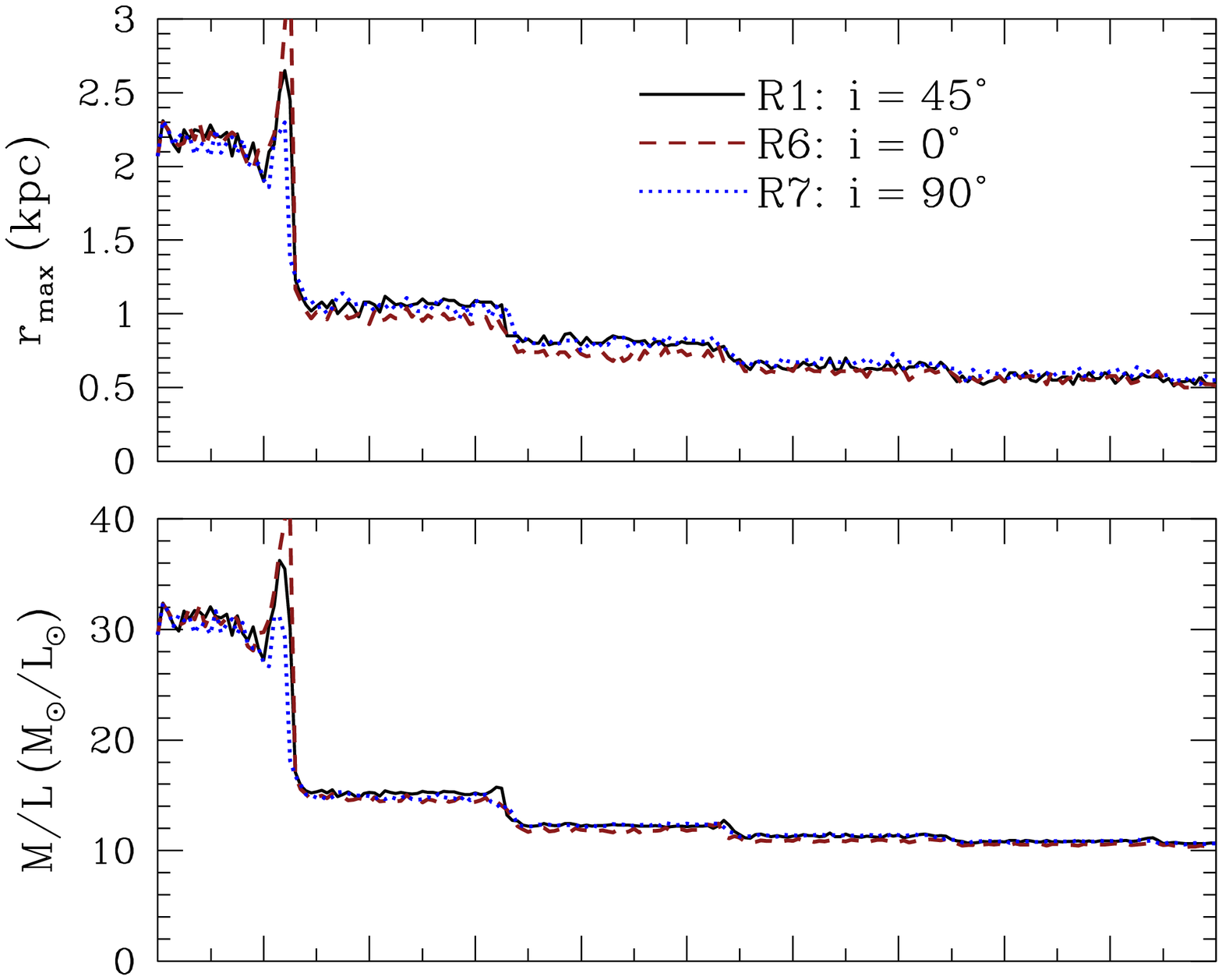}
  \includegraphics[scale=0.44]{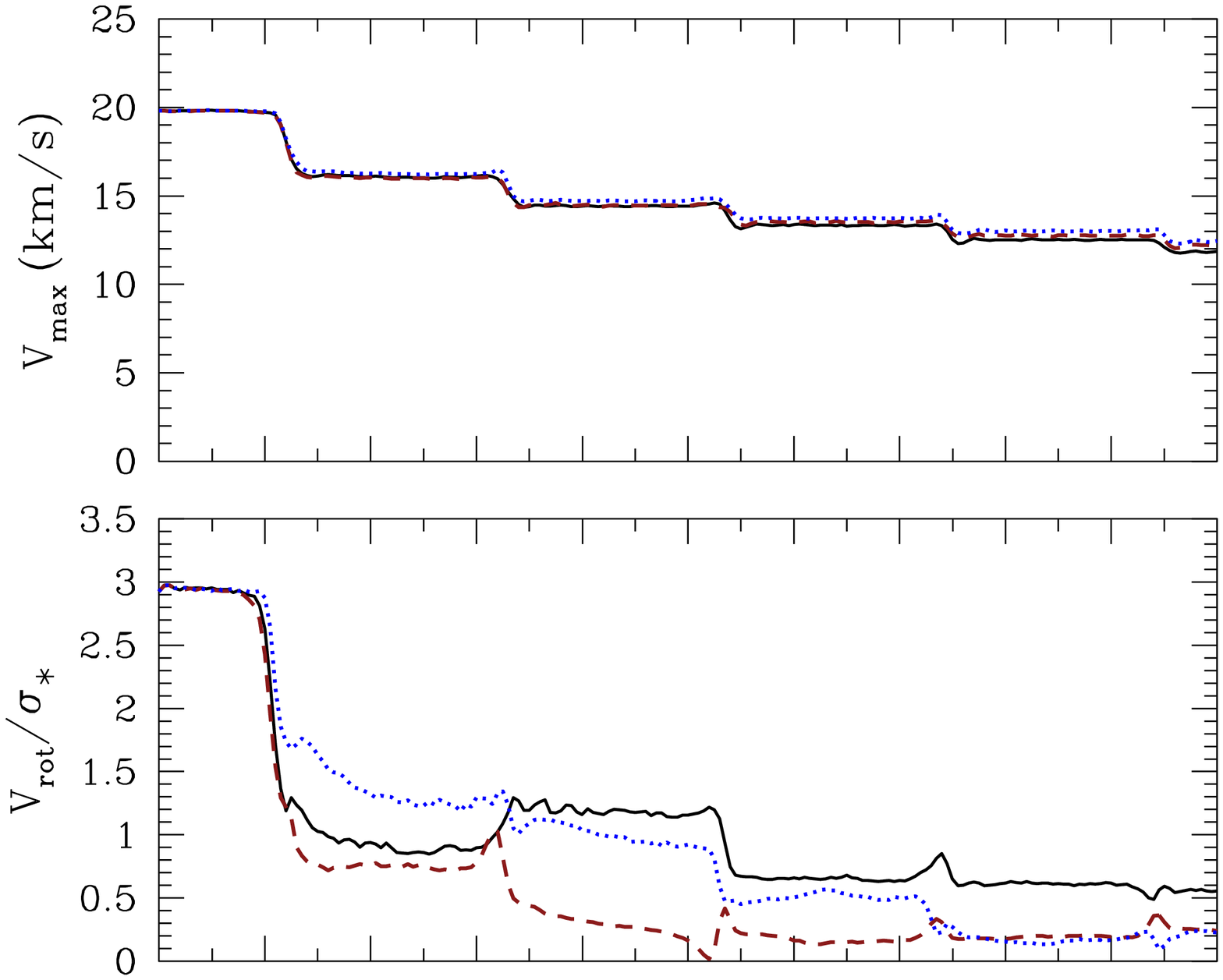}
  \includegraphics[scale=0.44]{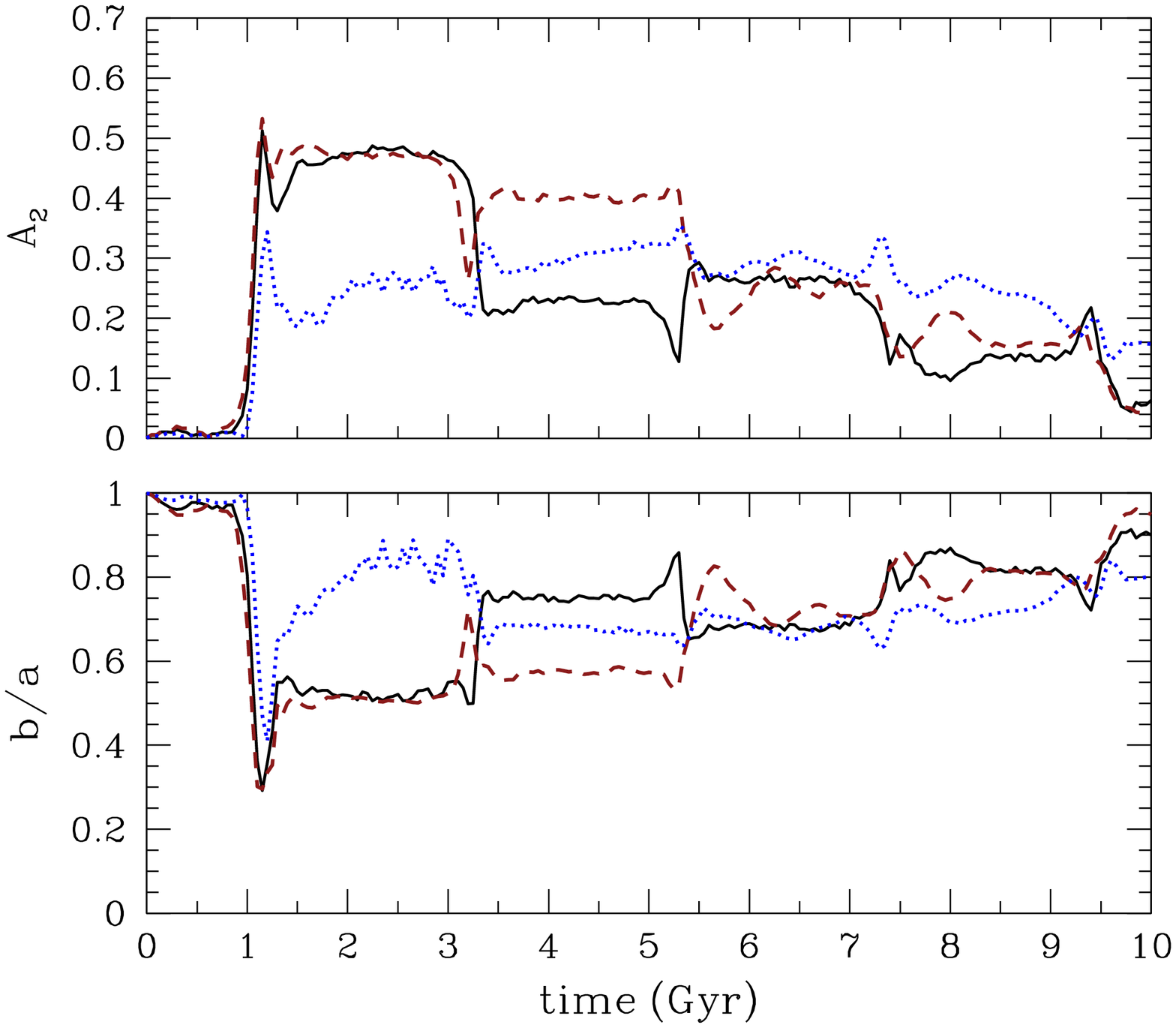}
  \includegraphics[scale=0.44]{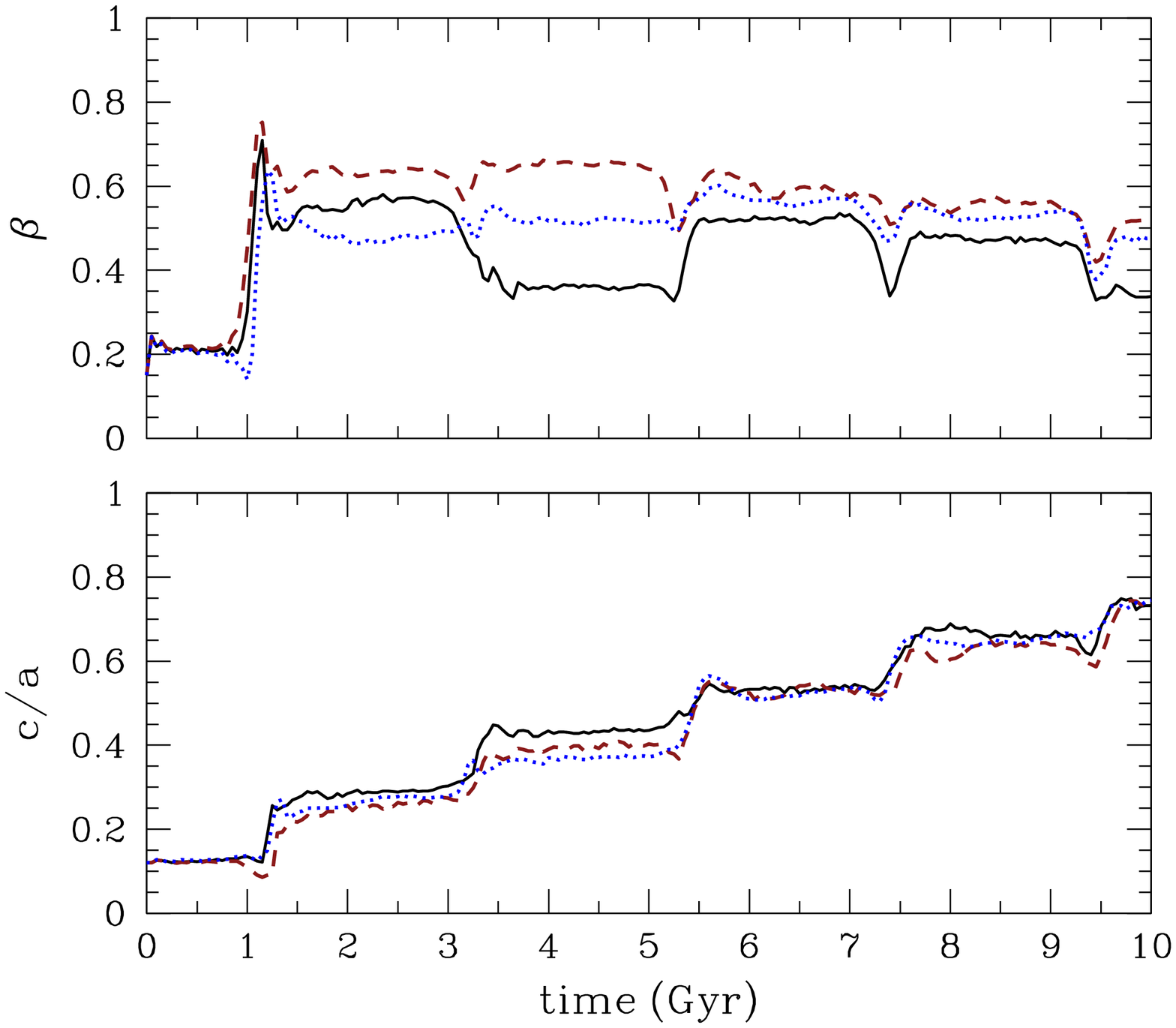}
\end{center}
\caption{Comparison of the evolution of various parameters as a function
  of time illustrating the dependence of the tidal transformation of
  disky dwarf galaxies on the inclinations of their disks with respect
  to the orbital plane, $i$. Results are presented for simulations R1,
  R6, and R7. For the specific choices of $i$ in this set of
  experiments, the tidal evolution of the rotationally-supported
  dwarfs inside their host galaxies and the efficiency of their
  transformation into dSphs depend very weakly on the disk
  inclination.
\label{fig5}}
\end{figure*}



\begin{figure*}[t]
\begin{center}
  \includegraphics[scale=0.44]{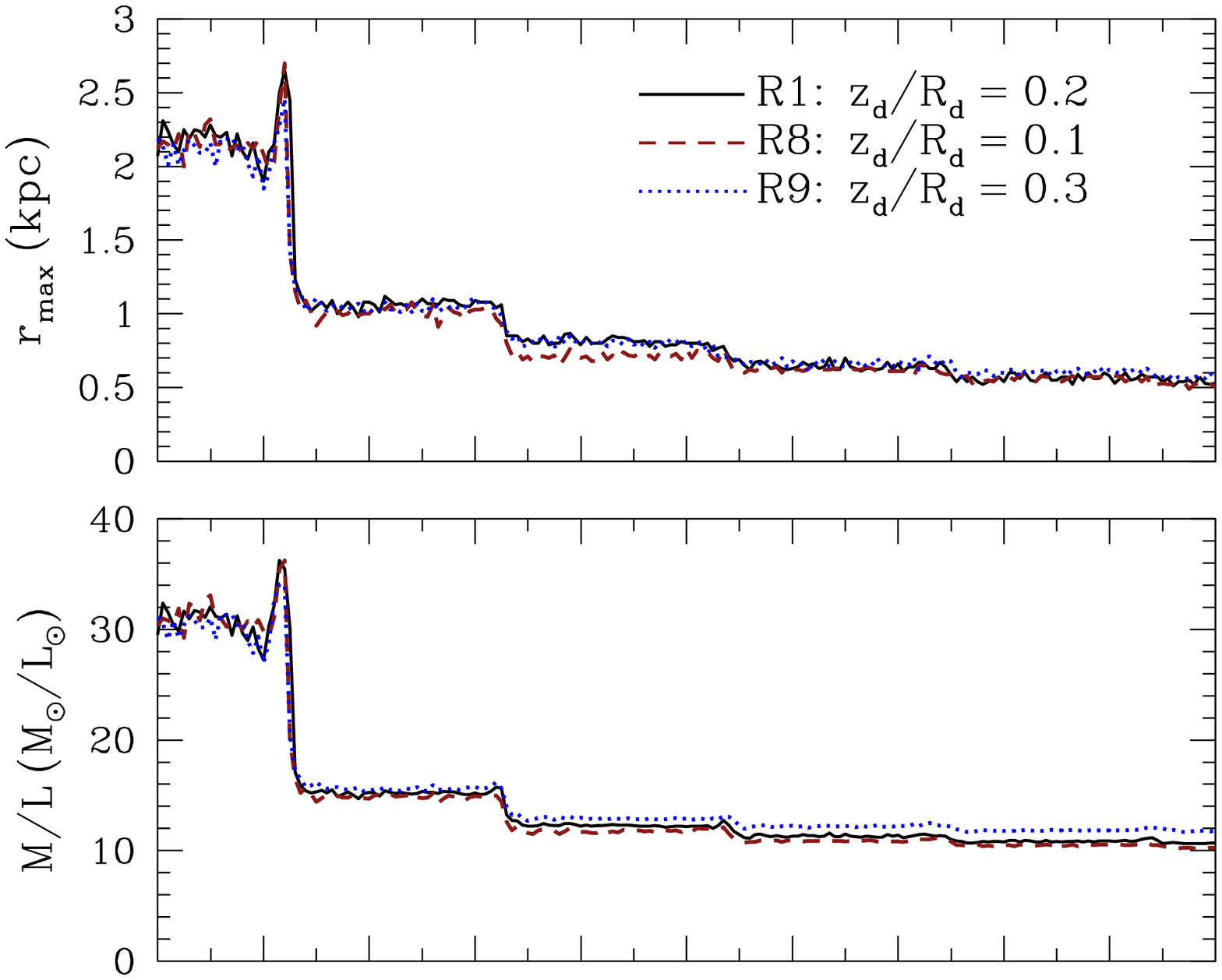}
  \includegraphics[scale=0.44]{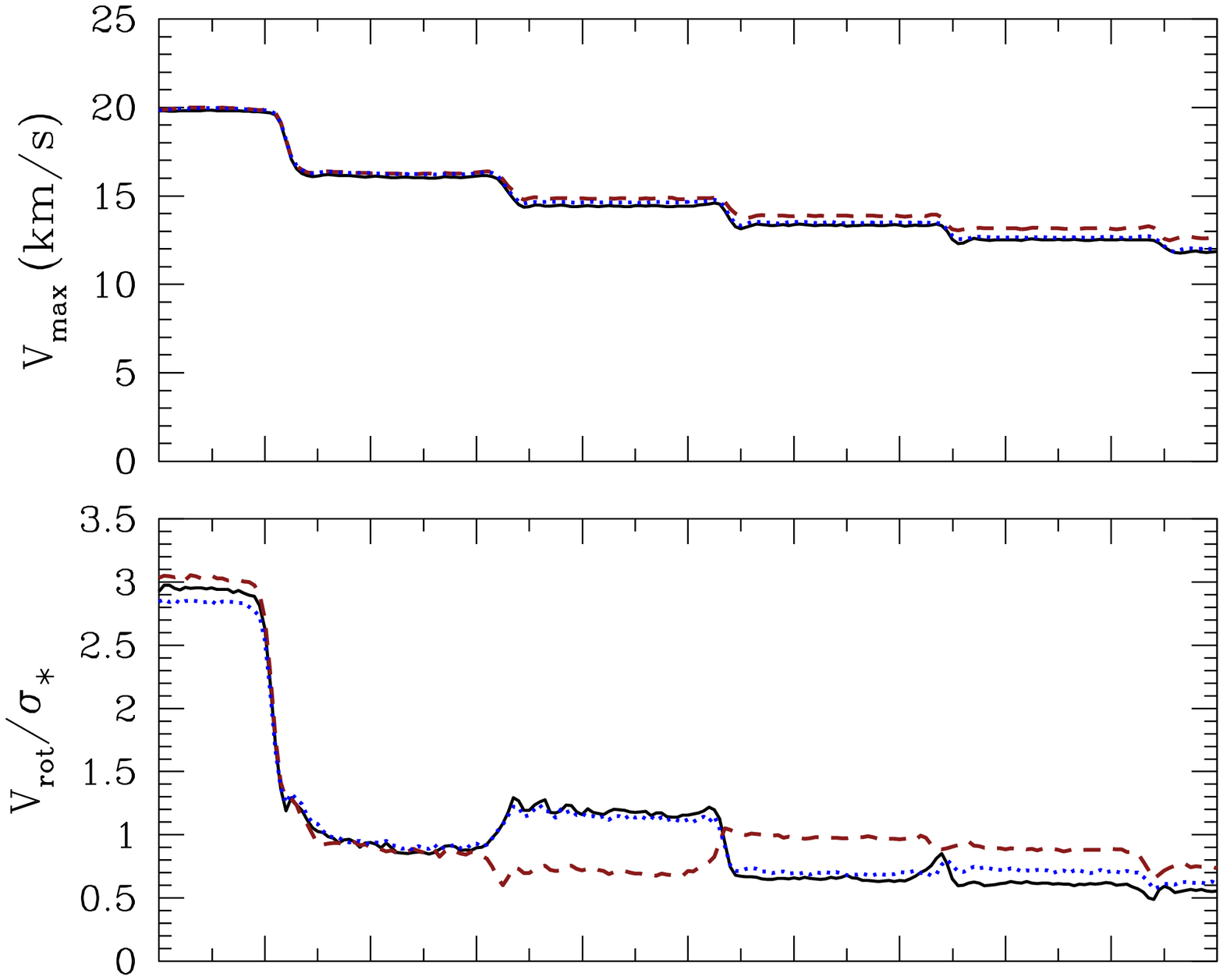}
  \includegraphics[scale=0.44]{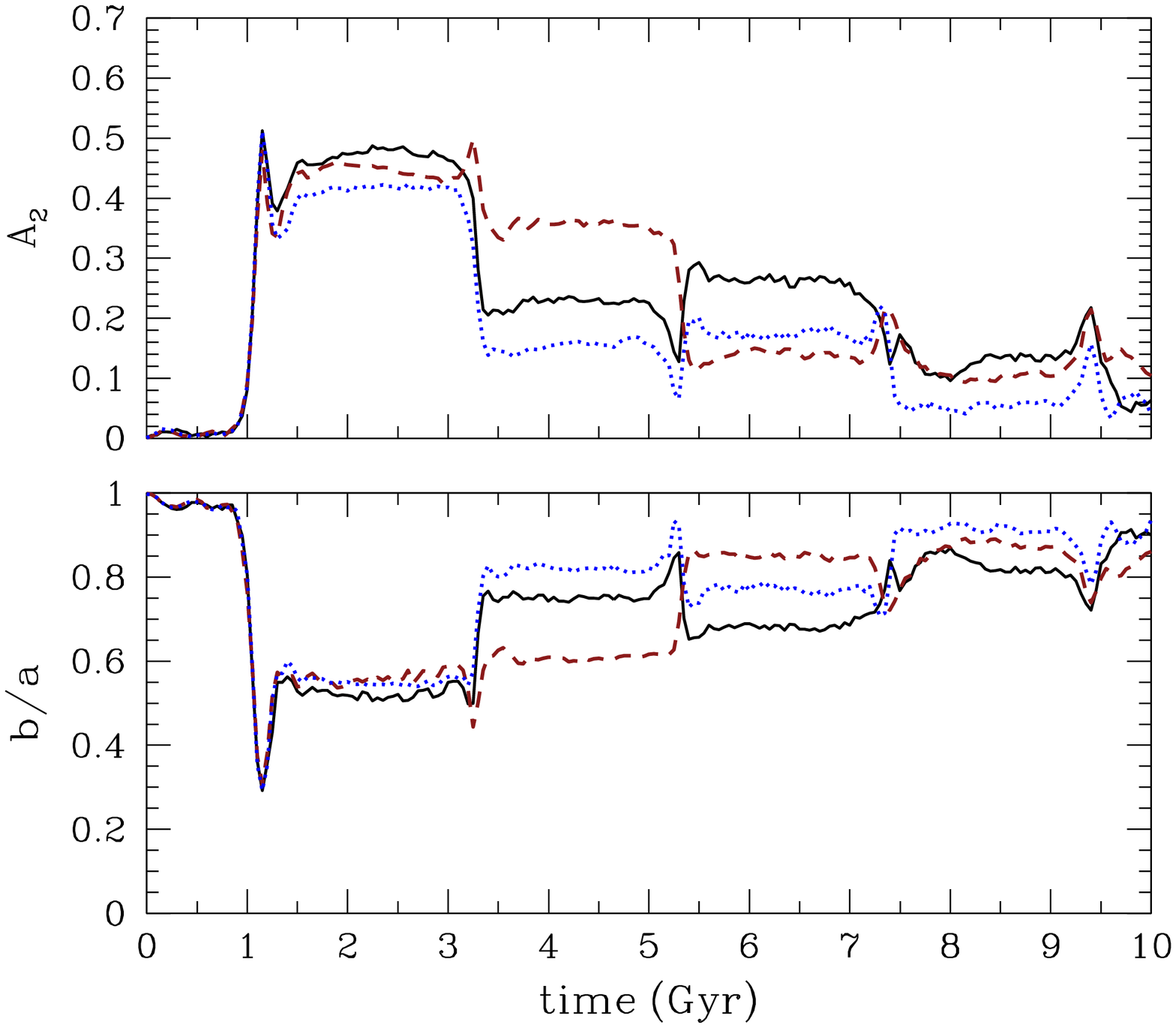}
  \includegraphics[scale=0.44]{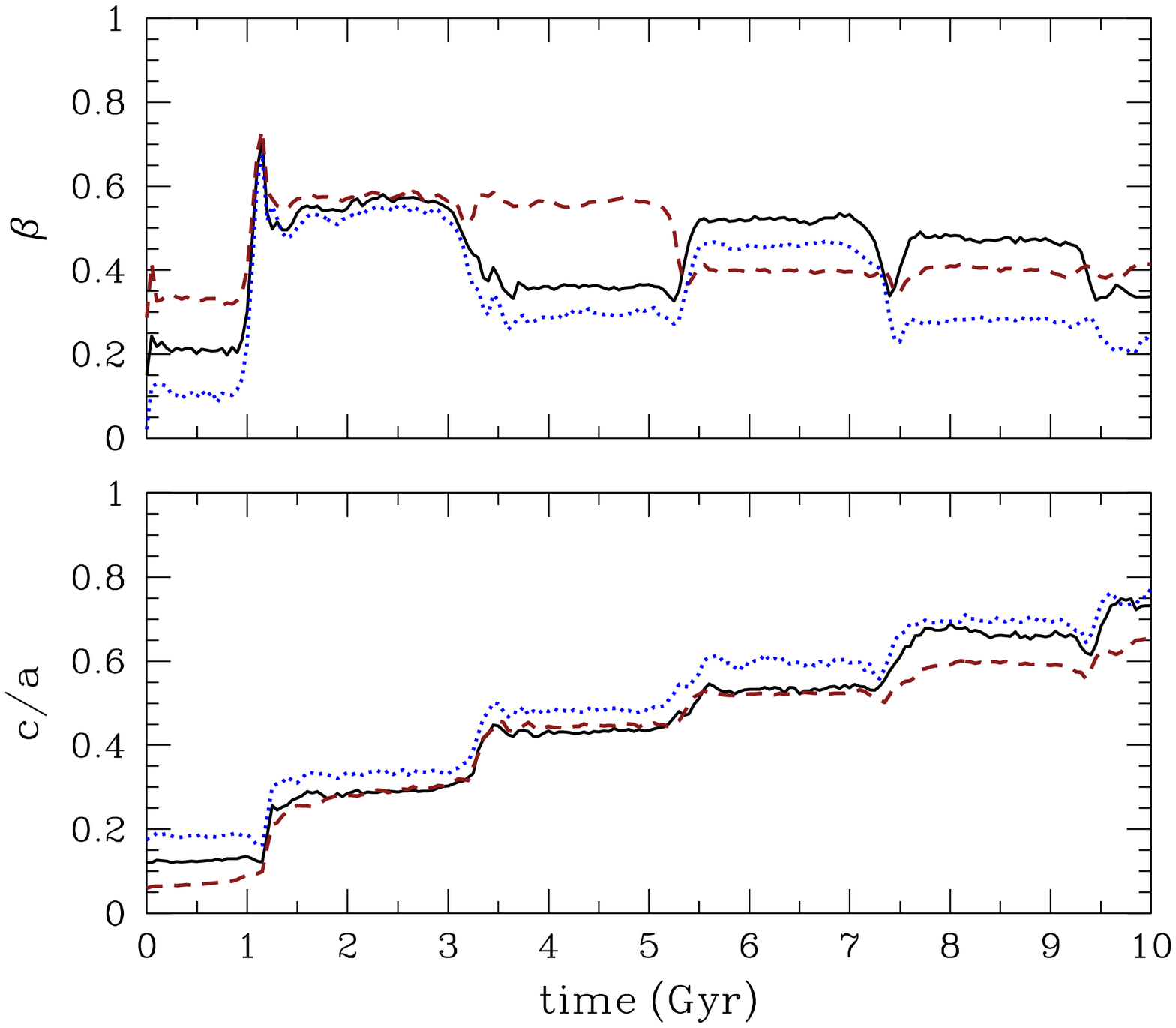}
\end{center}
\caption{Comparison of the evolution of various parameters as a function 
  of time illustrating the dependence of the tidal transformation of
  disky dwarf galaxies on the thicknesses of their disks. Disk
  thicknesses are parametrized as $z_d/R_d$ and results are presented
  for simulations R1, R8, and R9.  For the specific choices of
  $z_d/R_d$ in this set of experiments, the tidal evolution of the
  rotationally-supported dwarfs inside their host galaxies and the
  efficiency of their transformation into dSphs are essentially
  independent of the disk thickness.
\label{fig6}} 
\end{figure*}


\section{Efficiency of Tidal Stirring and Structural Parameters 
  of the Progenitor Disky Dwarfs}
\label{sec:struct_properties}

In this section, we explore how the initial structure of the
progenitor disky dwarfs could influence the outcome of their tidal
evolution inside the gravitational field of their host galaxies. We
begin by focusing on the parameters of the dwarf disk and discuss in
turn the effect of the disk inclination, thickness, mass, and scale
length. In \S~\ref{sec:halo_parameters}, we turn our attention to the
properties of the DM halo of the dwarf and investigate the extent to
which the effectiveness of tidal stirring is affected by the halo
concentration parameter and mass.

\subsection{Varying the Parameters of the Disk}
\label{sec:disk_parameters}

\subsubsection{Disk Inclination}
\label{subsec:disk_inclination}

We first investigate the extent to which the tidal evolution of a
late-type disky dwarf is affected by the initial inclination of its
disk with respect to the orbital plane. This is worth exploring as
stronger alignments between the orbital angular momentum of the dwarf
and the internal angular momentum of its disk can result in more
effective stripping \citep[e.g.,][]{Read_etal06b}. For this purpose,
we performed two additional simulations in which we placed the dwarf
galaxy model D1 on the same orbit as in simulation R1 after changing
the default inclination from $i=45\degrees$ to $i=0\degrees$ and to
$i=90\degrees$ (see Table~\ref{table:simulations}). Figure~\ref{fig5}
contains the results pertaining to this set of simulations.

Although the initial inclination of the dwarf disk is very different,
the size (in terms of $r_{\rm max}$), the mass (in terms of $V_{\rm
  max}$), and the $M/L$ ratio evolve fairly similarly.  This is a
consequence of two facts. First, DM halos were constructed with no net
angular momentum. Thus, their stripping should be independent of such
considerations and proceed in exactly the same way.  Second, the
dwarfs are not stripped down to such small scales that the alignment
between the stellar angular momentum and the orbital angular momenta
would have an important effect on the stripping of the stars
\citep[e.g.,][]{Read_etal06b}. Indeed, we find the decrease in mass
within $r_{\rm max}$ for both stars and DM to be independent of the
initial disk inclination.

It is important to note that \citet{Klimentowski_etal09a} reported
that disk inclination has a significant effect on the evolution of
$V_{\rm max}$ in similar tidal stirring experiments. Specifically,
they found that $V_{\rm max}$ decreased much more strongly in the
$i=0\degrees$ case compared to the $i=45\degrees$ and $i=90\degrees$
cases (see Figure~10 of \citealt{Klimentowski_etal09a}).  This was
because the DM halos in the \citet{Klimentowski_etal09a} experiments
were constructed with net angular momentum which was aligned with the
angular momentum of the dwarf disk. Moreover, the stellar disks were
much more extended compared to those of the present study and, as a
result, the stripping of the dwarfs down to the smallest scales of
stars did occur at some point during the evolution.  In fact, the mass
within $r_{\rm max}$ of both stars and DM in the
\citet{Klimentowski_etal09a} experiments decreased more in the
$i=0\degrees$ case lending support to the previous arguments.

The behavior of the bar strength amplitude $A_2$ in Figure~\ref{fig5},
demonstrates that all dwarfs possess a tidally-induced bar after the
first pericentric passage. However, the bar appears to be much weaker
in simulation R7, indicating that the tidal forces acting on the dwarf
disk were smaller in this case. During subsequent pericentric passages
the situation becomes progressively more complex as strong variations
in the values of $A_2$ are observed among the three simulations. At
the end of the evolution, the tidally-induced bars are diminished in
all cases.  Investigating the reasons for the very different evolution
of the bars in this set of experiments is clearly beyond the scope of
the present paper. However, it is interesting to note that recent
targeted numerical experiments have highlighted the importance of
stochasticity in the evolution of {\it isolated} disk galaxies leading
to macroscopic differences in the evolution of bars
\citep{Sellwood_Debattista09}.  Among the causes for this stochastic
evolution, \citet{Sellwood_Debattista09} identified interference
between multiple disk modes, amplification of noise, bending modes,
dynamical friction between the bar and the halo, and intrinsic chaos.
Obviously, further analysis would be required to determine whether the
observed behavior of the bars in this set of simulations can also be
attributed to stochasticity and/or possibly to other considerations
such as the different initial disk inclinations and resonances.

The final systems in simulations R6 and R7 are characterized by $c/a
\approx 0.7$ and $V_{\rm rot}/\sigma_{\ast} \approx 0.2$ (see
Table~\ref{table:summary}) and, therefore, would be classified as
dSphs according to the criteria described in \S~\ref{subsec:criteria}.
While the evolution of kinematics and shape is, in broad terms, fairly
similar among the three experiments, there are some interesting
differences that are worth mentioning. For example, the rotation is
lost more quickly in simulation R6 ($i=0\degrees$), where $V_{\rm
  rot}/\sigma_{\ast}$ drops almost to zero at $\sim 5$~Gyr and remains
constant thereafter. The substantial and rapid decrease of $V_{\rm
  rot}/\sigma_{\ast}$ in this experiment is explained by the presence
of a very strong bar instability between $\sim 1$ and $\sim 5$~Gyr. We
also note that in experiment R7 the axis ratio $b/a$ remains almost
constant with time and significantly different from unity, so that the
stellar component is triaxial rather than prolate at early times.

\subsubsection{Disk Thickness}
\label{subsec:disk_thickness}

In this section, we investigate the dependence of the tidal evolution
of a disky dwarf on the thickness of its disk. Such an investigation
is important for two reasons. First, feedback mechanisms and turbulent
motions in dwarf galaxies should be effective in producing thicker
systems.  Second, the efficiency of tidal heating and the strength of
bar instabilities that are both vital for tidal stirring should be
different in thicker stellar distributions
\citep[e.g.,][]{Kazantzidis_etal09}.  To this end, we generated two
additional dwarf galaxy models that were identical to D1 except for
their scale height, which was chosen equal to $z_d/R_d = 0.1$ and
$z_d/R_d = 0.3$. We refer to these models as ``D2'' and ``D3'',
respectively (see Table~\ref{table:init_param}).  The choice for the
larger thickness in model D2 reflects the greater importance of
pressure support in low-mass galaxies and was motivated by both
observations as well as results of recent numerical simulations of the
formation of isolated galaxies \citep{Kaufmann_etal07}, including
systems with $V_{\rm max} \approx 20 \kms$ as our default dwarf model.
On the other hand, model D3 with the thinner disk was constructed
mainly for completeness. After building these two new dwarf models, we
placed them on the same orbit as in simulation R1 and followed their
tidal evolution inside the host galaxy. We refer to these experiments
as ``R8'' and ``R9'' (see Table~\ref{table:simulations}) and present
the relevant results in Figure~\ref{fig6}.

Overall, for the specific choices of initial thickness in this set of
experiments, no significant difference in the evolution of the dwarfs
can be discerned. Had we adopted thicker and/or thinner disks, such
differences may have been more pronounced. The final properties of the
systems in simulations R8 and R9 indicate that the remnants can be
classified as dSphs (see Table~\ref{table:summary}).

Figure~\ref{fig6} demonstrates that $r_{\rm max}$, $V_{\rm max}$, and
the $M/L$ ratio evolve almost identically in the three experiments. A
similarly strong bar is induced in all dwarfs after the first
pericentric approach. While the thicker disk is initially more
isotropic by construction, all three dwarfs exhibit nearly identical
values of $\beta$ between the first and second pericentric approach.
At later times, the evolution is somewhat different, but generally the
the stellar component of the thicker disk (R9) has a more spherical
shape and is characterized by more isotropic orbits.  All dwarfs end
up mildly triaxial and very similar to the remnant in simulation R1,
but their paths to this state are slightly different.  Occasionally,
departures from the typical trend of decreasing rotational velocity
occur (see Figure~\ref{fig11}) but they happen at different
pericentric passages for different experiments. We come back to this
issue in \S~\ref{sec:discussion}.


\begin{figure*}[t]
\begin{center}
  \includegraphics[scale=0.44]{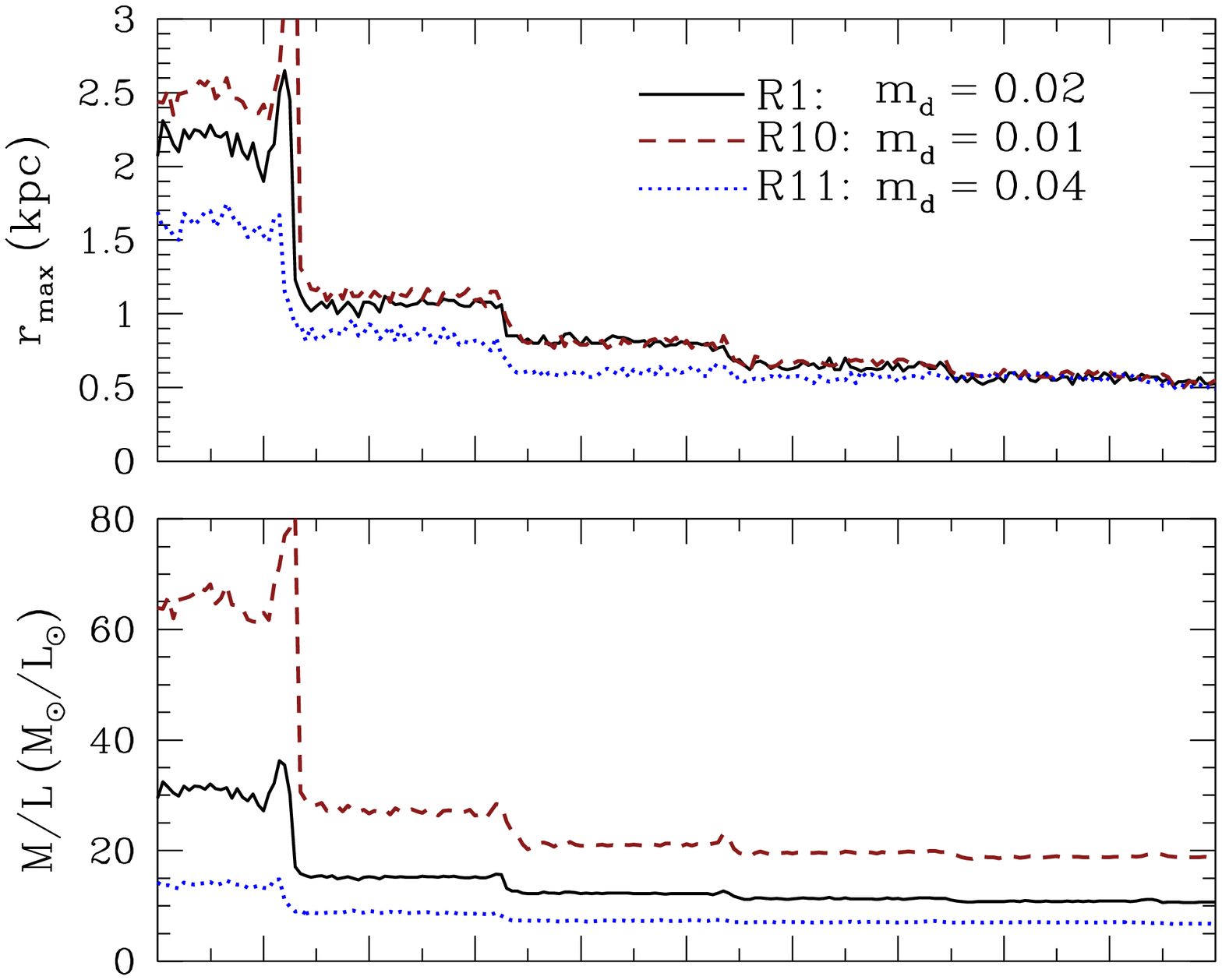}
  \includegraphics[scale=0.44]{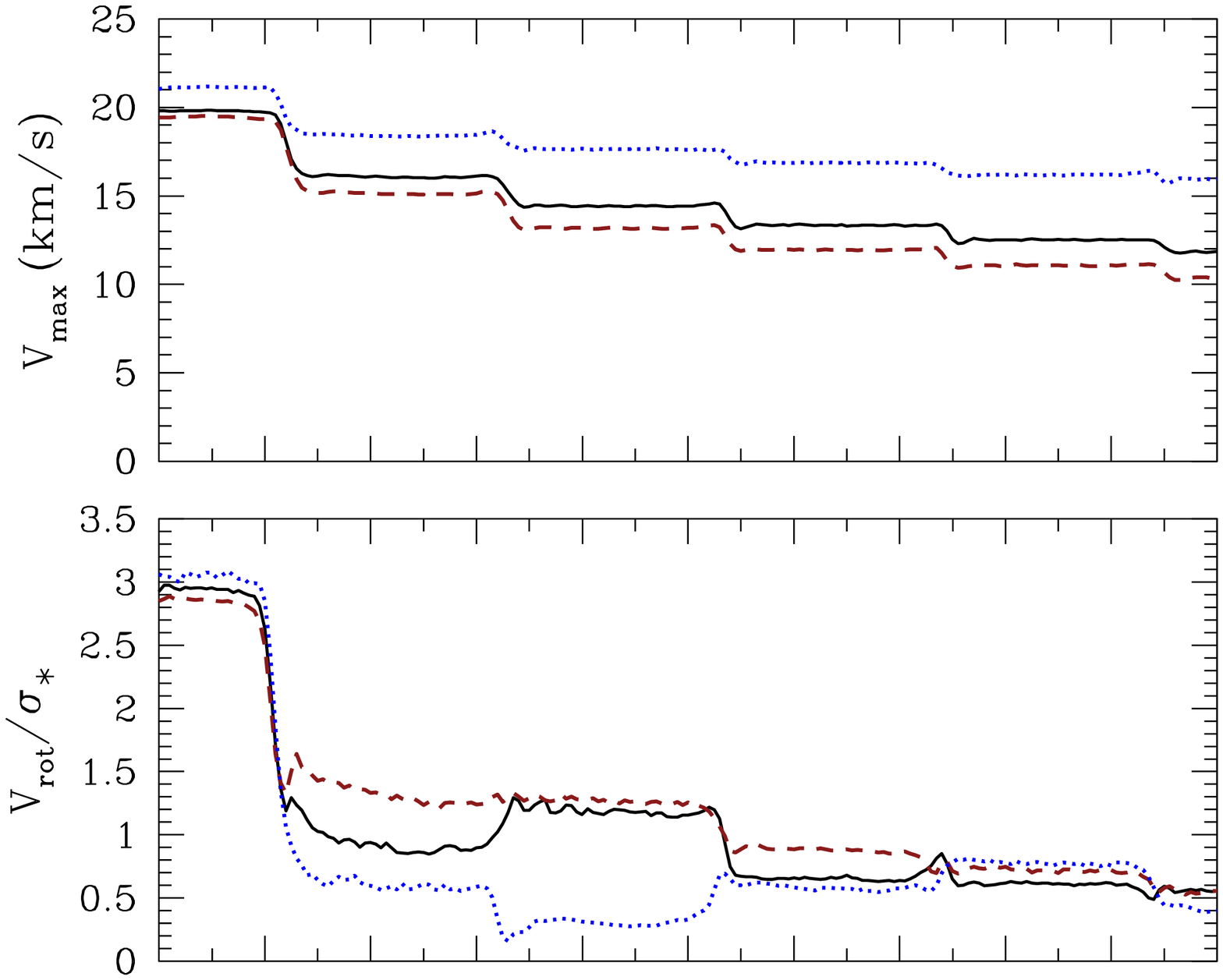}
  \includegraphics[scale=0.44]{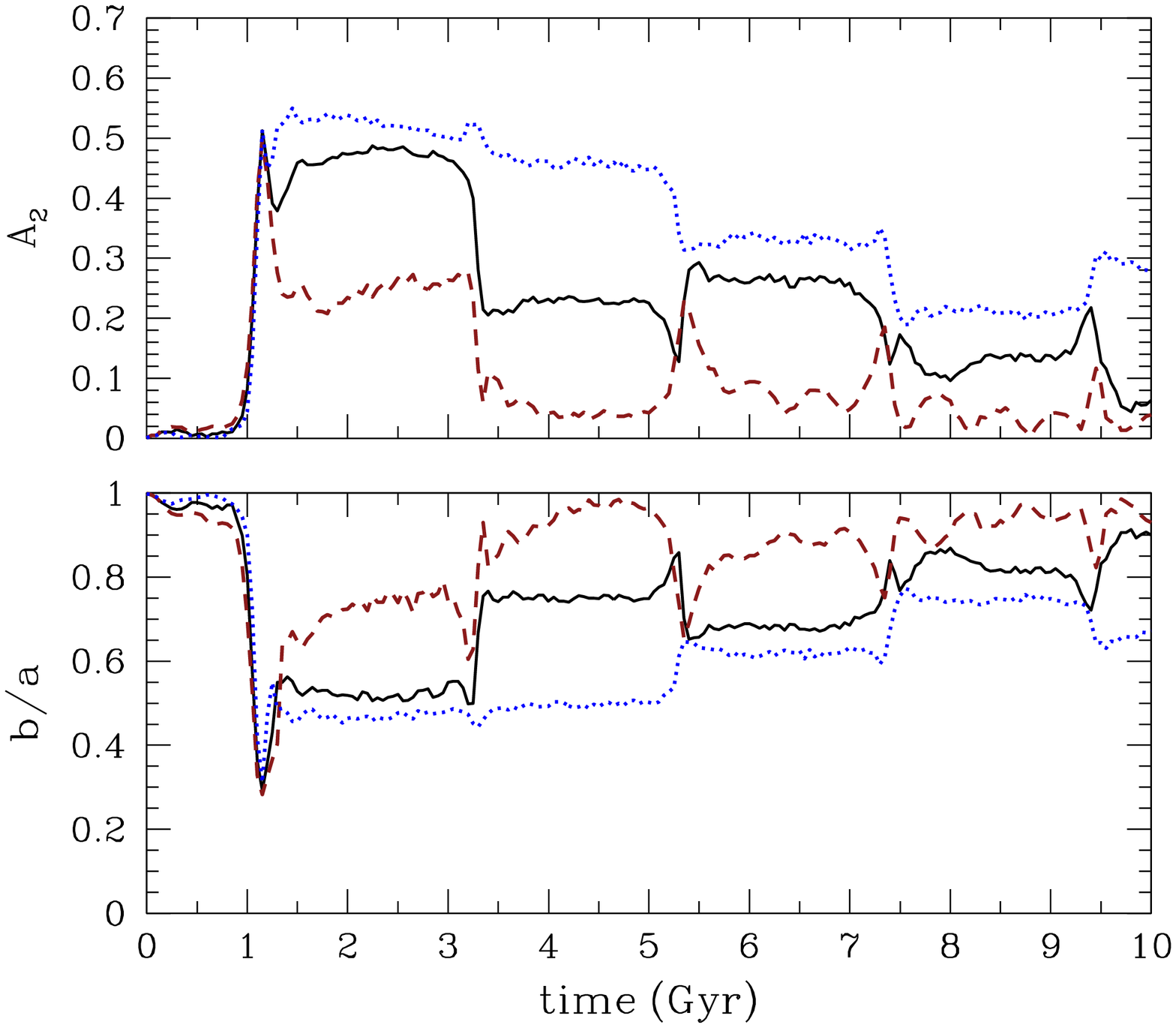}
  \includegraphics[scale=0.44]{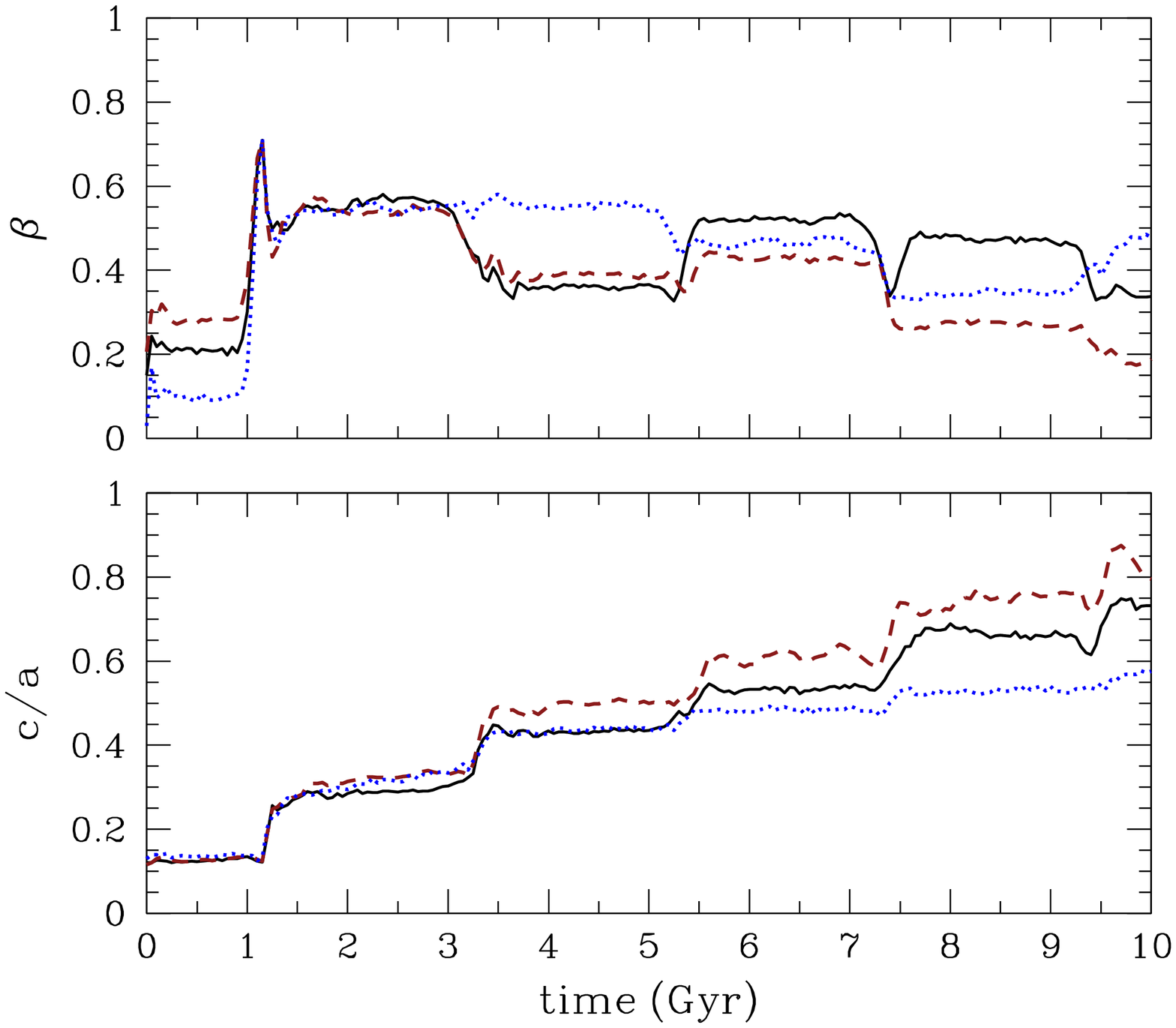}
\end{center}
\caption{Comparison of the evolution of various parameters as a function 
  of time illustrating the dependence of the tidal transformation of
  disky dwarf galaxies on the masses of their disks. Disk masses are
  parametrized as a given fraction, $m_d$, of the halo mass, $M_h$,
  and results are presented for simulations R1, R10, and R11.
  Rotationally-supported dwarfs with less massive disks exhibit
  stronger tidal evolution inside their host galaxies and the
  efficiency of their transformation into dSphs is augmented
  significantly.
\label{fig7}}
\end{figure*}


\subsubsection{Disk Mass}
\label{subsec:disk_mass}

In this section, we investigate the degree to which the tidal
evolution of a rotationally-supported dwarf is affected by the mass of
its disk. In order to ascertain this, we generated two additional
dwarf galaxy models that were identical to D1 except for their disk
mass, which differed by a factor of $2$: $m_d=0.01$ and
$m_d=0.04$\footnote{We note that formally, according to the model of
  \citet{Mo_etal98}, varying the disk mass and keeping the other
  relevant parameters ($M_h$, $\lambda$, and $c$) constant would
  result in different scale lengths for the dwarf disk.  However, for
  consistency and in order to isolate the effect of a single parameter
  on the tidal evolution of the dwarfs, we adopt identical disk scale
  lengths in this set of experiments. The same convention applies to
  the simulations described in \S~\ref{subsec:halo_concentration} in
  which we construct dwarf models with different halo concentration
  parameters.}. We refer to these models as ``D4'' and ``D5'',
respectively (see Table~\ref{table:init_param}). This range of $m_d$
values is consistent with that inferred from exquisite measurements of
the mass distribution of nearby dwarfs combining high resolution gas
kinematics in the THINGS survey and deep Spitzer photometry (Oh et al.
2010, in preparation).  It is also in agreement with results of
hydrodynamical simulations of dwarf galaxy formation
\citep[e.g.,][]{Tassis_etal03,Governato_etal10}.  After building these
two new dwarf models, we placed them inside the host galaxy on the
same orbit as in simulation R1 and followed their tidal evolution.  We
refer to these experiments as ``R10'' and ``R11'' (see
Table~\ref{table:simulations}) and present their results in
Figure~\ref{fig7}.

This figure demonstrates that the mass of their disk has a substantial
effect on the tidal evolution of the disky dwarfs.  Interestingly,
although the initial values of $r_{\rm max}$ are different and smaller
for the more massive disks due to the more concentrated mass
distribution, at the end of the evolution they converge to nearly the
same values.  The opposite is true for the mass, as quantified by
$V_{\rm max}$.  Indeed, the least massive disk increases the
susceptibility of the dwarf to tidal effects and the mass loss within
$r_{\rm max}$ is larger in this case.  Overall, the results presented
in Figure~\ref{fig7} indicate that the masses of the dwarfs are
differentiated as a result of the tidal evolution.

The $M/L$ ratios, which differed significantly in the beginning,
evolve in a fairly different way. The most massive disk, which had the
smallest value of $M/L$ initially, is affected most weakly and its
$M/L$ ratio decreases only slightly. The weak evolution of the $M/L$
ratio in this case can be attributed to the fact that the initial
$r_{\rm max}$ is very small (and remains so until the end of the
evolution).  As a result, $r_{\rm max}$ probes the central regions of
the dwarf, where the reasons that we have indicated in
\S~\ref{subsec:orbit_size} for the preferential stripping of DM over
stars do not apply. Indeed, DM particles within this inner region are
very unlikely to be on orbits of high enough eccentricity to have
apocenters larger than the tidal radius and be stripped. Moreover, the
slopes of the density profiles of stars and DM are similar below $\sim
1$~kpc, implying that the average binding energies of DM and stars do
not differ significantly in this region.  Interestingly, despite the
different evolution, the hierarchy of $M/L$ is preserved; the dwarf
which has the highest $M/L$ ratio initially also exhibits the highest
$M/L$ ratio in the end.

The mass of the disk affects the evolution of the stellar shape
substantially, and to some degree also influences that of the stellar
kinematics. Owing to its higher self-gravity, the most massive disk
(R11) develops the strongest tidally-induced bar after the first
pericentric approach, leading to a pronounced decrease in $V_{\rm
  rot}/\sigma_{\ast}$. We note that the values of $A_2$ remain the
largest in this case until the end of the evolution ($t=10$~Gyr). As a
result, the dwarf galaxy in simulation R11 retains a strongly prolate
shape with final values of $c/a \approx 0.6$ and $b/a \approx 0.7$. On
the other hand, the dwarf with the least massive disk (R10) remains
triaxial rather than prolate for most of the time.  Due to the
significantly lower disk self-gravity in this case, the tidal shocks
heat the disk much more effectively, producing a quite spherical
stellar component in the end ($c/a \approx 0.8$, $b/a \approx 0.9$).

However, the interpretation of the evolution of kinematics is less
obvious.  While the least massive disk loses its rotation
monotonically (see Figure~\ref{fig11}), the evolution of its most
massive counterpart is more chaotic with occasional strong increases
of the rotational velocity, even after two consecutive pericentric
passages (third and fourth). We note that this behavior is observed in
only one more simulation (R12; see next section) but is not as strong
as reported here. Overall, the values of $c/a$, in conjunction with
the fact that $V_{\rm rot}/\sigma_{\ast} \lesssim 0.6$, indicate that
the final systems in simulations R10 and R11 would be classified as
dSphs (see Table~\ref{table:summary}).


\begin{figure*}[t]
\begin{center}
  \includegraphics[scale=0.44]{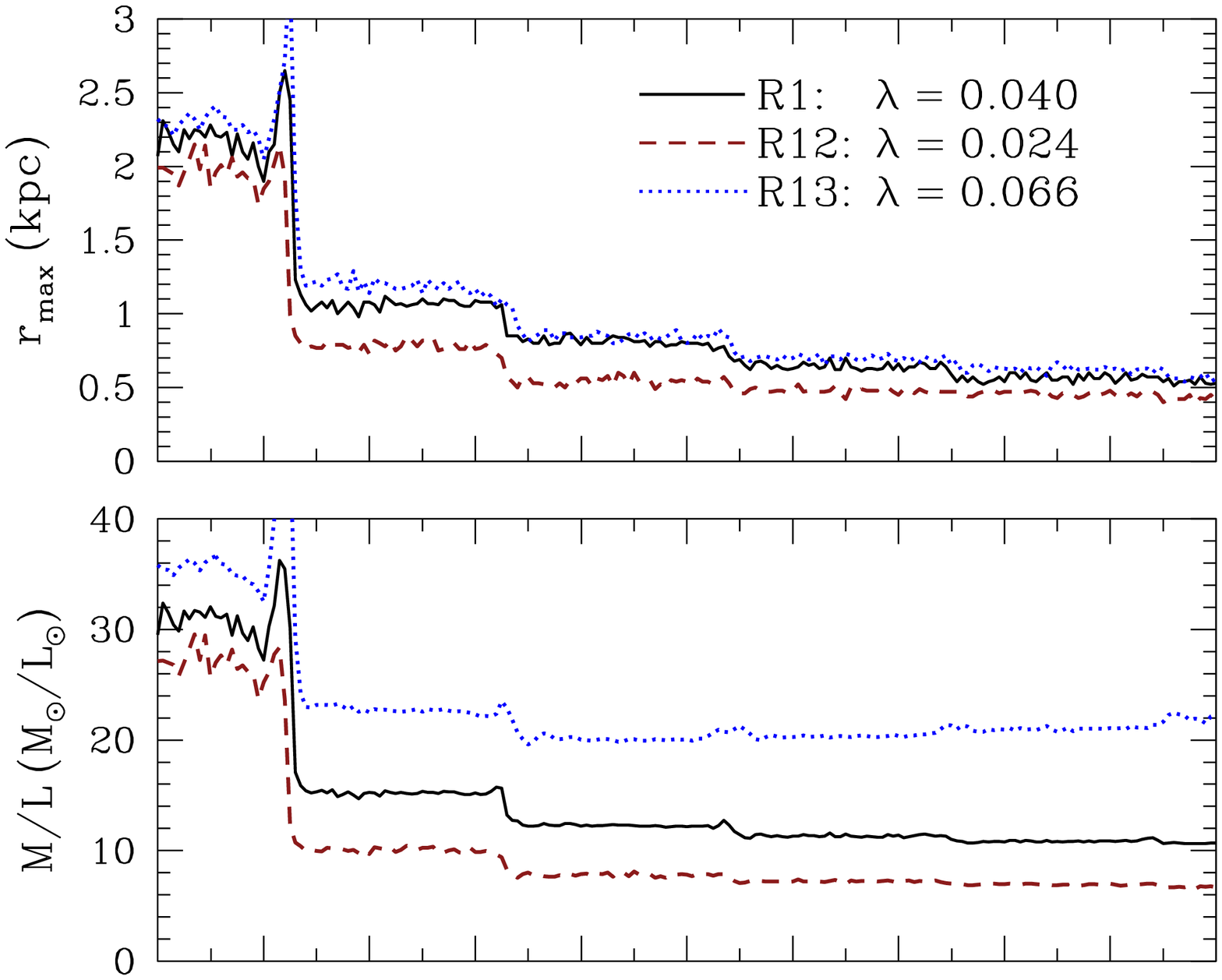}
  \includegraphics[scale=0.44]{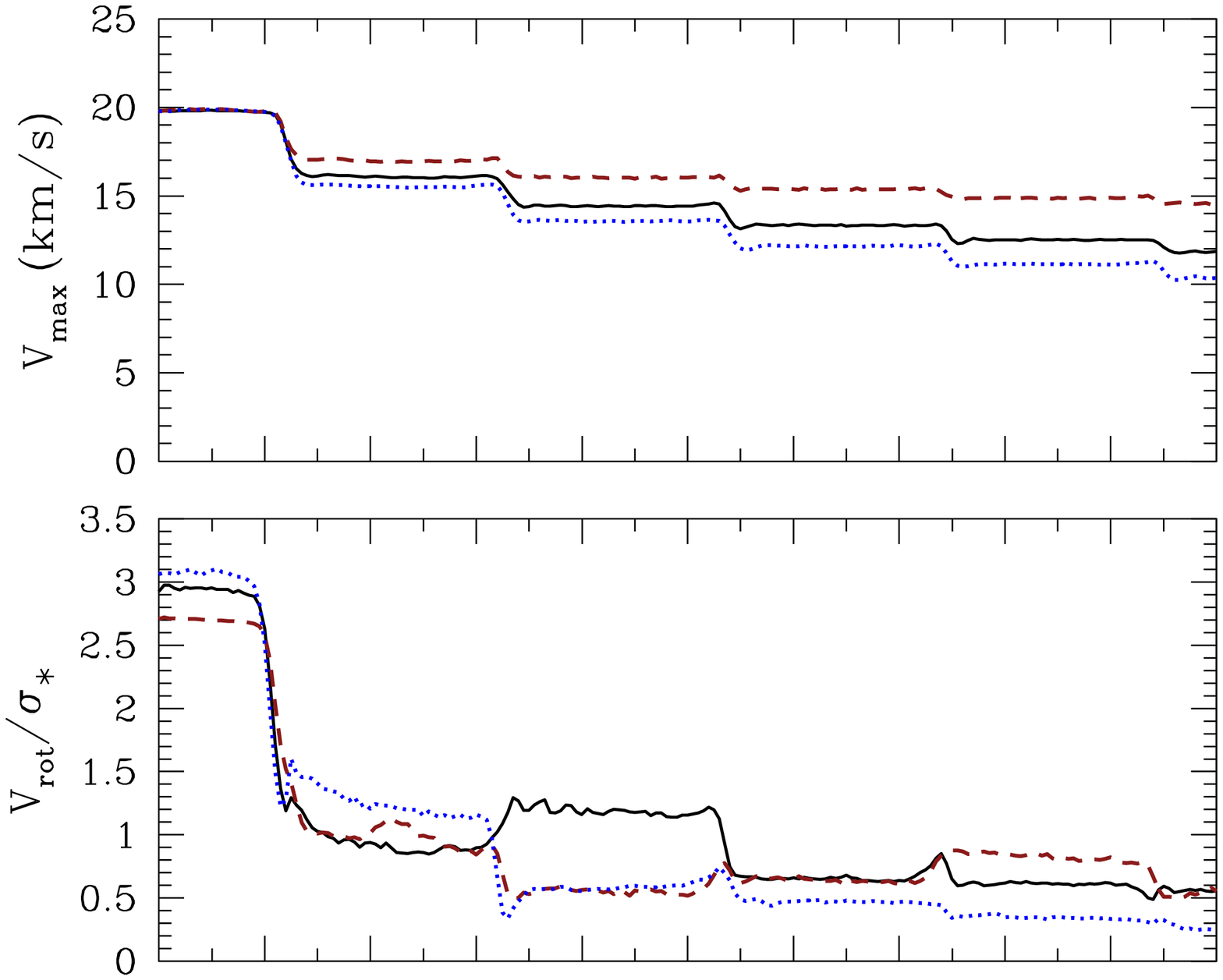}
  \includegraphics[scale=0.44]{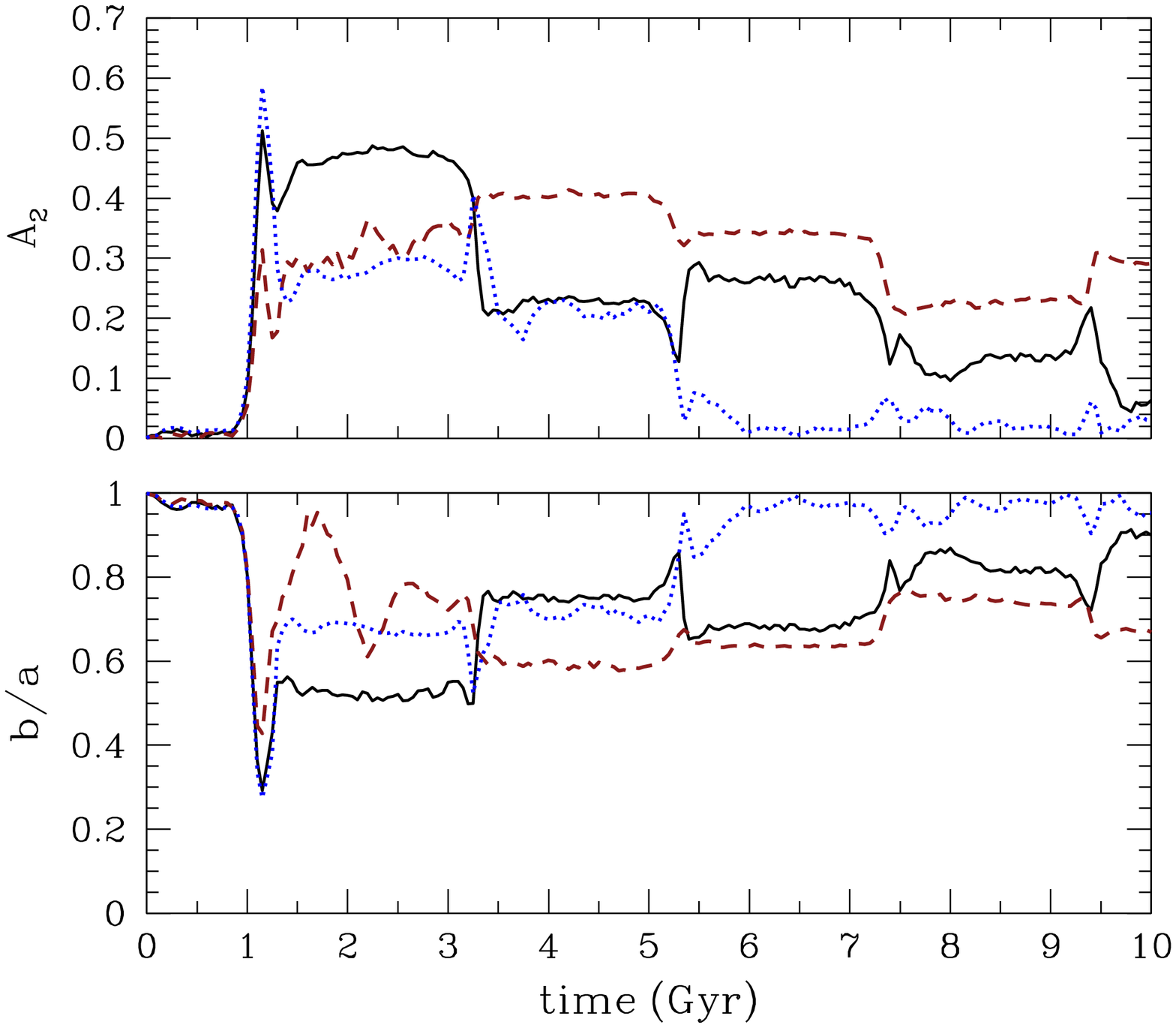}
  \includegraphics[scale=0.44]{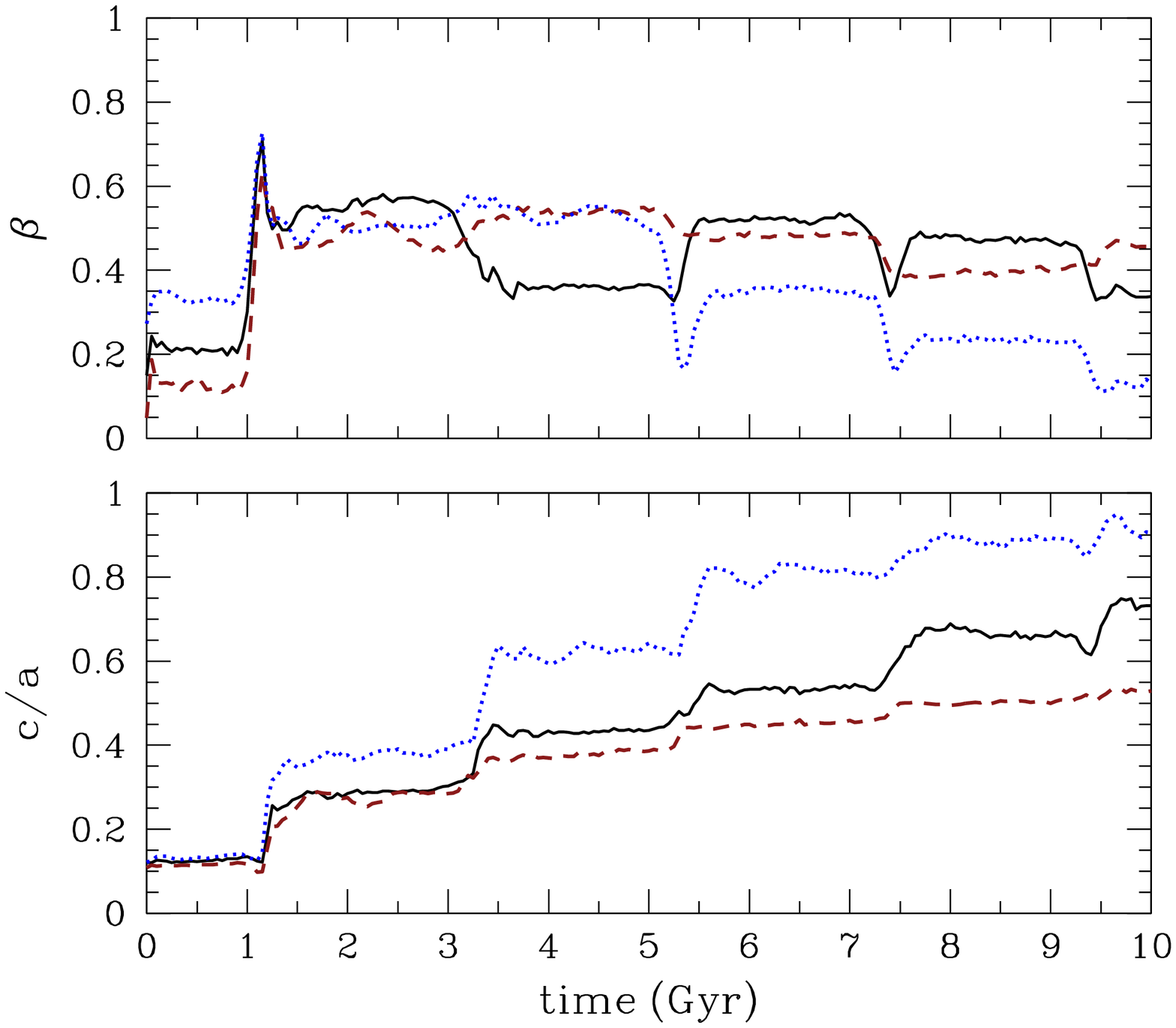}
\end{center}
\caption{Comparison of the evolution of various parameters as a function 
  of time illustrating the dependence of the tidal transformation of
  disky dwarf galaxies on the radial scale lengths of their disks,
  $R_d$.  Results are presented for simulations R1, R12, and R13. The
  spin parameters of the dwarf DM halos, $\lambda$, used to determine
  the disk scale lengths in this set of experiments are indicated in
  the labels (see text for details). The corresponding values of $R_d$
  for the three different choices of $\lambda$ are equal to $R_d =
  (0.41, 0.25, 0.66)$~kpc, respectively (see
  Table~\ref{table:init_param}).  Rotationally-supported dwarfs with
  more extended disks experience stronger tidal evolution inside their
  host galaxies and the efficiency of their transformation into dSphs
  is enhanced substantially.
\label{fig8}}
\end{figure*} 


\subsubsection{Disk Scale Length}
\label{subsec:disk_scale.length}

This section addresses the extent to which the tidal evolution of a
disky dwarf is influenced by the radial scale length of its disk,
$R_d$. For this purpose, we generated two additional dwarf galaxy
models that were identical to D1 except for the scale lengths of their
disks.  We reiterate that the disk scale length is not a free
parameter in our simulations, but rather is derived via the
semi-analytic galaxy formation model of \citet{Mo_etal98}. Because of
this choice our method requires us to assign values to the halo spin
parameter $\lambda$ in order to derive the scale length of the dwarf
disk, despite the fact that the halos of our dwarfs are constructed
with no net angular momentum.

Recall that we chose $\lambda = 0.04$ for our reference dwarf model
D1, close to the median value of halo spins found in cosmological
$N$-body simulations
\citep[e.g.,][]{Bullock_etal01b,Shaw_etal06,Maccio_etal07,Bett_etal07}.
To investigate the effect of the disk scale length, we initialized two
additional dwarf models with $\lambda = 0.024$ and $\lambda = 0.066$.
These values correspond to the $1\sigma$ deviations from the median
$\lambda$, assuming a log-normal distribution with $\lambda_{\rm med}
=0.04$ and $\sigma_{\lambda} = 0.5$, in agreement with the previous
cosmological studies.  We refer to these dwarf models as ``D6'' and
``D7'', respectively, and list their disk scale lengths in
Table~\ref{table:init_param}. In the corresponding experiments ``R12''
and ``R13'', we placed these dwarfs on the same orbit as in simulation
R1 and followed their tidal evolution inside the host galaxy. Note
that for consistency we have kept the thickness $z_d/R_d$ in this set
of experiments constant.  As a result, the disk becomes larger
(smaller) in both the radial and vertical direction when $\lambda$ is
increased (decreased).  The relevant results are presented in
Figure~\ref{fig8} which demonstrates that the remnant dwarfs in this
set of simulations would be classified as dSphs
(see Table~\ref{table:summary}).

This figure shows that the scale length of the dwarf disk is a
parameter that affects significantly the tidal evolution of the disky
dwarfs.  The initial value of $r_{\rm max}$ is smaller for the more
compact disk in simulation R12 and it remains so until the end of the
evolution. This experiment does not show the characteristic temporary
increase of $r_{\rm max}$ at the first pericentric approach observed
in the reference simulation R1. This is because the more compact disk
diminishes the ability of the host tidal field to distort the inner
regions of the dwarf galaxy. As expected, the mass loss within $r_{\rm
  max}$ measured by the decreasing value of $V_{\rm max}$ is more
pronounced for the more extended disks with a larger value of
$\lambda$. This is simply a consequence of the fact that larger disk
scale lengths make the potential well of the dwarf shallower and
decrease the total binding energy in the inner parts.

As expected, the dwarf with the most extended disk (R13) has a larger
initial value of the $M/L$ ratio within $r_{\rm max}$. Due to the
dissimilar values of $R_d$, the stripping of the stars proceeds
differently in this set of experiments and, as expected, is much more
effective in the case of the most extended disk.  This differential
stripping of the stars manifests itself in the evolution of the $M/L$
ratio.  Although the initial values of the $M/L$ ratios are not very
different, the final value of $M/L$ is much larger in simulation R13.
Interestingly, the $M/L$ ratio in this experiment starts to increase
during the intermediate stages of the evolution ($t\sim 5$~Gyr). Such
a behavior is similar to that reported in
\S~\ref{sec:orbital_parameters} for experiments R2 and R4, even though
this transition occurred earlier there and the final $M/L$ values
obtained were slightly lower.  We discuss this issue in
\S~\ref{sec:discussion}. Overall, Figure~\ref{fig8} indicates that the
$M/L$ ratios are differentiated as a result of the tidal evolution.

Evidently, the size of the disk affects substantially the evolution of
the stellar shape, and to some degree also influences that of the
stellar kinematics.  The behavior of the Fourier component $A_2$ after
the first pericentric passage is slightly misleading. Indeed, both
lower and larger $\lambda$ cases show similar values of this
parameter, but for different reasons.  While the most compact disk
(R12) retains an oblate shape, the most extended one (R13) is
triaxial, but neither shows such a strong bar mode as our reference
experiment R1. The situation changes radically after the second
pericentric passage.  On the one hand, owing to its higher
self-gravity, the low-$\lambda$ dwarf forms a fairly strong bar and
remains prolate until the end of the evolution with final values of
$c/a \approx 0.5$ and $b/a \approx 0.7$. In fact, the remnant in this
experiment exhibits the least spheroidal shape of all our classified
dSphs, highlighting the resilience of the most compact disk to tidal
heating.  This system could be the analogue of the most elongated
dSphs in the LG such as Ursa Minor \citep{Irwin_Hatzidimitriou95}. On
the other hand, tidal effects are quite efficient in the case of the
high-$\lambda$ disk with the lowest self-gravity and cause its rapid
transformation into a nearly isotropic ($\beta \approx 0.1$) and
spherical stellar component ($b/a \approx c/a \gtrsim 0.9)$ that
retains little rotation ($V_{\rm rot}/\sigma_{\ast} \lesssim 0.3$).


\begin{figure*}[t]
\begin{center}
  \includegraphics[scale=0.44]{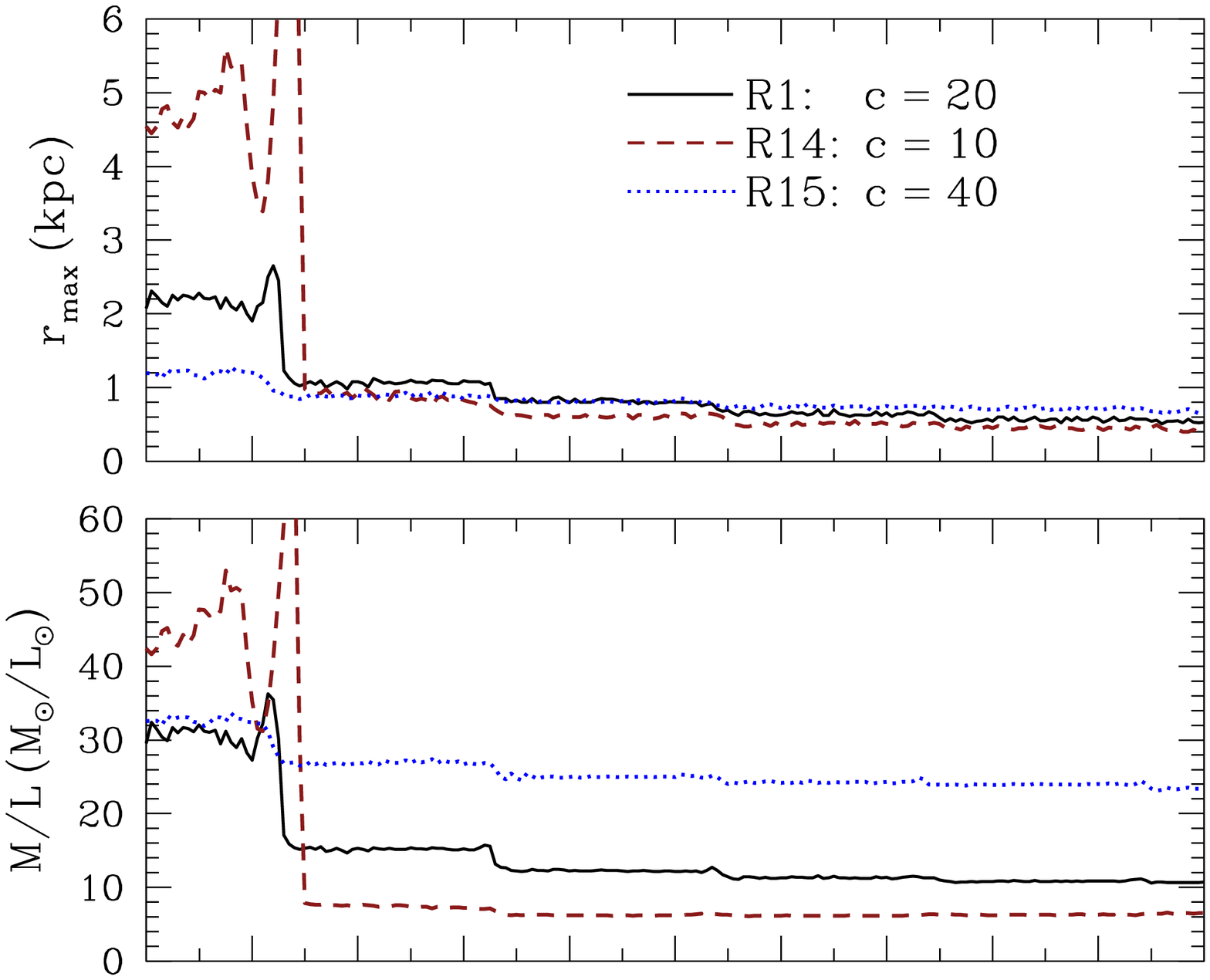}
  \includegraphics[scale=0.44]{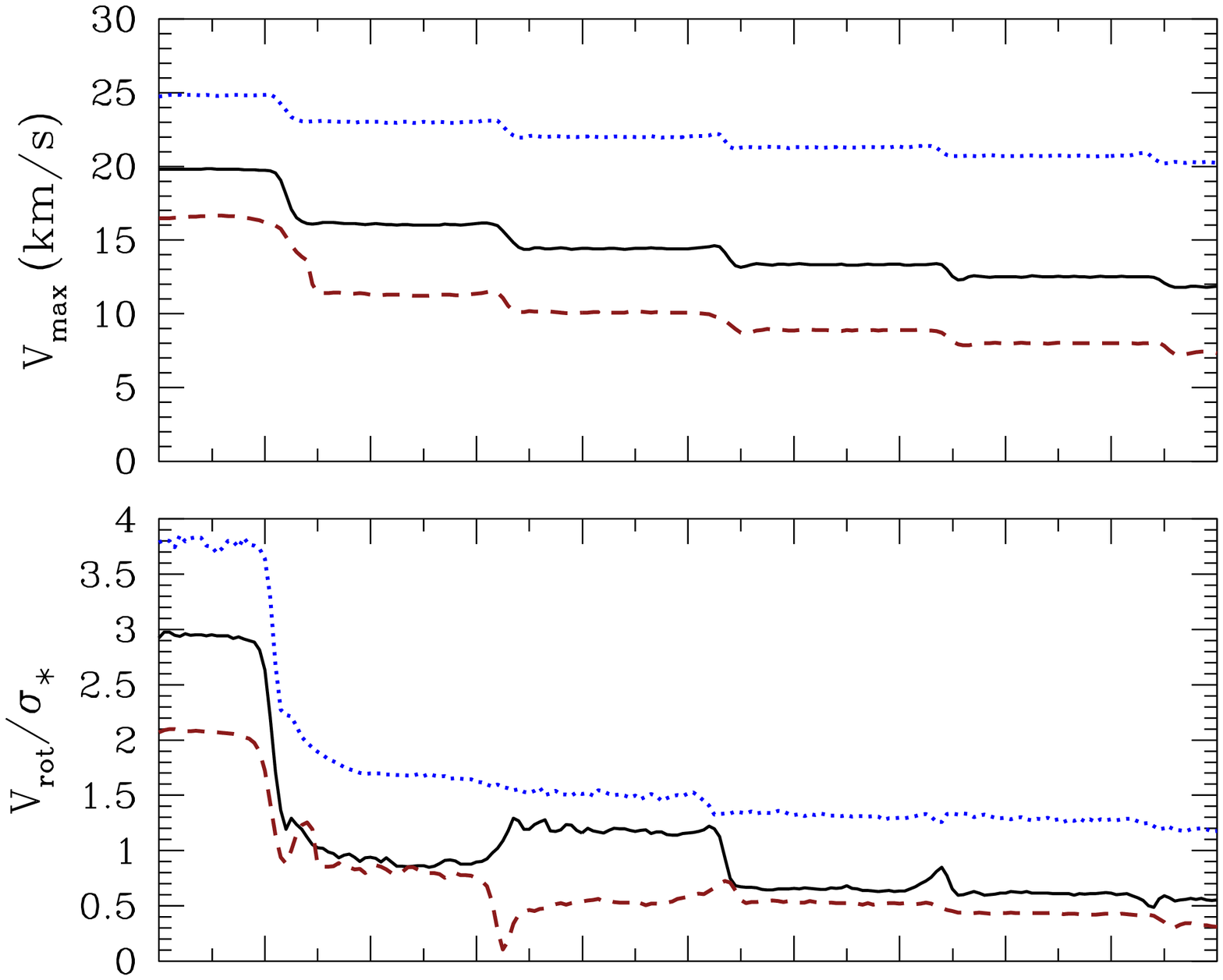}
  \includegraphics[scale=0.44]{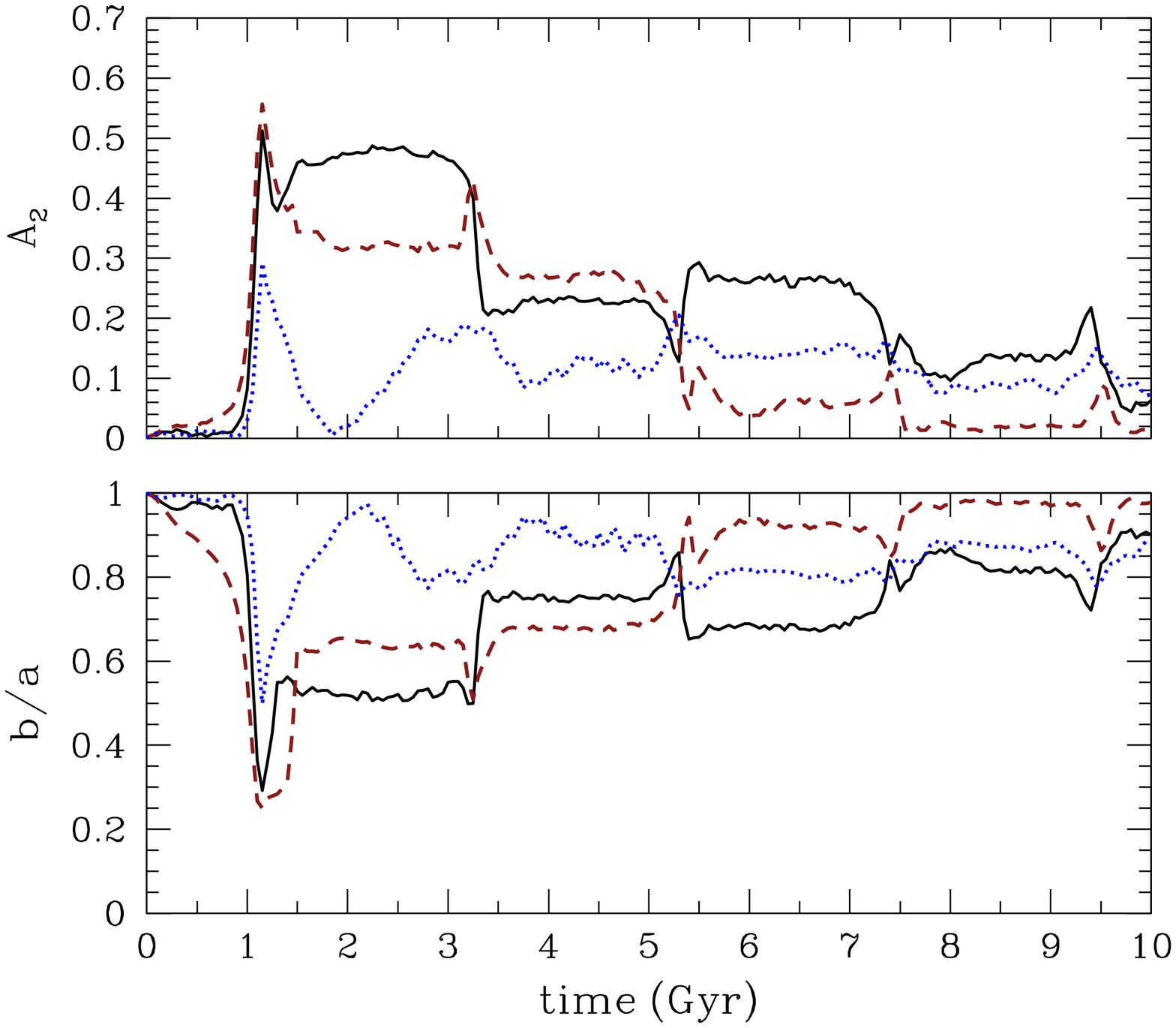}
  \includegraphics[scale=0.44]{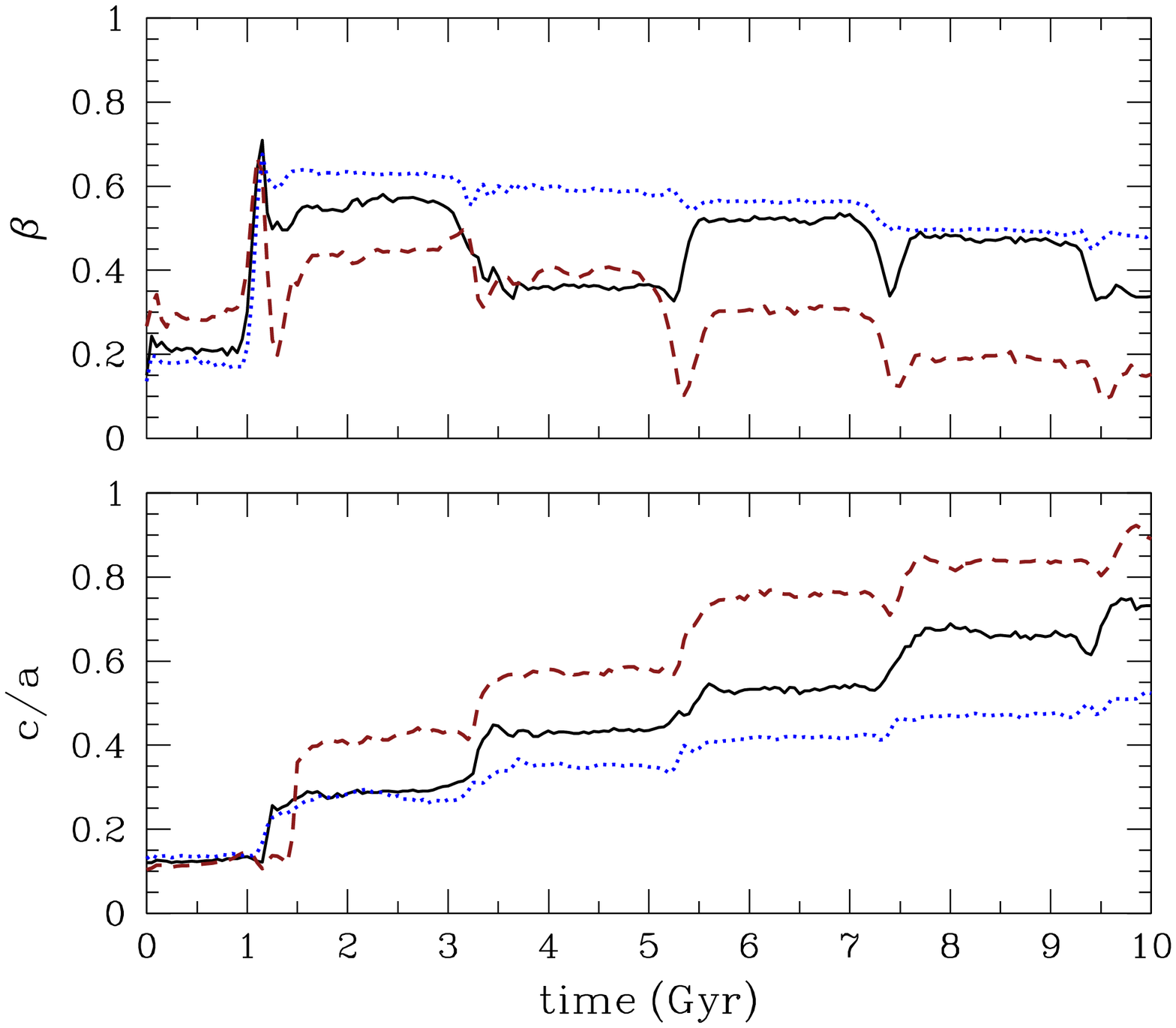}
\end{center}
\caption{Comparison of the evolution of various parameters as a function 
  of time illustrating the dependence of the tidal transformation of
  disky dwarf galaxies on the concentration parameters of their DM
  halos, $c$. Results are presented for simulations R1, R14, and R15.
  Rotationally-supported dwarfs embedded in less concentrated DM halos
  exhibit stronger tidal evolution inside their host galaxies and the
  efficiency of their transformation into dSphs is increased
  considerably.
\label{fig9}}
\end{figure*}


\subsection{Varying the Parameters of the Halo}
\label{sec:halo_parameters}

\subsubsection{Halo Concentration}
\label{subsec:halo_concentration}

In this section, we investigate the degree to which the tidal
evolution of a disky dwarf is affected by the concentration parameter
of its DM halo, $c$. To this end, we constructed two additional dwarf
galaxy models that were identical to D1 except for the halo
concentration parameters. Recall that we chose $c=20$ for our
reference model D1, close to the median concentration value for $z=0$
cosmological halos of mass $\sim 10^9 M_{\odot}$
\citep[e.g.,][]{Bullock_etal01a,Maccio_etal07}. To investigate the
effect of the halo concentration on our results, we constructed two
additional dwarf models with $c=10$ and $c=40$. These values roughly
correspond to the $2\sigma$ deviations from the median $c=20$, in
accordance with the previous cosmological studies\footnote{We note
  that $c=40$ is markedly incompatible with results of the
  distribution of DM in present-day LSB and dIrr galaxies through the
  modeling of rotation curves (see the recent review of
  \citealt{deBlok10} for a discussion on this issue).  Nevertheless,
  even though it is clearly inconsistent with observations, we decided
  to consider this limiting case simply for completeness.}. We refer
to these dwarf models as ``D8'' and ``D9'', respectively, and list
their structural parameters in Table~\ref{table:init_param}. In the
corresponding experiments ``R14'' and ``R15'', we placed these dwarfs
on the same orbit as in simulation R1 and followed their tidal
evolution inside the host galaxy.  Figure~\ref{fig9} contains the
results of these simulations.
 
As expected, the less concentrated dwarf is characterized by a larger
initial $r_{\rm max}$.  As in the case of the experiments with
different disk masses (R10, R11), after the first pericentric approach
the values of $r_{\rm max}$ converge. Interestingly, at late times the
hierarchy is even reversed (the dwarf with the most concentrated halo
exhibits the largest $r_{\rm max}$) but the differences are not
significant. As demonstrated by the weakest decrease in the value of
$V_{\rm max}$, the most concentrated halo increases the resilience of
the dwarf galaxy to tides and mass loss. Since the higher
concentration makes the potential well of the dwarf deeper, this
effect is similar to that of the larger disk mass discussed in
\S~\ref{subsec:disk_mass}.  The effect on the $M/L$ ratio is, however,
the opposite; although the $M/L$ ratio within $r_{\rm max}$ was
initially larger for the $c=10$ case it soon becomes the smallest.
This is because the DM particles are much more weakly bound in this
case and, as a result, are stripped fairly efficiently.  After a
period of $t=10$~Gyr inside the host galaxy, the $M/L$ ratios of the
dwarfs with the most and the least concentrated halo differ by a
factor of $\sim 4$. The weak evolution of the $M/L$ ratio in
simulation R15 is of similar origin to that of the experiment with the
most massive disk (R11).

The susceptibility to tides of the low-concentration dwarf in
experiment R14 also manifests itself in the evolution of the shape and
kinematics. The stellar component in this experiment evolves rapidly
and becomes nearly spherical rather quickly with a final $c/a \approx
0.9$.  The stronger effect of the tidal shocks in this case reflects
the fact that the dwarf responds more impulsively to the tidal
perturbation owing to its lower characteristic densities and,
correspondingly, longer internal dynamical times. After the first
pericentric passage, a strong bar develops in the dwarf disk ($A_2
\gtrsim 0.3$) and the $V_{\rm rot}/\sigma_{\ast}$ ratio starts to
decrease, reaching a final value of $V_{\rm rot}/\sigma_{\ast} \approx
0.3$.  The structure of the remnant in this case would resemble that
of observed dSphs.

On the other hand, the stellar component of the high-concentration
dwarf retains its disky shape, since it is less affected by tides.
This is similar to the cases of simulations R3 and R5 where the dwarf
galaxies experienced weaker tidal forces due to the larger pericentric
distances.  In experiment R15, the final axis ratio is higher ($c/a
\approx 0.5$) but still the stellar component remains oblate.  Most
importantly, as demonstrated by the very low values of $A_2$
throughout the evolution, a bar does not form in this case. As a
result, the angular momentum content of the dwarf does not decrease
efficiently and the $V_{\rm rot}/\sigma_{\ast}$ ratio remains larger
than unity until the end of the simulation. The absence of the bar in
this case is consistent with results of numerical experiments showing
that high values of halo concentration suppress the formation of bars
in isolated disk galaxies embedded in CDM halos
\citep{Mayer_Wadsley04}. To summarize, experiment R15 constitutes the
only case where a rotationally-supported dwarf is placed on the
reference orbit of simulation R1 and neither develops a bar nor yields
a dSph. This lends further support to the fact that bar formation and
the formation of dSphs are intimately linked in the context of the
tidal stirring model.


\begin{figure*}[t]
\begin{center}
  \includegraphics[scale=0.44]{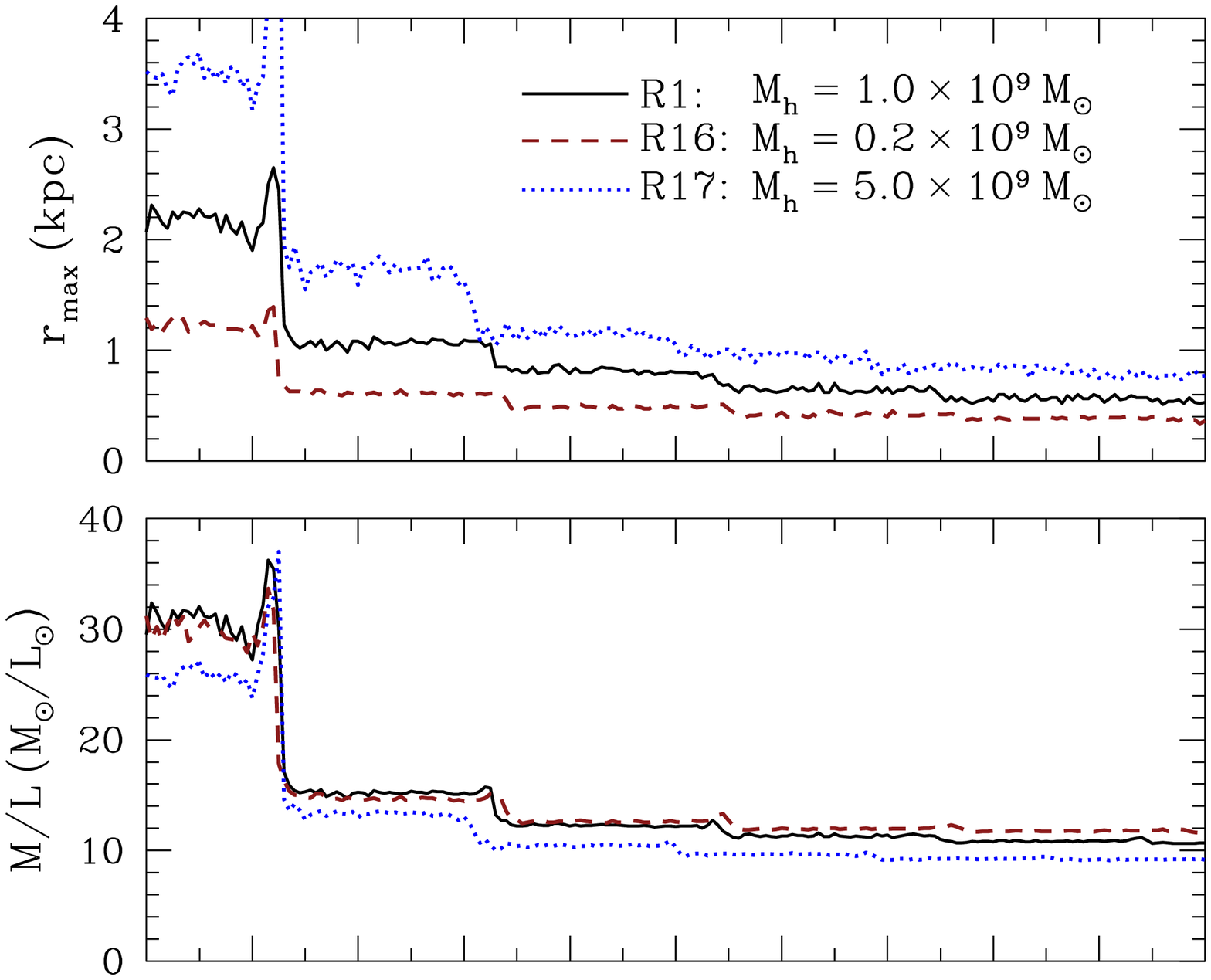}
  \includegraphics[scale=0.44]{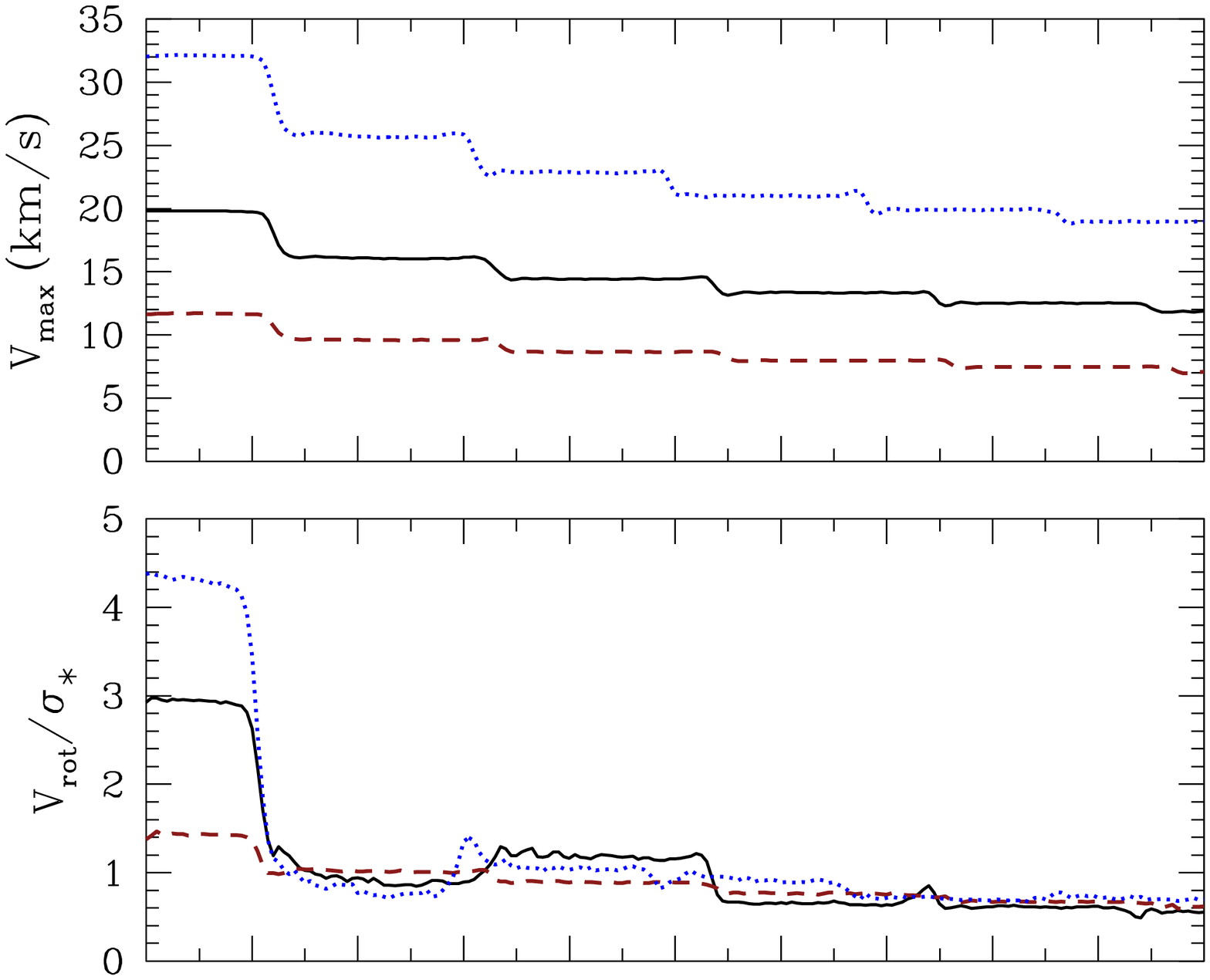}
  \includegraphics[scale=0.44]{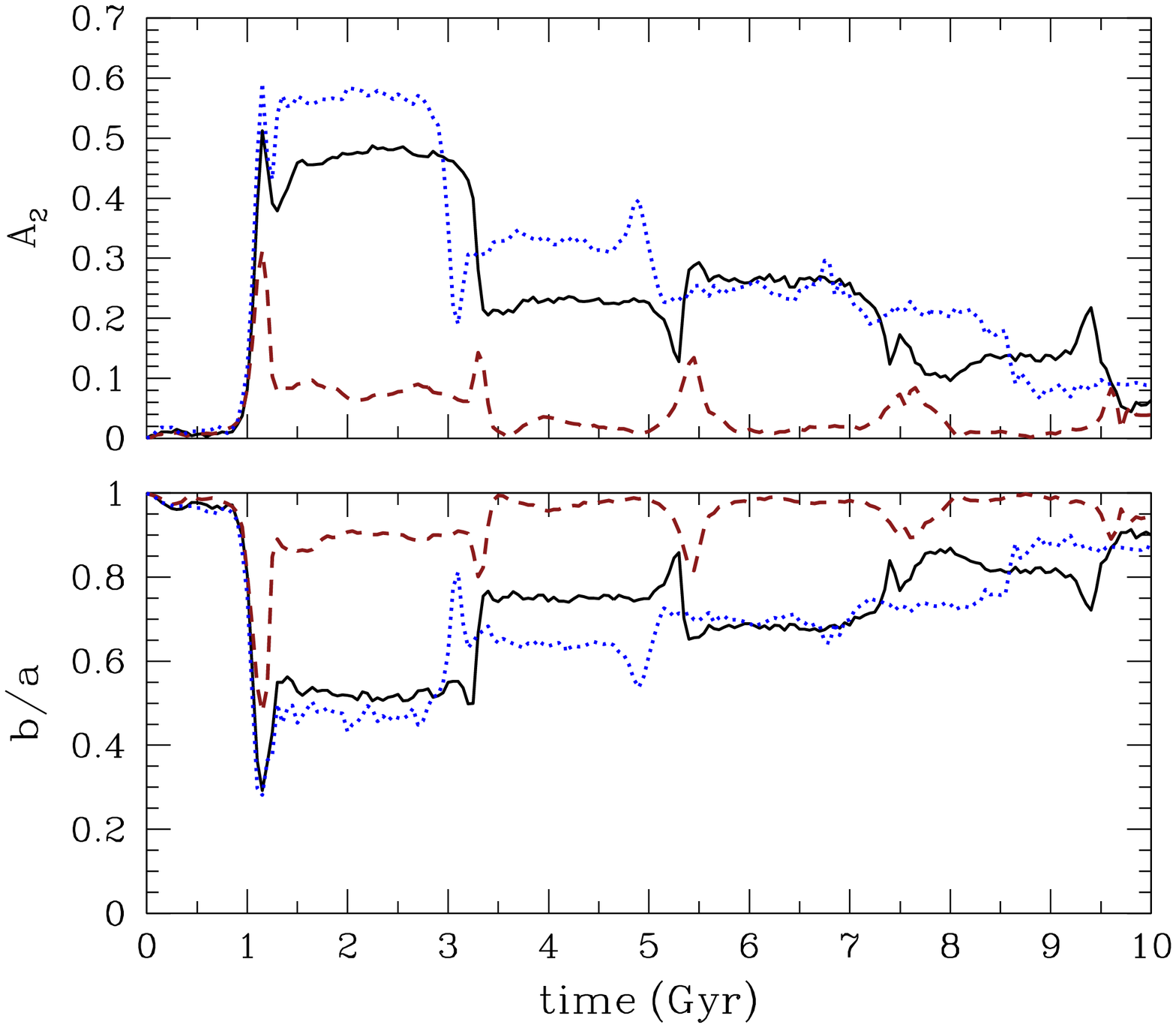}
  \includegraphics[scale=0.44]{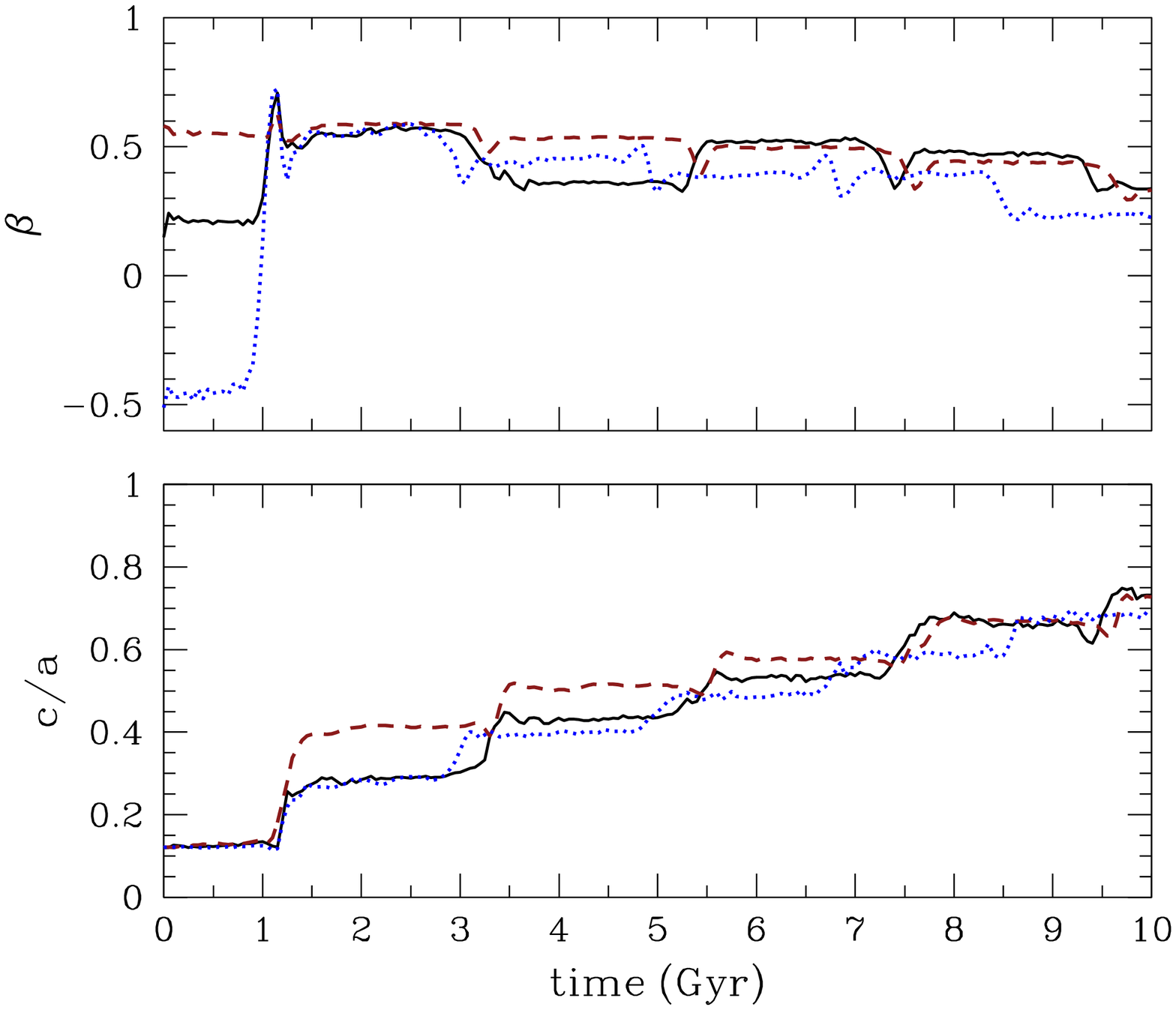}
\end{center}
\caption{Comparison of the evolution of various parameters as a function 
  of time illustrating the dependence of the tidal transformation of
  disky dwarf galaxies on the masses of their DM halos, $M_h$.
  Results are presented for simulations R1, R16, and R17.  Owing to
  the interplay between the effects of dynamical friction and the
  depth of the potential well, rotationally-supported dwarfs embedded
  in DM halos of substantially different mass display similar tidal
  evolution inside their host galaxies and the efficiency of their
  transformation into dSphs is of comparable magnitude.
\label{fig10}}
\end{figure*}



\begin{figure*}[t]
\begin{center}
  \includegraphics[scale=0.3]{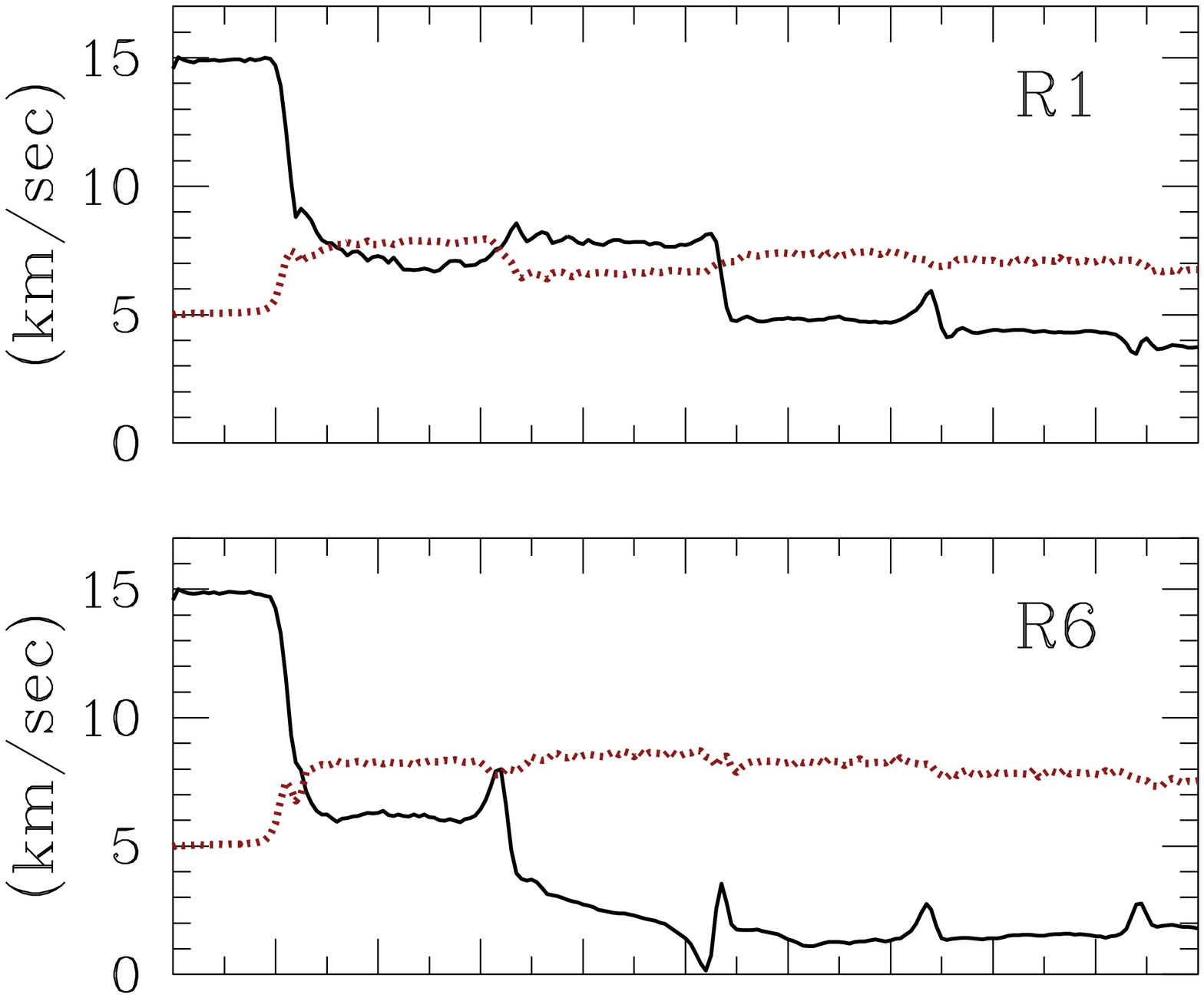}
  \includegraphics[scale=0.3]{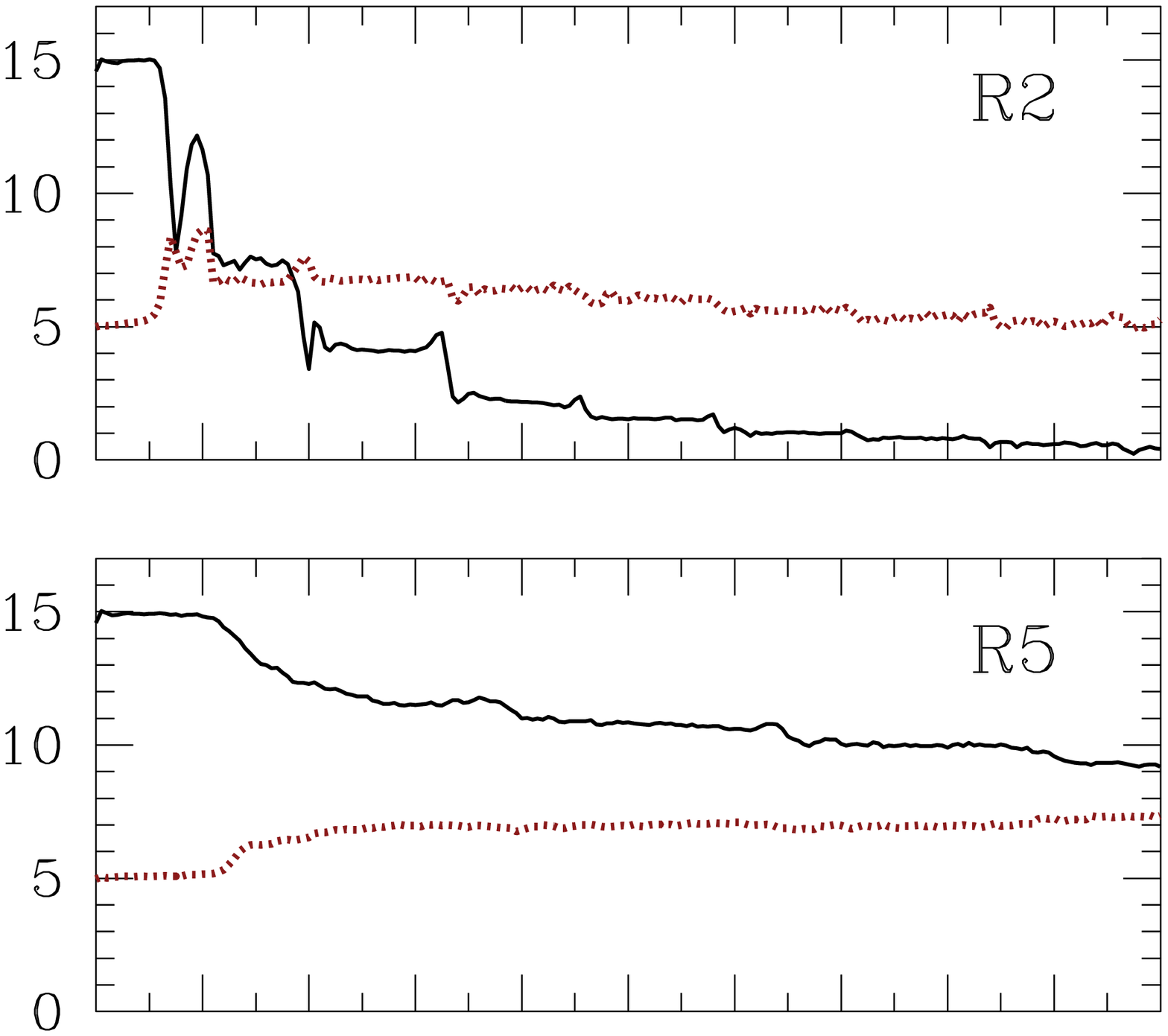}
  \includegraphics[scale=0.3]{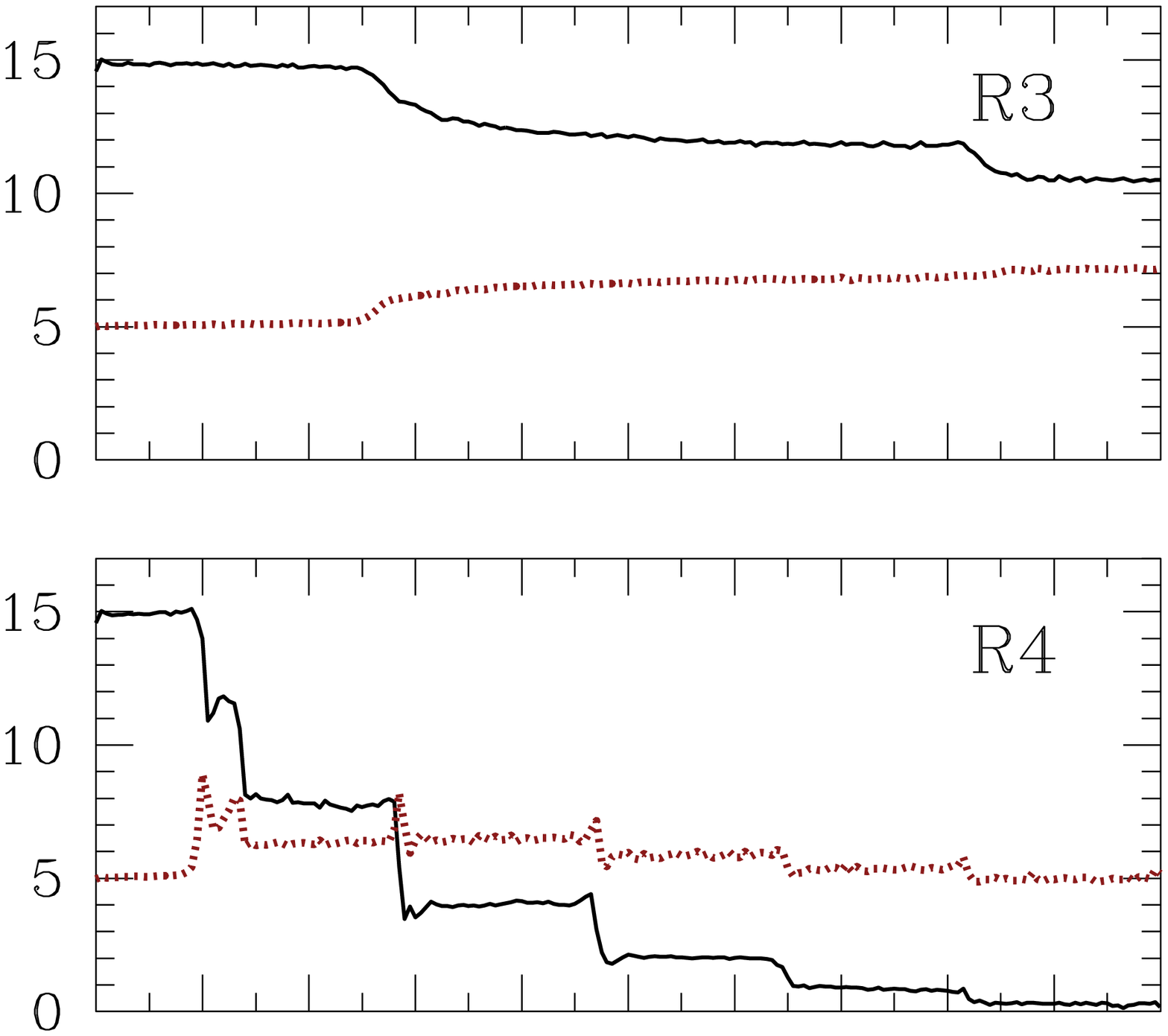}
  \includegraphics[scale=0.3]{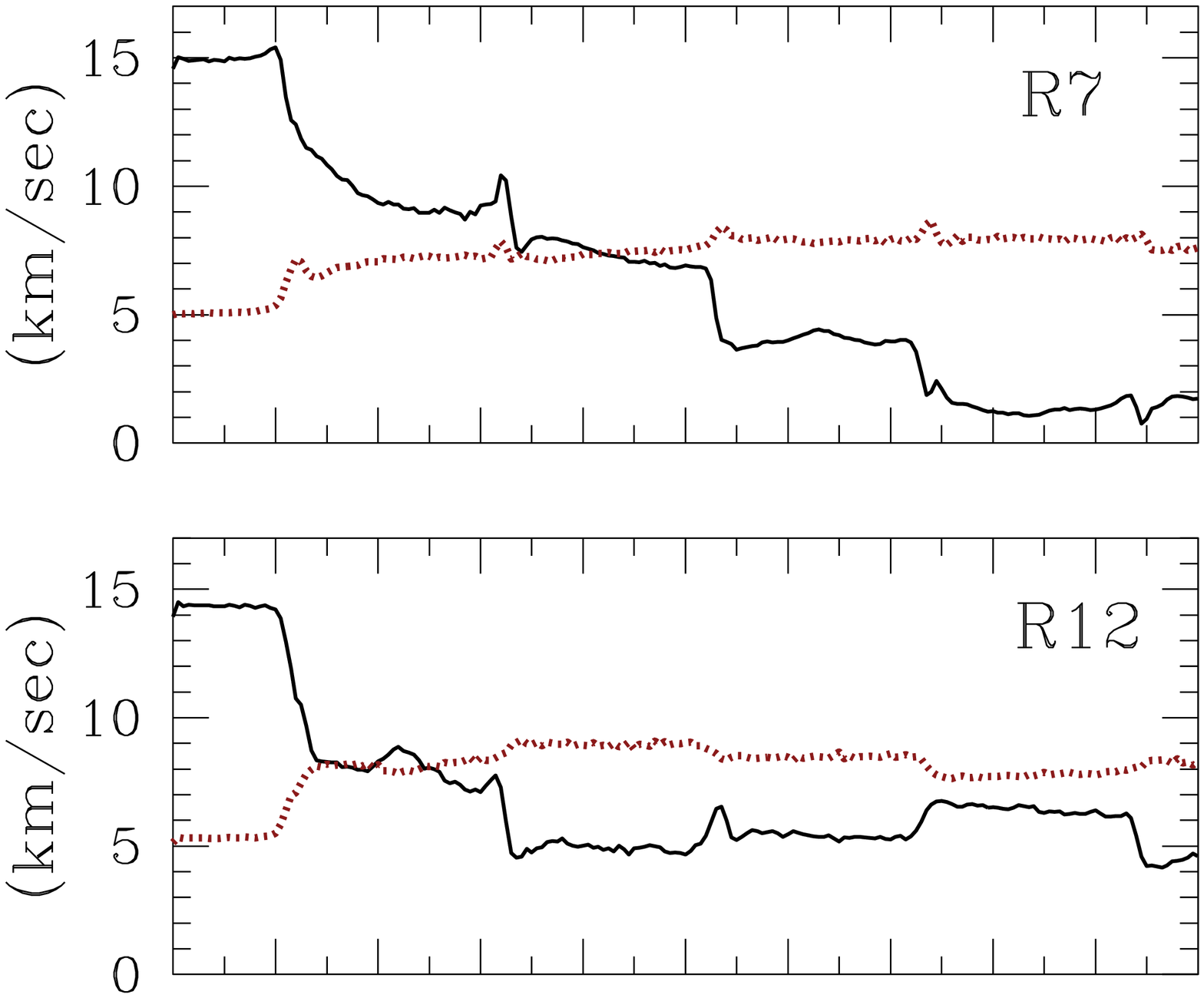}
  \includegraphics[scale=0.3]{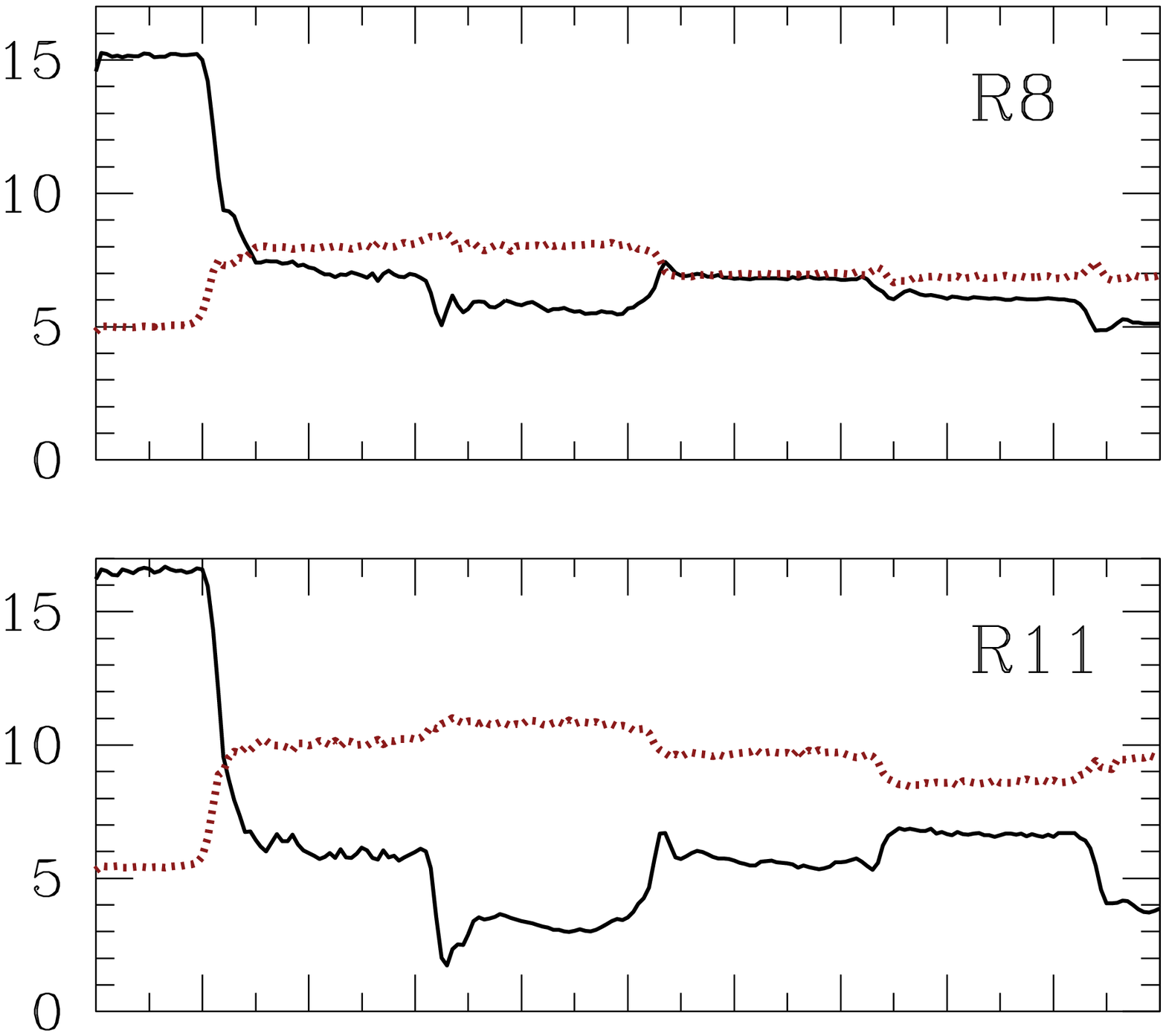}
  \includegraphics[scale=0.3]{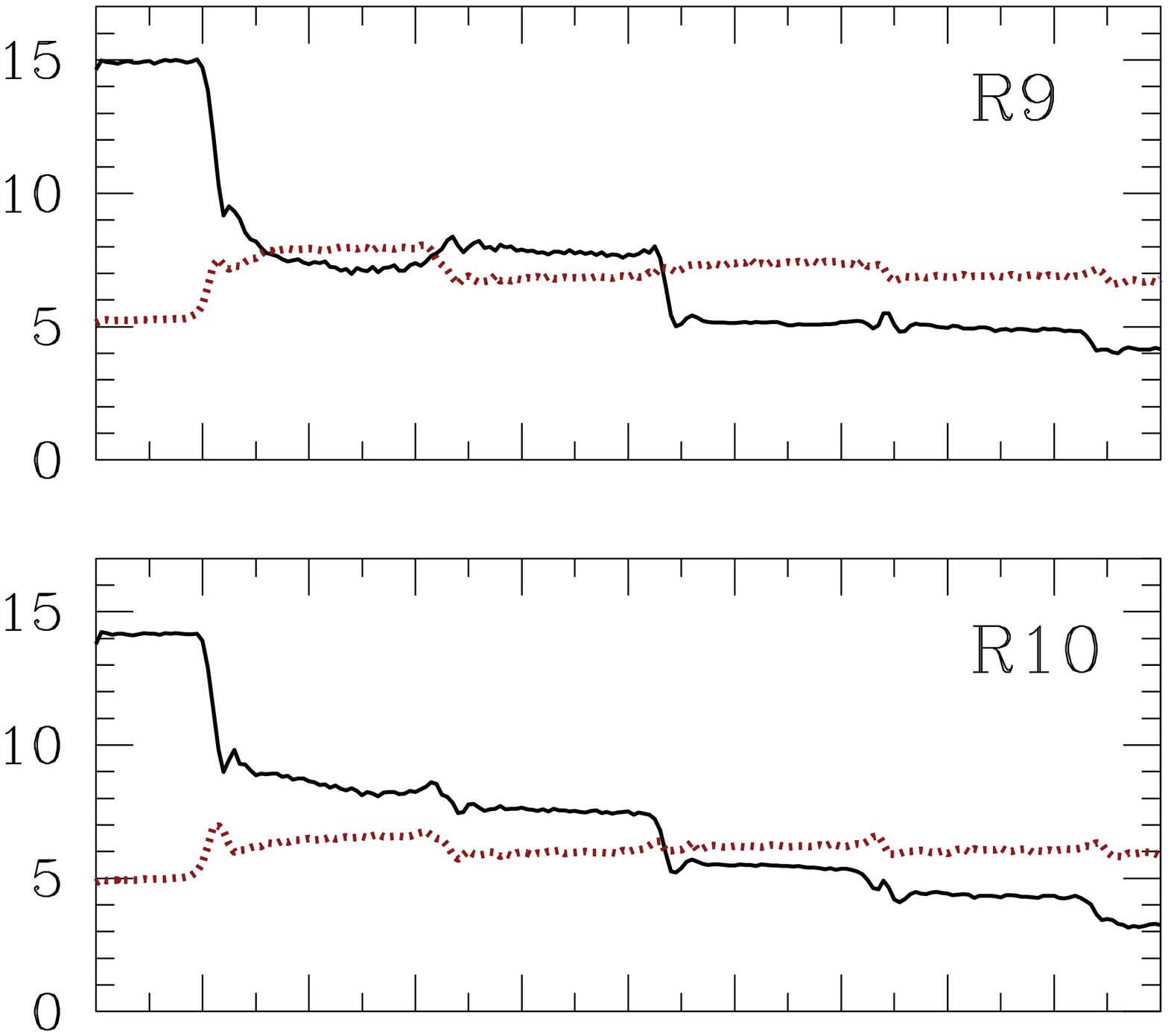}
  \includegraphics[scale=0.3]{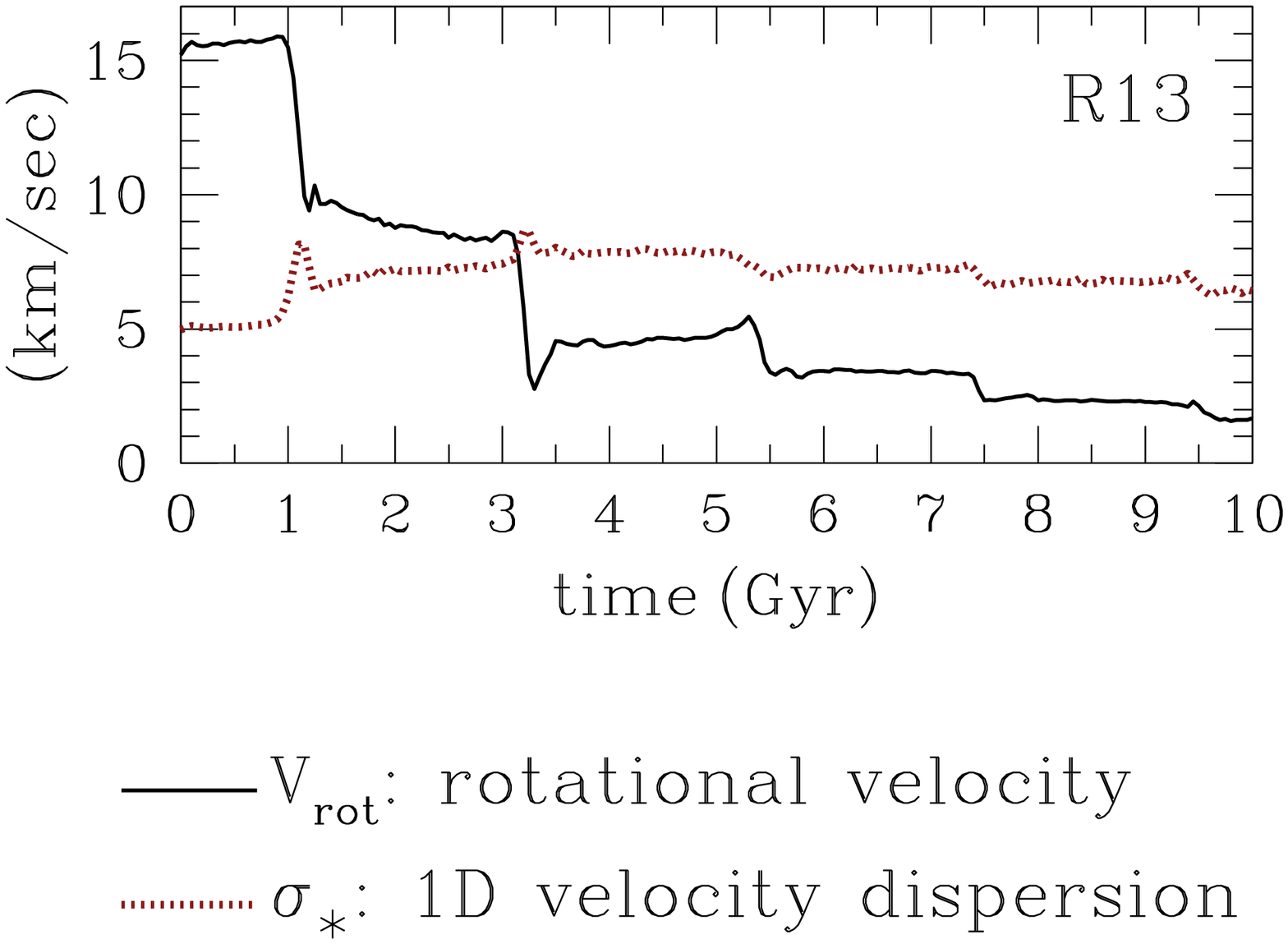}
  \includegraphics[scale=0.3]{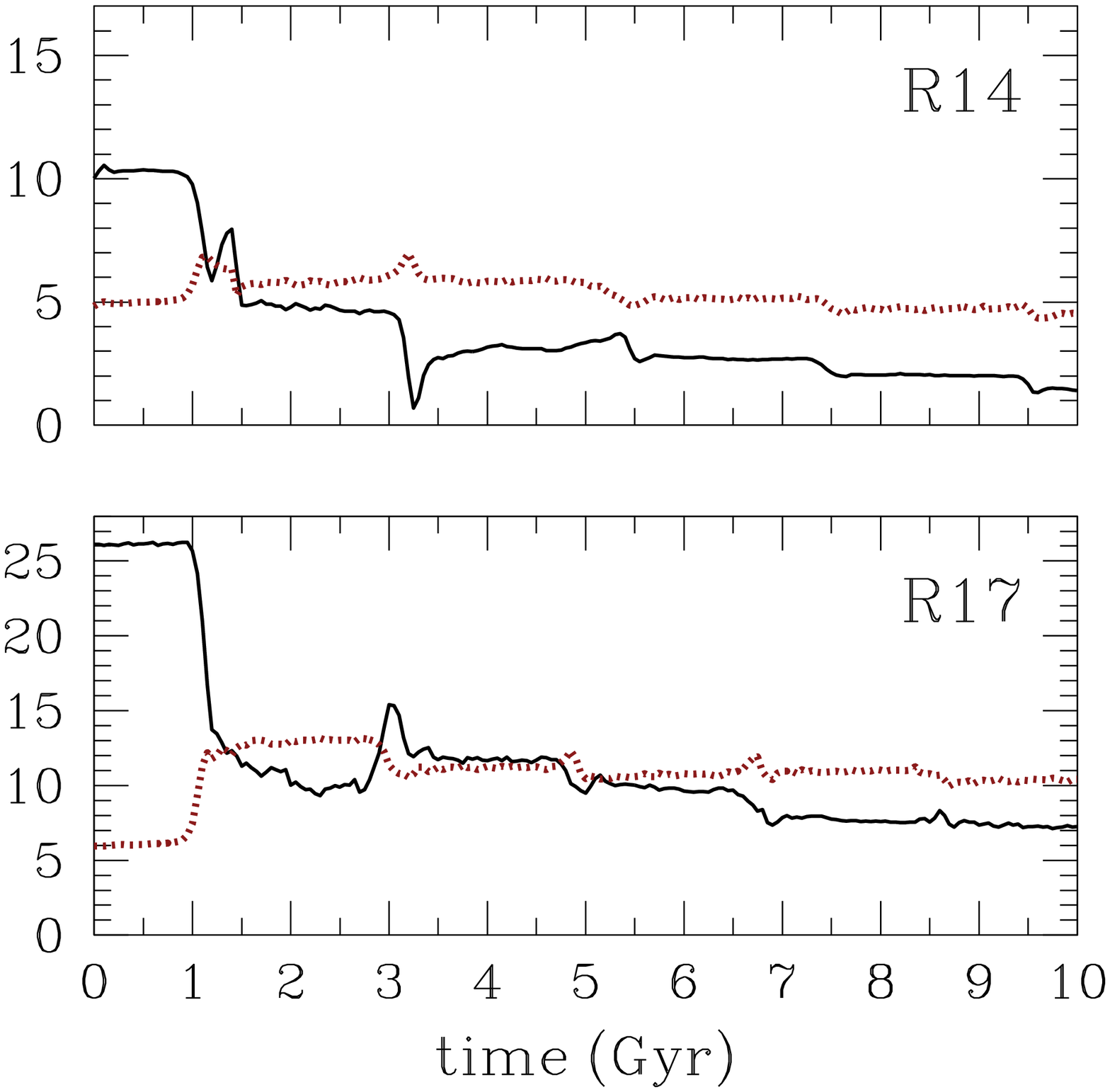}
  \includegraphics[scale=0.3]{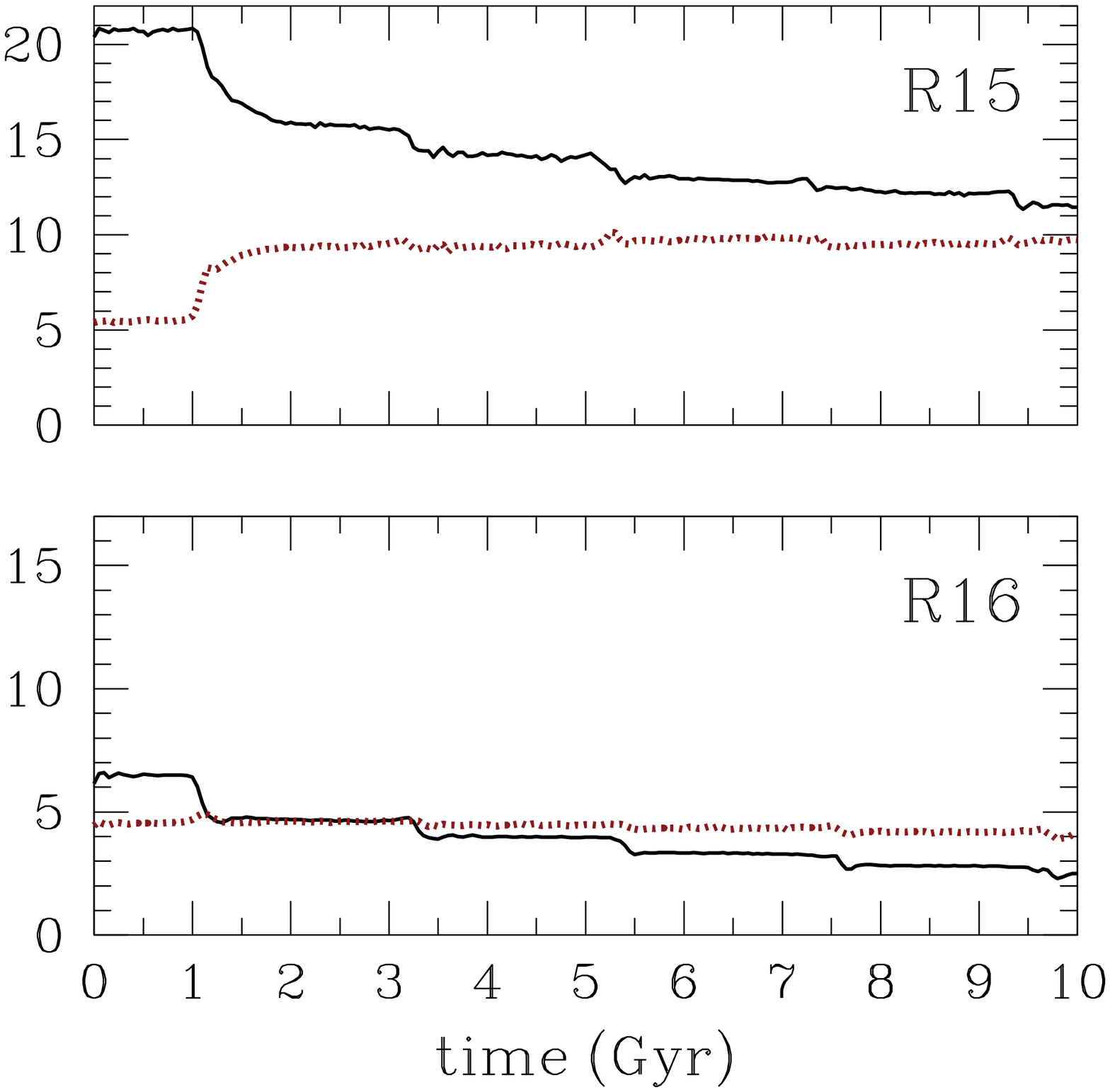}
\end{center}
\caption{Evolution of the stellar rotational velocity, $V_{\rm rot}$, 
  (solid lines) and the one-dimensional stellar velocity dispersion,
  $\sigma_{\ast}$, (dotted lines) of the dwarfs in simulations R1-R17.
  Parameters $V_{\rm rot}$ and $\sigma_{\ast}$ are computed within
  $r_{\rm max}$. The repeated action of tidal forces exerted by the
  host galaxy decreases the ratio of $V_{\rm rot}/\sigma_{\ast}$,
  gradually transforming the initially rotationally-supported disky
  dwarfs to pressure-supported stellar systems dominated by random
  motions.
\label{fig11}}
\end{figure*}


\subsubsection{Halo Mass}
\label{subsec:halo_mass}

In this section, we explore the degree to which the tidal evolution of
a disky dwarf depends on the mass of its DM halo, $M_h$.  Such a
dependence may be expected as various effects, including the strength
of tidal shocks as well as dynamical friction, will be affected by the
halo mass. For this purpose, we generated two additional dwarf galaxy
models that were identical to D1 except for the masses of their halos,
which differed by a factor of $5$: $M_h = 0.2 \times 10^9 M_{\odot}$
and $M_h = 5 \times 10^9 M_{\odot}$. This mass range roughly
corresponds to the range of initial subhalo masses that were accreted
since $z \lesssim 2$ and survived until the present time in the LG
simulation of \citet{Klimentowski_etal10}. We refer to these models as
``D10'' and ``D11'', respectively (see Table~\ref{table:init_param}).
The motivation behind using the same values for $m_d$, $z_d/R_d$, $c$,
and $\lambda$ in these models as in the reference one (D1) was imposed
mainly by simplicity. However, we do note that our choices are
reasonable given the scatter in $m_d$
\citep[e.g.,][]{Jimenez_etal03,Tassis_etal03}, $z_d/R_d$
\citep[e.g.,][]{Dalcanton_etal04,Kaufmann_etal07}, and $c$
\citep[e.g.,][]{Bullock_etal01a,Maccio_etal07} for galaxies in this
mass range, and the fact that the distribution of halo spins shows no
dependence on halo mass \citep[e.g.,][]{Maccio_etal07}. In the
corresponding experiments ``R16'' and ``R17'', we placed dwarf models
D10 and D11 on the same orbit as in simulation R1 and followed their
tidal evolution inside the host galaxy. Figure~\ref{fig10} presents
the results of these simulations.

As a result of our choices for the initial structural parameters of
the dwarf galaxies in this set of experiments, $r_{\rm max}$ is larger
for the dwarf with the most massive halo (R17). This is simply because
$r_{\rm max} \approx 2.16 R_h / c$ for the NFW profile, where $R_h$ in
our modeling corresponds to the cosmologically motivated virial radius
of the system ($R_h \approx r_{\rm vir}$). Figure~\ref{fig10} shows
that $r_{\rm max}$ remains larger for the most massive dwarf until the
end of the evolution. We emphasize that the pericentric passages which
are marked by sudden drops in $V_{\rm max}$ do not occur at the same
time for all three dwarfs, despite the fact that the orbital
parameters were initially identical.  Specifically, the orbit of the
most massive dwarf becomes progressively much tighter with
continuously reduced apocentric distances, leading to a shorter
orbital time (see Table~\ref{table:summary}).  This is due to the
effect of dynamical friction, whose strength depends strongly on the
mass of the moving body.

Interestingly, after $10$~Gyr of evolution inside the host, all three
dwarf galaxies lose approximately the same fraction of their initial
mass within $r_{\rm max}$ ($\sim 90\%$) and their $V_{\rm max}$
decreases by approximately the same factor of $\sim 1.7$. This is
intriguing given that mass loss is generically affected, among other
things, by the depth of the potential well of the dwarf and the
details of the dwarf orbit (e.g., the pericentric distance) around the
primary galaxy. We note that by construction the three dwarfs have
approximately equal average densities within $r_{\rm max}$, and
therefore comparable dynamical times at $r_{\rm max}$.  Using this
fact coupled with some order-of-magnitude calculations it is possible
to gain qualitative insight as to the relative evolution of $V_{\rm
  max}$ and mass within $r_{\rm max}$ is similar in simulations R1,
R16, and R17.

According to the tidal approximation of the impulse (which should be
valid in our simulations for the purposes of dimensional analysis),
the energy injected at each pericentric passage is given by $\Delta E
\propto M_{\rm host}^2 M R^2 V_{\rm rel}^{-2}$, where $M_{\rm host}$
is the mass of the host, $M$ denotes the mass of the dwarf, $R$ is a
characteristic radius of the dwarf (related to $r_{\rm max}$), and
$V_{\rm rel}$ is the relative velocity between the two galaxies at the
pericenter of the orbit \citep{Spitzer58,Binney_Tremaine08}. Since, as
noted above, the average density within $r_{\rm max}$ is roughly
constant among the dwarfs, we have $R \propto M^{1/3}$ in the three
different models.  Given that $M_{\rm host}$ is the same and $V_{\rm
  rel}$ is fairly similar in the three experiments considered here, it
follows that $\Delta E \propto M^{5/3}$. By virtue of the virial
theorem, the energy content of the dwarf should scale as $E \propto
GM^2/R \propto M^{5/3}$, and also $ E \propto M V^2$, where $V$ is a
characteristic velocity of particles within the dwarf (related to
$V_{\rm max}$).  Hence, the fractional increase in energy caused by
the tidal shocks, $\Delta E / E$, is roughly constant, explaining why
the $V_{\rm max}$ and $M_{\rm max}$ of the dwarfs in simulations R1,
R16, and R17 evolve in a fairly similar way. We stress that similar
conclusions would be reached had we used extensions of the standard
impulse approximation appropriate for perturbers with extended mass
distributions as well as for satellites that move on either
straight-path or eccentric orbits \citep{Gnedin_etal99}. This is
because such extensions are characterized by analogous scalings of the
relevant variables in our calculations above.

The $M/L$ ratios are similar in all cases and decrease monotonically
with time reaching a value of $\sim 10 M_{\odot}/L_{\odot}$ in the end
of the evolution.  The behavior of kinematics shows an interesting
feature. While the values of $V_{\rm rot}/\sigma_{\ast}$ and $\beta$
were quite different for the three models initially, they become very
similar after the first pericentris passage and remain so until the
end of the evolution. The evolution of the shape shows a more clear
trend with mass. After the first pericentric approach, the bar is
stronger in the most massive dwarf compared to that of the default
case and it remains so until the end of the evolution. This simply
reflects the higher self-gravity of the disk in simulation R17.  On
the other hand, while the least massive dwarf remains disky during the
entire evolution, its axis ratio $c/a$ increases significantly with a
final value of $c/a \approx 0.75$.  The dwarf remnants in simulations
R16 and R17 are characterized by $c/a \approx 0.7$ and $V_{\rm
  rot}/\sigma_{\ast} \lesssim 0.7$ (see Table~\ref{table:summary}) and
hence would be classified as dSphs.

A last point of interest is that simulation R16 constitutes the only
case where the formation of a dSph is not associated with the presence
of a tidally-induced bar. Indeed, the increase of the bar amplitude,
$A_2$, in this case is only temporary and occurs at the first
pericentric passage when the dwarf is strongly elongated by tidal
forces.  The absence of a bar is consistent with the fact that the
disk of this dwarf exhibits the highest value of the Toomre stability
parameter (see Table~\ref{table:init_param}).  Bar formation causes
rotational motion to be transformed into radial motion and thus, in
conjunction with tidal heating which also increases the velocity
dispersion, is intimately linked to the decrease of $V_{\rm
  rot}/\sigma_{\ast}$ required for the transformation to a
pressure-supported dSph \citep{Mayer_etal01a,Debattista_etal06}. In
the case of the least massive dwarf, the initial $V_{\rm
  rot}/\sigma_{\ast}$ within $r_{\rm max}$ is rather low ($V_{\rm
  rot}/\sigma_{\ast} \lesssim 1.5$) and therefore the bar stage is not
necessary to decrease it considerably to levels that are appropriate
for dSphs ($V_{\rm rot}/\sigma_{\ast} \lesssim 1$). The decrease of
$V_{\rm rot}/\sigma_{\ast}$ in the dwarf of experiment R16 occurs
almost exclusively at pericentric approaches and is driven by the
decrease in $V_{\rm rot}$ (see Figure~\ref{fig11}). The decrease in
rotational velocity within $r_{\rm max}$ is simply caused by tidal
heating which perturbs the circular orbits of the disk stars
converting part of the rotation (ordered motions) in the disk to
random motions.  It is worth noting that in the reference simulation
R1, for example, the combination of tidal heating and the
tidally-induced bar lead to a decrease in the $V_{\rm
  rot}/\sigma_{\ast}$ of the dwarf by approximately a factor of $5$,
illustrating the importance of bar formation in reducing the initial
high values of $V_{\rm rot}/\sigma_{\ast}$ appropriate for disky
dwarfs to those characteristic of dSphs.


\begin{figure*}[t]
\hspace{0.1cm}
\begin{tabular}{c}
  \includegraphics[scale=0.45]{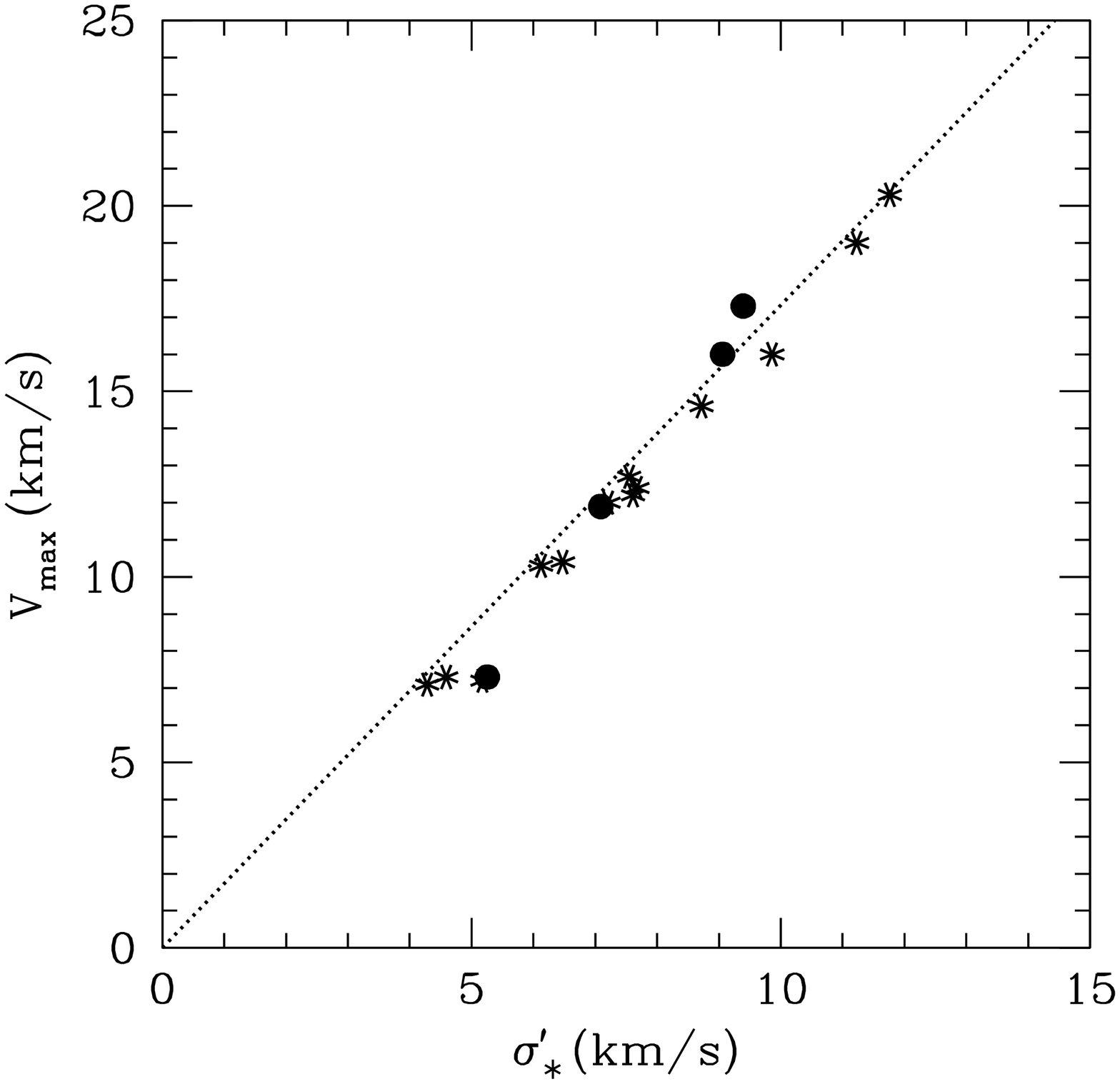}
\end{tabular}
\hspace{-0.5cm}
\begin{tabular}{c}
  \includegraphics[scale=0.45]{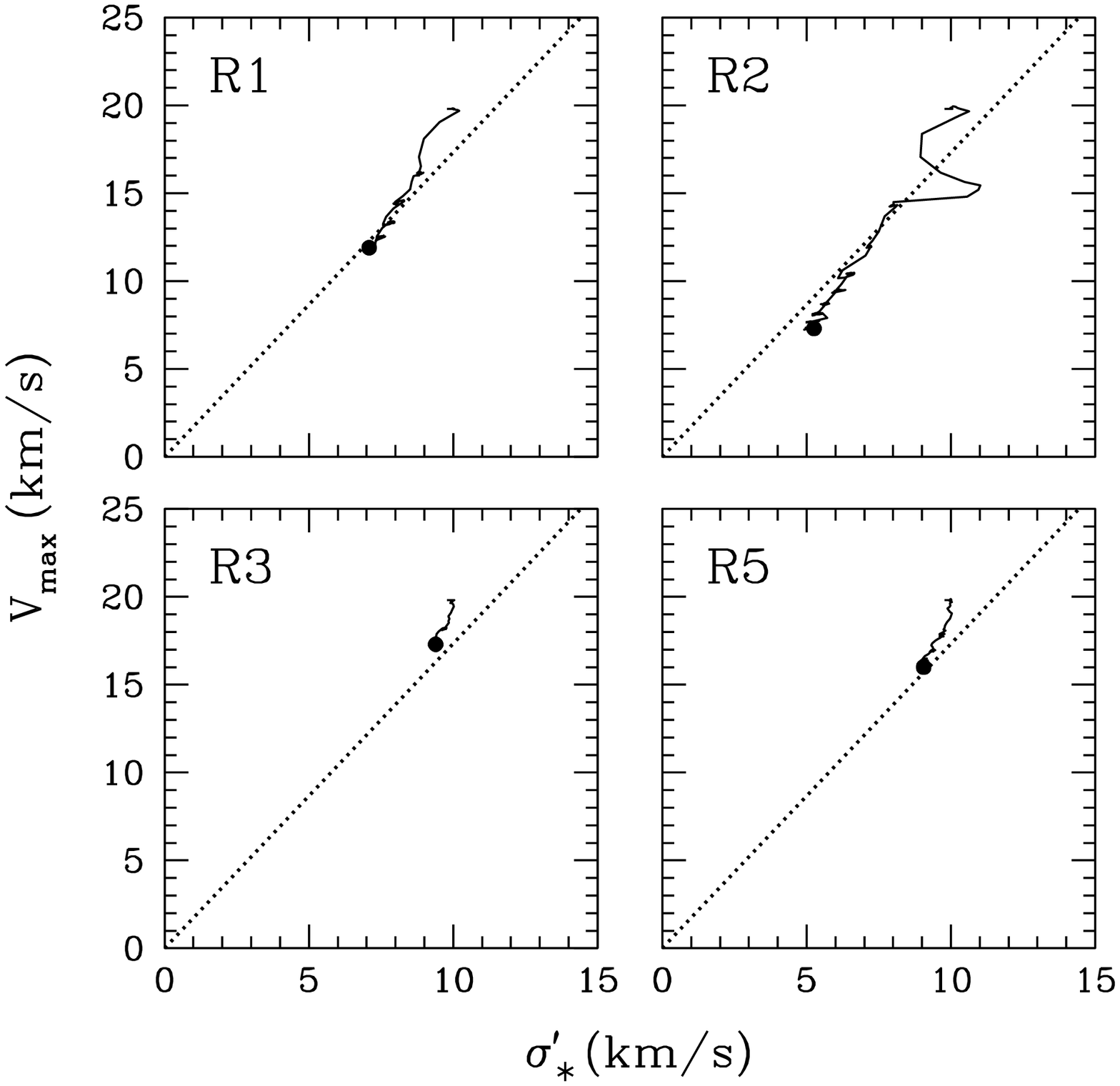} 
\end{tabular}
\caption{Relation between the maximum circular velocity, $V_{\rm max}$,
  and the one-dimensional, central stellar velocity dispersion,
  $\sigma_{\ast}^{\prime}$, of the dwarfs in our simulations.  The
  latter quantity corresponds to the line-of-sight central velocity
  dispersion measured in the observations and includes the
  contribution of rotation (see text for details).  Left: The data
  points indicate the values calculated in the final stages of the
  dwarfs ($t=10$~Gyr) in experiments R1-R17, and the dotted line is
  intended to identify a slope of $\sqrt{3}$ and is not a fit to the
  data. The formal best-fit slope differs by less than $\sim 4\%$ from
  $\sqrt{3}$. Circles show results for specific experiments R1, R2,
  R3, and R5 discussed in the right panel. The relation $V_{\rm max} =
  \sqrt{3}\,\sigma_{\ast}$, adopted by \citet{Klypin_etal99} to map
  the observed central stellar velocity dispersion of dSphs to the
  maximum circular velocities of their halos, is well justified within
  the context of the tidal stirring model. Right: Four examples of the
  evolution of individual dwarf galaxies on the $V_{\rm
    max}-\sigma_{\ast}^{\prime}$ plane (solid lines).  Results are
  shown for representative simulations R1, R2, R3, and R5, where the
  same dwarf model (D1) was placed on different obits.  R1 is our
  reference experiment.  R2 produces one of the most strongly evolved
  dwarfs in our simulations, while R3 and R5 correspond to the least
  tidally-transformed systems in our experiments.  In all panels, the
  filled circles mark the end of the evolution inside the host
  ($t=10$~Gyr). As in the left panel, dotted lines correspond to
  slopes of $\sqrt{3}$. The dwarf galaxies exhibit fairly different
  types of trajectories depending on their orbital parameters and can
  move substantially on the $V_{\rm max}-\sigma_{\ast}^{\prime}$ plane
  over the course of their evolution inside their hosts.
\label{fig12}}
\end{figure*}


\section{Discussion}
\label{sec:discussion}

We begin this section by investigating the $V_{\rm max}-\sigma_{\ast}$
relation in our simulations, which is of particular interest for the
missing satellites problem. In \S~\ref{subsec:Vrot} and
\S~\ref{subsec:M/L}, we explore the reasons for the increasing
rotational velocity and the increasing $M/L$ ratio, respectively,
observed in some of our experiments.  In \S~\ref{subsec:Torb_rperi},
we elucidate the fundamental orbital parameter that influences the
transformation of disky dwarfs into dSphs and thus the efficiency of
the tidal stirring mechanism. In \S~\ref{subsec:implications}, we
discuss some additional implications related to the findings of this
work. Lastly, in \S~\ref{subsec:caveats_directions} we conclude with a
few words of caution and a discussion of promising directions for
future work that may lead to more conclusive statements about the role
of environmental processes in shaping the nature of dwarf galaxies in
environments such as that of the LG.

\subsection{$V_{\rm max}-\sigma_{\ast}$ Relation}
\label{subsec:Vmax_sigma}

Here we investigate the relation between the one-dimensional, central
stellar velocity dispersion, $\sigma_{\ast}$, and the maximum circular
velocity, $V_{\rm max}$, in our simulated dwarf galaxies.  The
knowledge of this relation is crucial, particularly in the context of
the missing satellite problem \citep{Moore_etal99,Klypin_etal99}, as
it reflects the link between observable properties and dark halo
properties in dwarf galaxies. One of the main difficulties intrinsic
to this problem lies in the fact that the mapping between $V_{\rm
  max}$ and $\sigma_{\ast}$, the latter being a directly observable
quantity, depends sensitively on the assumed density structure of the
DM halo of the dwarf.  For example, adopting an isothermal halo model
with a flat circular velocity profile to estimate $V_{\rm max}$ would
yield $V_{\rm max} \sim \sqrt{2}\,\sigma_{\ast}$
\citep[e.g.,][]{Moore_etal99}, while assuming an isotropic stellar
velocity dispersion tensor would result in $V_{\rm max}\sim
\sqrt{3}\,\sigma_{\ast}$ \citep[e.g.,][]{Klypin_etal99}.  Given the
uncertainties regarding the conversion of $\sigma_{\ast}$ to $V_{\rm
  max}$ (see \citealt{Kravtsov10} for a thorough discussion on this
issue), it is interesting to check the general validity of these
assumptions directly in our simulations. The relevant analysis is
presented in Figure~\ref{fig12}.

The left panel of this figure shows the relation between the
one-dimensional stellar velocity dispersion and the maximum circular
velocity $V_{\rm max}$ in the final stages of all simulated dwarfs
($t=10$~Gyr).  We remind the reader that the velocity dispersion
profiles of our remnants do not strongly vary with radius and thus our
dispersion measurements are close to the central values commonly used
by observers. For the purposes of this presentation, we have also
replaced $\sigma_{\ast}$ used throughout the paper by
$\sigma_{\ast}^{\prime} \equiv [(\sigma_r^2 + \sigma_{\theta}^2 +
\sigma^{\prime\,2}_{\phi})/3]^{1/2}$, where $\sigma^{\prime\,2}_{\phi}
= \sigma_{\phi}^2 + V_{\rm rot}^2$. This is because rotation was found
to be important in some of the remnants, specifically those of
simulations R3, R5, and R15 which do not produce dSphs, and thus it
had to be included in the calculation of their velocity dispersion. We
stress that this modification is not necessary in the majority of
cases where dSphs with minimal intrinsic rotation are formed.

The panel demonstrates a strong correlation between
$\sigma_{\ast}^{\prime}$ and $V_{\rm max}$ in all dwarf remnants
regardless of their morphology and kinematics. This is not entirely
unexpected as both of these quantities are measures of the depth of
the potential of the dwarfs. Interestingly, the relation $V_{\rm max}
= \sqrt{3}\,\sigma_{\ast}$, formally valid for a tracer stellar
population with an isotropic velocity dispersion tensor, reproduces
the results of the simulations remarkably well \citep[see
also][]{Klimentowski_etal09a}. Indeed, the formal best-fit slope of
the $V_{\rm max}-\sigma_{\ast}^{\prime}$ relation in our simulations
differs by less than $\sim 4\%$ from $\sqrt{3}$. Given the wide range
of orbital and structural parameters that we adopted for the
progenitor disky dwarfs in our experiments, this finding is
particularly noteworthy. We stress that the relation between
$\sigma_{\ast}$ and $V_{\rm max}$ is derived by directly probing the
stellar kinematics of the simulated dwarfs. It may thus be reliably
used to map the observed central stellar velocity dispersions to halo
maximum circular velocities in dwarf galaxies when addressing the
missing satellites problem.

Overall, we conclude that moderate conversion factors between
$\sigma_{\ast}$ and $V_{\rm max}$ such as those originally adopted by
\citet{Moore_etal99} and \citet{Klypin_etal99} to formulate the
missing satellites problem were reasonable. Recently, using controlled
simulations of subhalo evolution, \citet{Penarrubia_etal08} have
argued in favor of much larger conversion factors in observed MW
satellites ($\approx 2-3$; see, however, \citealt{Kravtsov10} for a
discussion of possible caveats in the approach of
\citealt{Penarrubia_etal08}). We stress that our results should at
least hold for dwarf galaxies that have moderate $M/L$ ratios as our
final systems (e.g., Leo I or Fornax).  Dwarfs that are much more DM
dominated (e.g., Draco or Ursa Minor) may, in principle, exhibit a
different relation between $\sigma_{\ast}$ and $V_{\rm max}$ owing to
their (possibly) different formation and evolutionary histories
\citep[e.g.,][]{Mayer_etal07}.

Furthermore, the relation between $V_{\rm max}$ and $\sigma_{\ast}$
reported here is valid for dwarfs that were originally embedded in
cuspy NFW-like DM halos, like the ones employed in the present study,
and would likely be different with other halo density structures.
Indeed, if the halo density profiles in the progenitor disky dwarfs
were shallower than NFW in the inner regions that are probed by the
stars \citep[e.g.,][]{Governato_etal10}, the resulting circular
velocity profiles will still be slowly rising within the luminous
radius, and the conversion factor between $\sigma_{\ast}$ and $V_{\rm
  max}$ can be substantially larger ($\gtrsim 2-4$; see, e.g.,
\citealt{Stoehr_etal02}).

Our findings are corroborated by a number of studies.  For example,
applying Jeans modeling and varying the anisotropy in the velocity
distribution of stars, \citet{Kazantzidis_etal04a} found that the
stellar kinematics of Fornax ($\sigma_{\ast} \sim 11-13 \kms$) and
Draco ($\sigma_{\ast} \sim 8-10 \kms$), can be reproduced in halos
with $V_{\rm max} \sim 20-30 \kms$, which would imply
$\sigma_{\ast}/V_{\rm max} \sim 0.3-0.65$.  Similar results were
obtained by \citet{Zentner_Bullock03} and this range of $V_{\rm max}$
was also confirmed by \citet{Madau_etal08} using the cosmological Via
Lactea simulation of the formation of a MW-sized halo. Another
investigation of Draco by \citet{Lokas_etal05}, which considered the
contamination of the stars by tidal tails and included the fourth
velocity moment in the analysis, found $\sigma_{\ast} \sim 8 \kms$ and
$V_{\rm max} \sim 16 \kms$, again in agreement with our results.
Lastly, \citet{Walker_etal09} have recently analyzed the observed
velocity dispersion profiles of a number of classic dSphs.  Treating
the anisotropy of the stellar velocity distribution as a free
parameter, these authors estimated the $V_{\rm max}$ of the host halos
in the range $\sim 10-25 \kms$, similar to those that would be
obtained assuming the relation between $V_{\rm max}$ and
$\sigma_{\ast}$ reported here.

The right panel of Figure~\ref{fig12} illustrates four examples of the
{\it evolution} of individual dwarfs on the $V_{\rm
  max}-\sigma_{\ast}^{\prime}$ plane.  Results are presented for
representative experiments R1, R2, R3, and R5, where the same dwarf
galaxy (D1) was placed on different obits, but the general trends are
confirmed in all simulations. The combination of tidal stripping and
loss of angular momentum dominating over tidal heating dictates the
characteristic evolution toward lower values of $V_{\rm max}$ and
$\sigma_{\ast}^{\prime}$ seen in all cases. The panel also shows that
the dwarf galaxies exhibit different trajectories depending on their
orbital parameters and can move substantially on the $V_{\rm
  max}-\sigma_{\ast}^{\prime}$ plane during the course of their
evolution inside their hosts. Both facts imply that the final $V_{\rm
  max}-\sigma_{\ast}^{\prime}$ relation in our simulations is not a
consequence of the initial conditions.


\begin{figure}[t]
  \centerline{\epsfxsize=3in \epsffile{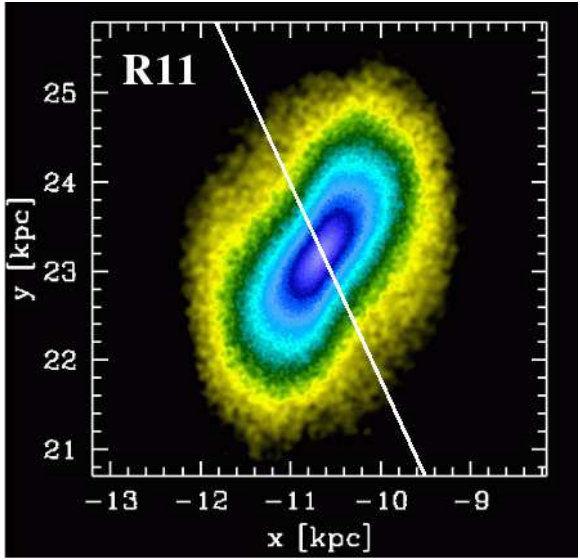}}
\vspace{0.1cm}
\caption{Surface density map of the stellar distribution of the dwarf in
  simulation R11. Particles are color-coded on a logarithmic scale,
  with hues ranging from yellow to blue indicating increasing stellar
  density. Local density is calculated using an SPH smoothing kernel
  of $32$ neighbors. The distribution of stars is projected onto the
  orbital plane and is visualized at the third pericentric approach
  ($t = 5.25$~Gyr from the start of the simulation). The solid line
  indicates the direction towards the center of the host galaxy,
  located at $(x,y) = (0,0)$. A bar is evident in the stellar
  component and the rotational velocity of the stars, $V_{\rm rot}$,
  is anticlockwise in this image. The tidal forces point in the same
  direction as the rotational velocity vectors of the stars in the
  dwarf, resulting in an increase of $V_{\rm rot}$.
\label{fig13}}
\end{figure}


\subsection{Increasing Rotational Velocity}
\label{subsec:Vrot}

In the majority of cases we studied, the rotational velocity of the
stars within $r_{\rm max}$, $V_{\rm rot}$, decreased systematically
with time leading from the initial rotationally-supported dwarfs to
pressure-supported stellar systems dominated by random motions. There
are instances in the evolution, however, where $V_{\rm rot}$ increases
at pericentric passages.  As seen in Figure~\ref{fig11}, this occurs
in simulations R1, R9, and R17 (after the second pericentric
approach), R8 (after the third pericentric approach), and R11 and R12
(after the third and fourth pericentric passage).

Investigating for the reason behind the temporary increases of the
rotational velocity we checked the orientation of the elongated
stellar component at the pericentric approaches at which such
increases occur.  We found that the typical orientation of the bar is
similar to that illustrated in Figure~\ref{fig13} where we see the
stellar component of simulation R11 at the third pericentric approach
projected onto the orbital plane. The rotational velocity of the stars
is anticlockwise in this plot. Therefore, the tidal forces which act
along the direction towards the host galaxy (solid line) point in the
same direction as the rotational velocity vectors of the stars in the
dwarf, effectively speeding up the rotation. Note that the case shown
in Figure~\ref{fig13} is the only instance in the entire evolution of
the whole suite of $17$ experiments where the total angular momentum
of the stars within $r_{\rm max}$ increases significantly. In all
other cases of increasing $V_{\rm rot}$ the angular momentum does not
increase because it is controlled not only by the rotational
velocities of the stars but also by $r_{\rm max}$, which decreases at
each pericentric passage.

Lastly, we note that the observed increases in $V_{\rm rot}$ are
always associated with changes in the shape of the stellar component
from a more prolate to a more triaxial configuration, or equivalently
with an increase of the difference between the axis ratios $b/a$ and
$c/a$.  This is most evident in experiment R11, where the prolate
shape with $b/a \approx c/a \approx 0.4$ at the third pericentric
passage transforms into a triaxial one (with $b/a > c/a$). A similar
phenomenon occurs at the subsequent (fourth) pericentric approach.
These changes are also accompanied by changes in other parameters. As
verified in Figure~\ref{fig2} and Figures~\ref{fig3}$-$\ref{fig10},
when the rotational velocity (or $V_{\rm rot}/\sigma_{\ast}$)
increases, there is always a drop in the anisotropy parameter $\beta$
and the bar strength amplitude $A_2$.  This means that at these
instances the shape of the stellar component becomes less prolate and
the orbital structure of the stellar distribution changes as the
stellar orbits become less radial.

\subsection{Increasing $M/L$ Ratio}
\label{subsec:M/L}

In most of our simulations, the $M/L$ ratio computed within $r_{\rm
  max}$ decreases monotonically as a function of time for the reasons
we discussed in \S~\ref{subsec:orbit_size}. In three of our $17$
experiments, however, the $M/L$ ratio starts to increase at some point
during the evolution. This happens most prominently in simulations R2
and R4, which are characterized by the smallest pericentric distances
and produce the most strongly tidally stirred dwarfs.  Although much
smaller, such an increase in the $M/L$ ratio is also observed in
experiment R13 where the stellar disk is more extended due to the
larger initial value of the radial scale length, $R_d$, adopted in
this case.  We stress at the outset that these increases of the $M/L$
ratio are rather moderate and thus cannot account for the exceptional
DM content in some of the classic dSphs such as Draco and Ursa Minor.
This implies that the very high values of $M/L$ inferred for some dSph
galaxies may not be a consequence of tidal evolution, but could be the
result of either the formation process of their progenitors or other
mechanisms that affected the baryonic mass fraction of dSphs. For
example, the presence and subsequent removal of gas by the combination
of tides and ram pressure has been shown to be vital in producing
systems with extreme DM content \citep{Mayer_etal06,Mayer_etal07}.

\begin{figure}[t]
\begin{center}
  \includegraphics[scale=0.42]{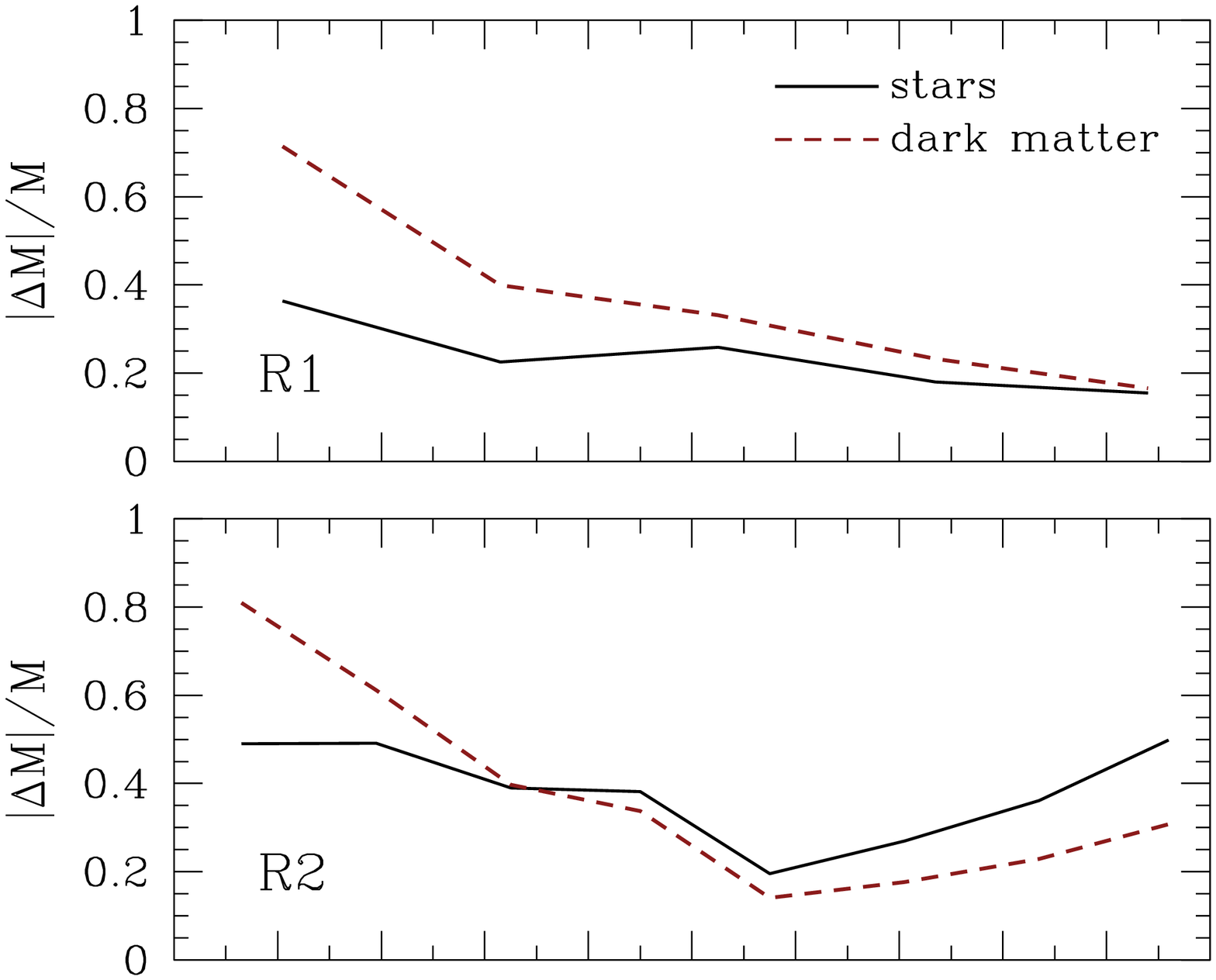}
  \includegraphics[scale=0.42]{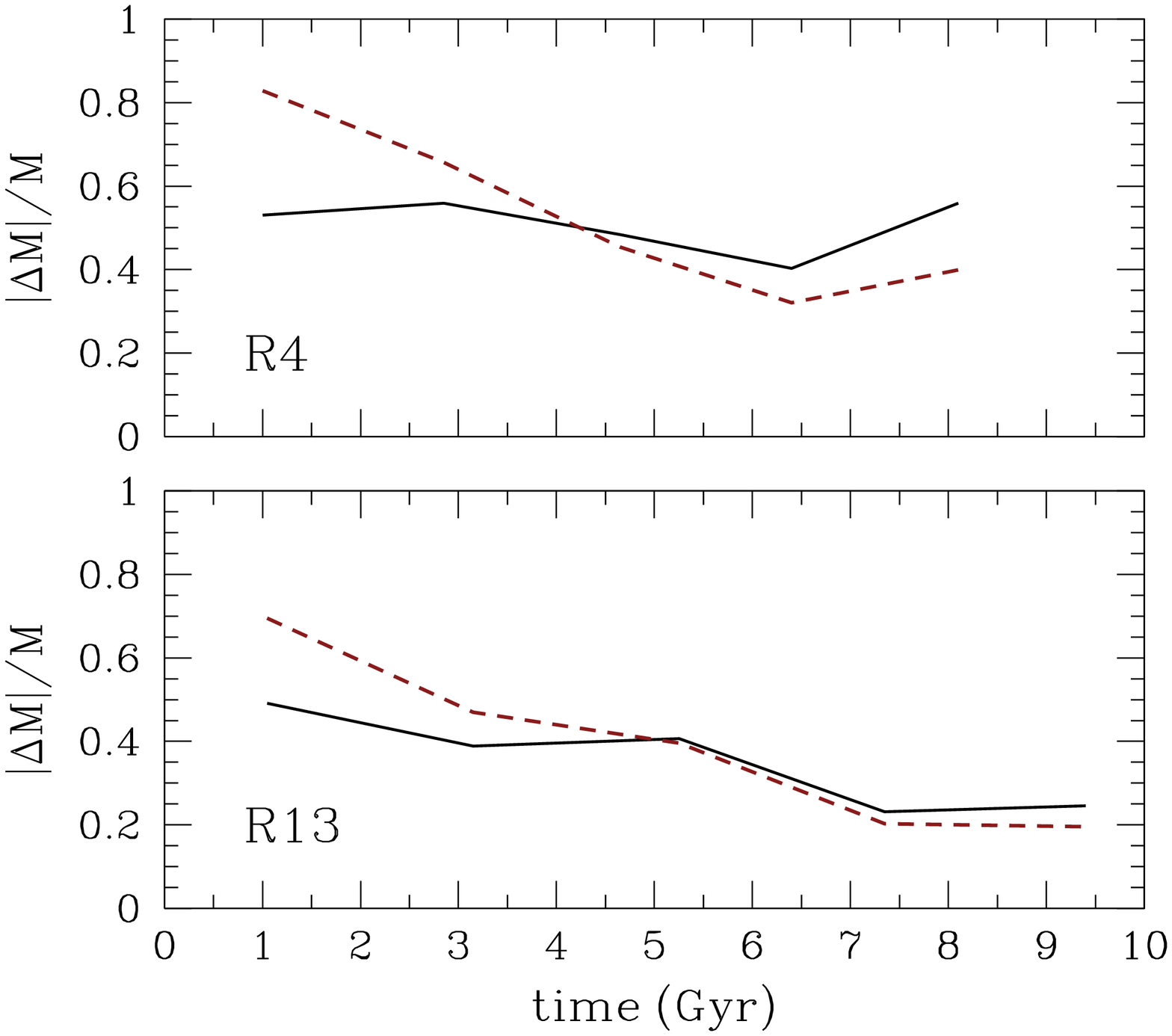}
\end{center}
\caption{Relative mass loss, $|\Delta M|/M$, of the stellar (solid
  lines) and DM component (dashed lines) in simulations R1, R2, R4,
  and R13. The values of $\Delta M/M$ are assigned at pericenters and
  are computed as $|\Delta M|/M \equiv |(M_{i+1} - M _i)|/M_i$, where
  $M_{i+i}$ and $M_i$ are the masses within $r_{\rm max}$ at the
  following and preceding apocenter, and $i$ denotes the apocenter
  number. The relative mass loss of the DM initially exceeds that of
  the stars in all cases, reflecting the more efficient tidal
  stripping of the extended DM halos of the dwarfs. While the mass
  loss in DM dominates at all times in the reference experiment R1, in
  simulations R2, R4 and R13 the stars begin to be stripped more
  effectively at some point during the evolution. This preferential
  stripping of the stars leads to the increase of the $M/L$ ratio
  reported at intermediate and late times in
  Figures~\ref{fig2},~\ref{fig4} and~\ref{fig8}.
\label{fig14}}
\end{figure}

Figure~\ref{fig14} investigates the reason behind the increasing $M/L$
ratio. This figure shows the relative mass loss, $|\Delta M|/M$, of
stars and DM separately in simulations R2, R4, and R13.  For
comparison, we also present the same quantities in our reference
experiment R1.  To avoid noise in the data, the values of $|\Delta
M|/M$ were assigned at pericenters by taking the masses within $r_{\rm
  max}$ at the following and preceding apocenter, namely $|\Delta M|/M
= |M_{i+1} - M_i|/M_i$, where $i$ is the apocenter number.
Figure~\ref{fig14} demonstrates that in all four simulations the
relative mass loss of the DM component is initially much larger than
that of the stars, and that the former always dominates in experiment
R1. In simulations R2, R4, and R13, however, the stars begin to be
stripped more effectively at some point during the late stages of the
evolution, leading to the increase of the $M/L$ ratios reported in
Figures~\ref{fig2},~\ref{fig4}, and~\ref{fig8}.

The reason for this behavior can be attributed to the fact that in
these three cases the dwarf galaxies are stripped down to the scales
where the alignment between the angular momenta of the stars and the
orbital angular momenta of the dwarfs start to have an important
effect on stellar stripping. In experiments R2 and R4, with the
smallest pericentric distances, this is due to the very strong tidal
forces.  In simulation R13, the disk is much more extended and,
consequently, the dwarf galaxy is stripped down to the stellar scales
faster than, for example, in the reference experiment R1. Because in
all experiments the stars retain a certain amount of angular momentum
(while the DM halos were non-rotating by construction) and the
orientation of the angular momentum vector does not strongly vary with
time, the stars begin to get stripped more efficiently compared to the
DM as their angular momenta are (on average) more strongly aligned
with the orbital angular momentum of the dwarf
\citep[e.g.,][]{Read_etal06b}. It is important to emphasize, however,
that the details of how the $M/L$ ratio would evolve depends, among
other things, on the initial angular momentum content of the DM halos,
the amount of angular momentum retained by the stars in the late
stages of the evolution, and the degree of alignment between the
angular momenta of stars and DM and the orbital angular momentum.


\subsection{Orbital Time or Pericentric Distance?}
\label{subsec:Torb_rperi}

\begin{figure}
\begin{center}
  \includegraphics[scale=0.42]{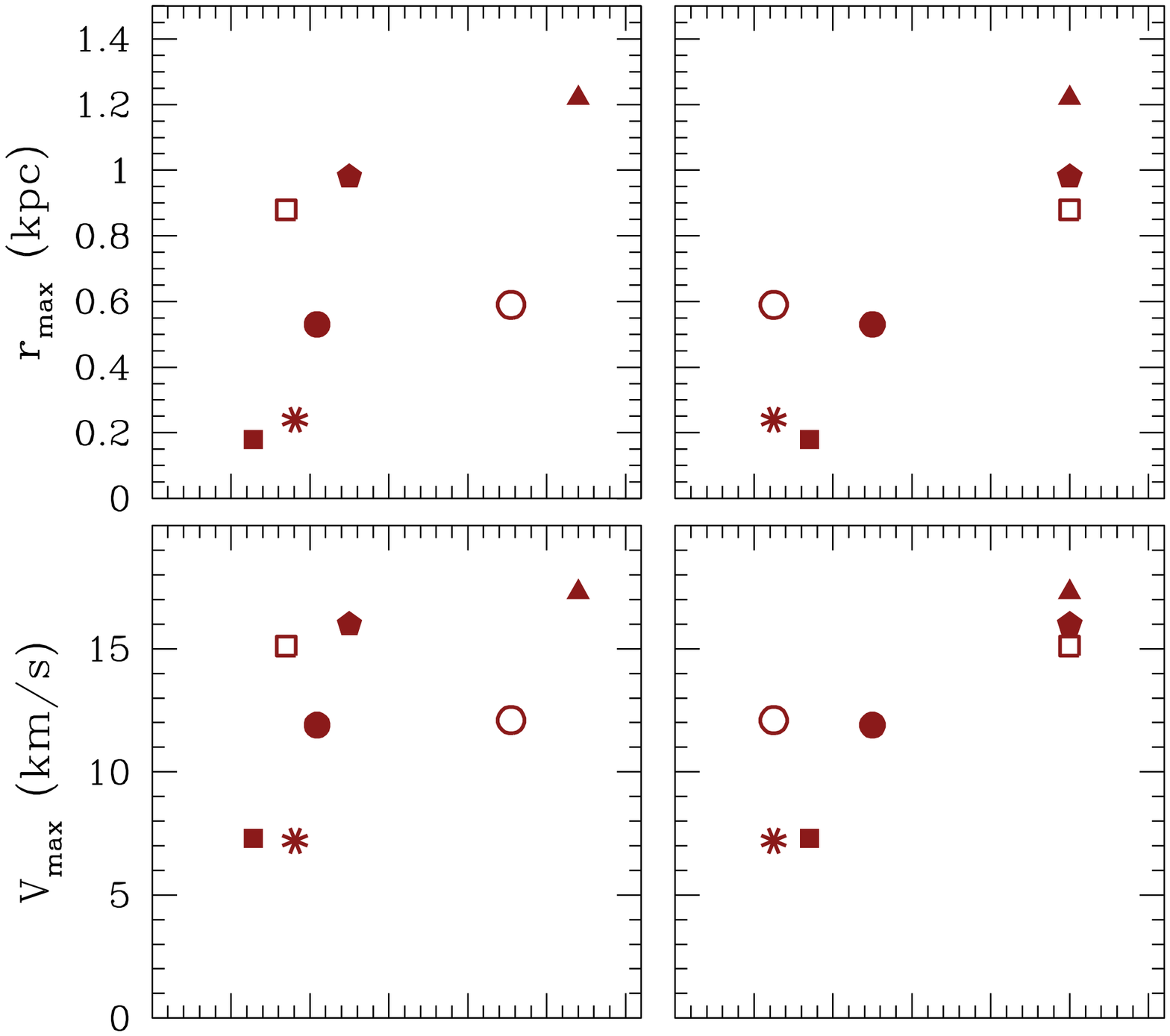}
  \includegraphics[scale=0.42]{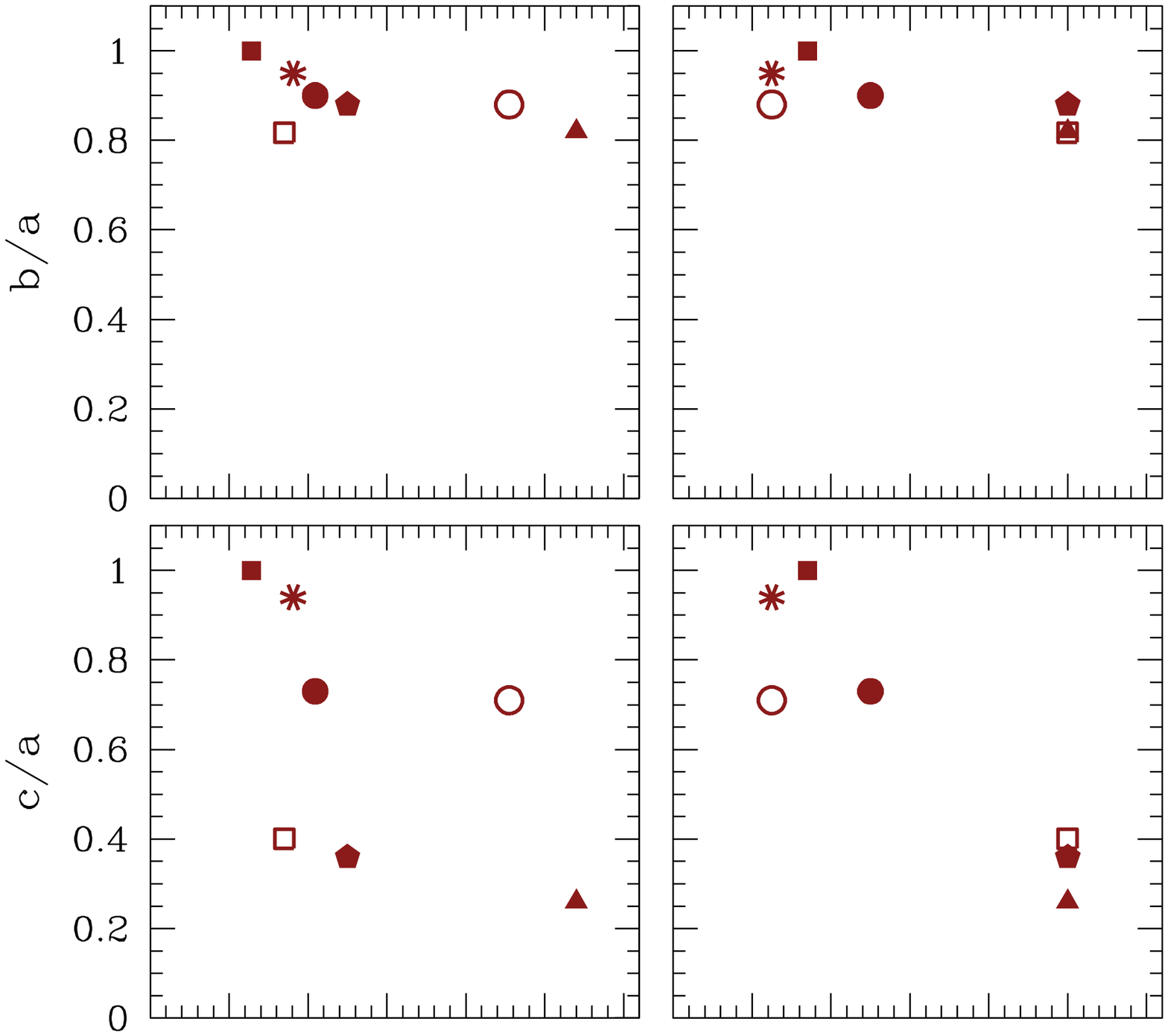}
  \includegraphics[scale=0.42]{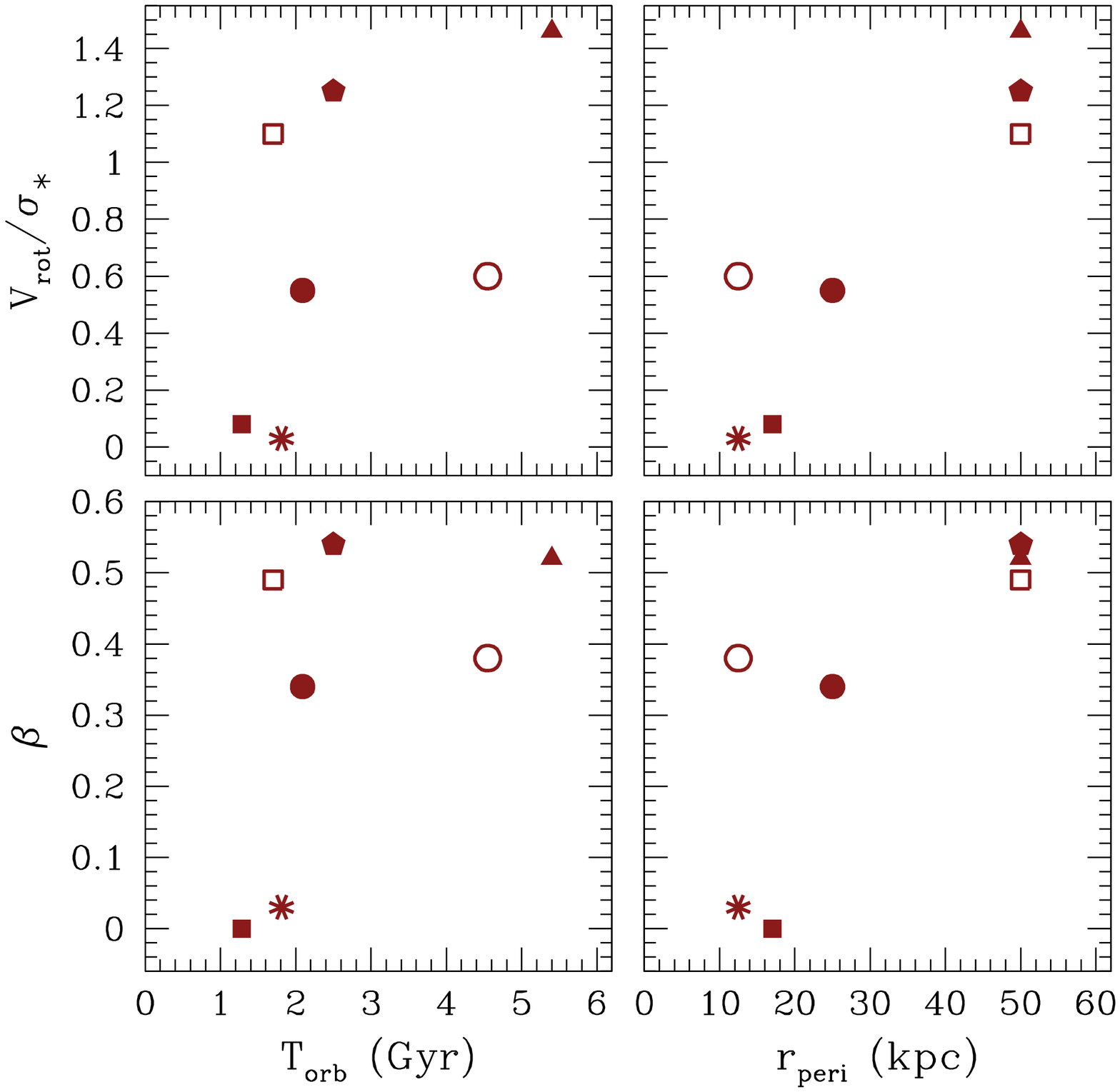}
\end{center}
\vspace{-0.4cm}
\caption{Final parameters of the simulated dwarfs as a function 
  of orbital time (left column) and pericentric distance (right
  column). Filled symbols show results for simulations R1-R5 (see
  Table~\ref{table:simulations}). Open symbols correspond to
  experiments where the dwarf galaxy model D1 used in simulations
  R1-R5 was placed on two additional orbits inside the host galaxy
  (see text for details). The pericentric distance which determines
  the strength of the tidal shocks constitutes the key factor
  responsible for the effective metamorphosis of a disky dwarf into a
  dSph. The combination of short orbital times and small pericentric
  distances, characteristic of dwarfs being accreted by their hosts at
  high redshift, induces the strongest and most complete
  transformations.
  \label{fig15}}
\end{figure}

In this section, we discuss the role of different orbital parameters
in influencing the transformation of disky dwarfs into dSphs and thus
the efficiency of the tidal stirring mechanism.  Specifically, we
focus on the relative importance of the orbital time, $T_{\rm orb}$,
and the pericentric distance, $r_{\rm peri}$, which determine the
number and the strength of the tidal shocks, respectively. The
relevant analysis is presented in Figure~\ref{fig15}. This figure
shows, as a function of $r_{\rm peri}$ and $T_{\rm orb}$, the final
values for some of the parameters that we used throughout this study
to describe the global evolution of the dwarf galaxies inside the
tidal field of their hosts. In particular, we consider the parameters
$r_{\rm max}$, $V_{\rm max}$, $b/a$, $c/a$, $V_{\rm
  rot}/\sigma_{\ast}$, and $\beta$.

The filled symbols in Figure~\ref{fig15} present results for
simulations R1-R5 which have been discussed in
\S~\ref{sec:orbital_parameters} (see also Table~\ref{table:summary}).
Open symbols correspond to two additional experiments, which were
designed to complement the existing simulations R1-R5. In this new set
of experiments, we placed the dwarf galaxy model D1 used in
experiments R1-R5 on the following orbits: (i) a short orbital
time-large pericentric distance orbit with $r_{\rm peri}= 50$~kpc,
$r_{\rm apo} = 80$~kpc, and $T_{\rm orb} \sim 1.7$~Gyr and (ii) a long
orbital time-small pericentric distance orbit with $r_{\rm peri}=
12.5$~kpc, $r_{\rm apo} = 250$~kpc, and $T_{\rm orb} \sim 4.6$~Gyr.
This choice of parameters ensures that the pericentric distances and
orbital times in these new simulations are not related in the same way
as in experiments R1-R5, a fact that would be important for the
interpretation of the results.

Figure~\ref{fig15} confirms that the final properties of the dwarfs
vary dramatically depending on the parameters of the orbit.  Although
the behavior of the various quantities clearly shows that the final
outcome of the transformation depends on both $T_{\rm orb}$ and
$r_{\rm peri}$, a closer inspection of the results in this figure
illuminates the more crucial role of the pericentric distance compared
to that of the orbital time in shaping dSphs via tidal stirring.

Let us first consider the filled symbols in Figure~\ref{fig15}.  Dwarf
galaxies with the same pericentric distances but with orbital times
that differ by approximately a factor of $2$ (triangle and pentagon)
end up with a similar set of final properties. On the other hand,
dwarfs with comparable orbital times but with pericentric distances
that differ by a factor of $2$ (pentagon, circle, and star) display a
wide range of final properties, a fact which reflects the difference
in pericentric distances.

Focusing now on the open symbols which correspond to the new
experiments, Figure~\ref{fig15} shows that the dwarf galaxy on the
long orbital time-small pericentric distance orbit (circle) is
characterized by $c/a \gtrsim 0.5$ and $V_{\rm rot}/\sigma_{\ast}
\lesssim 1$ and hence would be classified as a dSph. We note that this
dwarf has experienced only $2$ pericentric passages. On the contrary,
although it has suffered a much larger number of tidal shocks, the
dwarf on the short orbital time-large pericentric distance orbit
(square) has $c/a \approx 0.4$ and $V_{\rm rot}/\sigma_{\ast} > 1$ and
thus did not undergo a transformation into a dSph.

Overall, the results in Figure~\ref{fig15} lead to the following set
of conclusions. First, the degree of the tidal transformation depends
strongly on the combination of pericentric distance and orbital time.
As expected, small pericentric distances and short orbital times,
corresponding to orbits associated with a large number of strong tidal
shocks and also characteristic of dwarfs being accreted by their hosts
at high redshift, produce the most substantial tidal evolution.
Furthermore, the larger the number of strong tidal shocks a disky
dwarf suffers, the stronger and thus more complete is its
transformation.

Second, in order to simply transform a disky dwarf into a dSph
(according to the criteria of \S~\ref{subsec:criteria}) without
necessarily inducing the strongest transformation, the pericentric
distance is a more salient parameter than the orbital time. We stress
that this conclusion depends on the condition that $T_{\rm orb}$ is
short enough to allow the disky dwarfs to conclude at least two
pericentric passages\footnote{As discussed in \citet{Mayer_etal01a},
  disky dwarfs which due to their very long orbital times experience
  just one tidal shock during their orbital evolution would not be
  transformed, even if their pericentric distances are fairly small
  (except maybe in cases of very rare, nearly radial orbits). In such
  circumstances, the orbital time will play the key role in
  determining the dynamical evolution of infalling dwarfs.}.  Such a
requirement should be generally satisfied in the LG environment, as a
significant fraction of LG dwarfs have likely had enough time to
complete at least two orbits inside their hosts.  Indeed, in the
context of the CDM cosmogony, galaxy-sized halos form early and, in
particular, the MW halo was probably already in place at $z \sim 2$
\citep[e.g.,][]{Governato_etal07,Governato_etal09}. In addition,
structure formation in hierarchical models indicates that satellites
which are found closer to the center of the primary system at $z=0$
are typically those that fell into the host potential earlier
\citep[e.g.,][]{Diemand_etal07}. According to this, the present
distances of most dSphs would suggest that their progenitors should
have been accreted by their hosts at relatively early epochs.
Assuming a MW-type host potential and the range of orbital parameters
of LG satellites that survived until the present time
\citep{Klimentowski_etal10}, results in a typical orbital time of
$T_{\rm orb} \sim 2$~Gyr. Combining this value with a conservative
redshift of accretion of $z \sim 1$ suggests that dwarf galaxies in
the LG should have been able to complete $3$ or even $4$ passages
inside their hosts. This line of reasoning establishes the pericentric
distance, and not the orbital time, as the fundamental factor that
controls the tidal transformation of disky dwarfs into dSphs in the LG
and similar environments.


\subsection{Implications}
\label{subsec:implications}

In the present study we have established that the strong tidal field
of a MW-sized host galaxy can transform late-type,
rotationally-supported dwarfs into stellar systems with the kinematic
and structural properties of the classic dSphs in the LG and similar
environments. Of course, tidal stirring does not constitute the only
{\it environmentally-driven} mechanism proposed for the origin of
dSphs. An alternative promising model incorporates direct
gravitational interactions between dwarf galaxies. Indeed,
cosmological simulations of structure formation have indicated that
some dwarfs may have entered their Galaxy-sized hosts as members of
infalling groups \citep{Kravtsov_etal04,D'Onghia_Lake08}. It is thus
plausible that these dwarfs could have experienced significant tidal
perturbations before being accreted by the primary galaxies.
\citet{Kravtsov_etal04} did demonstrate this possibility in their
{\LCDM} cosmological simulations of MW-sized halos.

Using controlled $N$-body experiments, \citet{D'Onghia_etal09} have
recently shown how the interaction between a pair of disky dwarfs with
mass ratios ranging from $1$:$10$ to $1$:$100$ can result in the rapid
formation of a dSph. As the less massive member of the pair plunges
close to the disk of the more massive companion, its stellar component
can experience substantial mass loss and can be tidally heated into a
spheroidal configuration, even after only one close pericentric
passage. This mechanism, termed ``resonant stripping'', is effective
when the spin angular frequency of the dwarf disk and the angular
frequency of the orbit of the dwarf around the more massive system are
comparable in magnitude, and the spin and orbital angular momenta are
nearly aligned (that is, in a nearly prograde encounter).  Such
individual, close encounters between dwarf galaxies may have been more
common during the assembly process of the host halos from accreting
groups \citep{Kravtsov_etal04,D'Onghia_Lake08}. Given that the typical
velocity dispersions in such groups should have been quite low (with a
dominant group member of the size of the LMC), it is also possible
that some of these interactions may have resulted in actual mergers
between the dwarf galaxies.

While all these different tidal mechanisms which rely on {\it
  pre-processing} the dwarfs before they are accreted by the primary
galaxy may play a role at some level, they must also be consistent
with the morphology-density relation, a crucial constraint that the
tidal stirring model can naturally satisfy. At present, there is no
definitive observational or theoretical evidence to rule out any of
the scenarios for the origin of dSphs. This also includes models that
advocate the {\it ab initio} formation of dSphs from cosmological
initial conditions \citep[e.g.,][]{Ricotti_Gnedin05,Sawala_etal10}.
In all likelihood, all mechanisms do operate simultaneously to
differing degrees in producing the population of dSphs in the LG and
similar environments.

Our investigation of the role of different pericentric distances and
orbital times on the efficiency of tidal stirring has interesting
implications for the expected properties of dSphs. Our results show
that in order to transform a disky dwarf with an orbital apocenter in
excess of $200$~kpc into a dSph, an orbit with an eccentricity that is
substantially larger than the median expected eccentricity is required
(see Figures~\ref{fig2} and~\ref{fig15}).  Given that the current
distances of dwarf galaxies in the LG are suggestive of their orbital
apocenters, the distant dSph Leo I (currently at a distance of $\sim
250$~kpc from the MW; e.g., \citealt{Caputo_etal99}) should be on a
very wide, nearly radial orbit, assuming that it is the product of
tidal stirring. Furthermore, even on such eccentric orbits the
remnants possess some residual rotation at the level of $V_{\rm
  rot}/\sigma_{\ast} \approx 0.5$ (see open circle in
Figure~\ref{fig15}). We thus predict that Leo I should have a higher
$V_{\rm rot}/\sigma_{\ast}$ compared to the dSphs located closer to
the primaries. Interestingly, albeit inconclusive, there are
signatures of intrinsic rotation in Leo I
\citep{Sohn_etal07,Mateo_etal08}.  We note that such rotation has been
shown to explain the observed kinematics of this dwarf
\citep{Lokas_etal08}.

Our prediction that in the context of tidal stirring wider orbits
should be associated with less evolved dSphs, and thus with higher
values of $V_{\rm rot}/\sigma_{\ast}$, should apply, even more so, to
the oddly isolated galaxies Cetus and Tucana, which are located in the
outskirts of the LG at more than $500$~kpc from M31 and the MW,
respectively.  Although measurements of kinematics are extremely
challenging due to the very large distances involved, recent
spectroscopic studies have presented tentative evidence for rotation
in Cetus \citep{Lewis_etal07} and Tucana \citep{Fraternali_etal09}
with $V_{\rm rot}/\sigma_{\ast}$ values in the range of $0.5 \lesssim
V_{\rm rot}/\sigma_{\ast} \lesssim 1$. If confirmed by future
observations, detection of rotation at such levels would be consistent
with the predictions of the present study.

Furthermore, at distances in the range $\sim 300-600$~kpc from the MW
and M31, we find the dwarf galaxies Phoenix and LGS3, which are
classified as ``transition-type'' dwarfs \citep{Mateo98}.  While these
objects exhibit structural properties consistent with those of dSphs,
they are associated with some residual gas (see, e.g.,
\citealt{Grcevich_Putman09} and references therein).  On the other
hand, these transition-type dwarfs exhibit negligible or no ongoing
star formation (no detection of $H_{\scriptsize II}$ regions has been
reported), a fact that differentiates them from standard gas-rich,
star-forming dIrrs \citep[e.g.,][]{Grebel99,Skillman05}.  Being at
intermediate distances between dSphs and dIrrs and combining
properties of both classes of objects, these transition-type dwarfs
can provide valuable tests for the predictions of tidal stirring
regarding the different stages of the transformation process.

One possibility for their origin is that their progenitors were disky
satellites that were accreted only recently by their hosts on wide,
fairly radial orbits, and are still in the process of being
transformed into dSphs by tidal stirring. It is then plausible that
these transition-type dwarfs have only concluded one pericentric
passage (or they are on the way to their first pericentric approach),
having lost most of their gas by ram pressure stripping. This is a
reasonable assumption given that gas removal by the combination of
tides and ram pressure proceeds faster than the morphological
transformation of the stellar component \citep{Mayer_etal06}. If the
scenario proposed here is correct and their transformation has only
been partial, transition-type dwarfs should possess significant
rotation in their stellar component (as in the case of the remote
dSphs Leo I, Tucana, and Cetus discussed above).  Verifying
observationally this basic theoretical prediction would be important
-- unfortunately, the stellar kinematics of these systems is poorly
known. Moreover, provided that they are at the intermediate stages of
their transformation, our results suggest that they should also show
signs of bar-like distortions.  Definitive conclusions on all of these
issues would require a synergy between detailed photometric and
kinematic measurements as well as exhaustive comparisons with
theoretical models. To this end, extending the search for evidence of
tidal stirring in systems with more recent assembly histories than the
LG may reveal a population of satellites in the process of being
transformed, and thus offer unique opportunities to constrain the
tidal stirring model. Interestingly, recent studies using the SDSS
database have uncovered a population of dwarf galaxies in the Virgo
cluster that exhibit disk-like features and bars, appearing to be in a
transitional stage between a disk and a spheroid
\citep{Lisker_etal07}. The existence of this class of objects is, in
broad terms, consistent with the predictions of tidal stirring for the
different stages of the transformation.

Several alternative scenarios to tidal stirring exist for explaining
the nature of the spheroidal morphology of the distant LG dwarfs.  For
example, as a result of three-body interactions, satellites can
acquire extremely energetic orbits with apocenters beyond the virial
radius of the primary and be ejected to large distances
\citep{Sales_etal07}.  In this model, ejected subhalos are typically
the less massive members of a pair of satellites that is tidally
disrupted during the first approach onto the host. Such ejections can
also occur during the tidal disruption of a bound system of {\it
  multiple} subhalos that is accreted as a single unit by the primary
galaxy \citep{Ludlow_etal09}.  In both of these scenarios, strong
tidal interactions with the more massive companions can took place
before the satellites are accreted by their hosts. Such encounters may
eventually transform the dwarfs into dSphs, either via resonant
stripping \citep{D'Onghia_etal09} or other gravitational processes
such as mergers. Interestingly, \citet{Kravtsov_etal04} reported in
their cosmological simulations the existence of a few satellite
systems at distances of $\sim 1000$~kpc from their MW-sized hosts.
Based on their final $V_{\rm rot}/\sigma_{\ast}$ values, these objects
would be classified as dSphs and could plausibly represent the
counterparts of Cetus and Tucana in the context of the LG. The tidal
heating of these systems did occur in small groups that were accreted
by the primary halo at the present epoch. As the tidal forces unbind
these accreting groups, energy redistribution can increase the orbital
energy and apocentric distance for some of the satellites, providing
an explanation for the presence of isolated dSphs in the periphery of
the host galaxy. The extreme radial velocity of Leo I
\citep[e.g.,][]{Mateo_etal08} and the relatively high recession
velocity of Tucana \citep{Fraternali_etal09} may already suggest that
these systems have been propelled into their highly energetic orbits
through the type of interactions suggested by \citet{Sales_etal07} and
\citet{Ludlow_etal09}.

Binary mergers between individual satellite galaxies taking place
outside of infalling groups may offer another alternative explanation
for the puzzling presence of the isolated dSphs in the LG.  Indeed, it
has already been demonstrated that interactions and direct mergers of
subhalos can lead to their very strong evolution
\citep[e.g.,][]{Knebe_etal04,Knebe_etal06,Klimentowski_etal10}.  For
example, \citet{Klimentowski_etal10} found in their constrained
simulation of the LG that $\sim 10\%$ of all surviving subhalos in the
MW and M31 have undergone a major encounter with another subhalo in
the past.  Most of these events occurred at very early times, between
$z\sim 1$ and $z\sim 3$, while the interacting subhalos have not yet
become satellites and are still outside their hosts.  Whether binary
mergers between dwarfs can explain the existence of dSphs orbiting far
from the primary galaxies, such as Tucana and Cetus, is currently
under investigation via a combination of cosmological simulations and
controlled numerical experiments (Kazantzidis et al., 2010a, in
preparation). Ongoing investigations aimed at studying in detail the
stellar components of these isolated dSphs
\citep[e.g.,][]{Bernard_etal09,Monelli_etal10a}, including their star
formation histories \citep[e.g.,][]{Monelli_etal10b}, may soon provide
useful constraints on the various competing models that have been
proposed so far for their origin.

In this paper, we have investigated the tidal evolution of dwarf
galaxies comprising exponential stellar disks. Recent cosmological
simulations do support the idea that isolated dwarfs are rotating
disks (\citealt{Governato_etal10}; see, however,
\citealt{Sawala_etal10}). The morphological and dynamical
transformation of such disky dwarf galaxies into dSphs under the
action of tidal forces is a rather rich process which involves several
stages and is characterized by different events.  Consequently,
numerical simulations where a spheroidal stellar system is postulated
from the beginning \citep[e.g.,][]{Munoz_etal08,Penarrubia_etal08} may
be inadequate to describe how dSphs have evolved to the present time
subject to the tidal field of their hosts.  Indeed, if the picture
presented here is correct, dSph galaxies should have formed relatively
late in most cases and only after concluding a number of pericentric
passages inside the primary galaxies.

More specifically, \citet{Penarrubia_etal08} performed a series of
$N$-body simulations to study the dynamical evolution of dSphs in a
host potential assuming a spherical King model for the stellar
component embedded within an NFW halo. These authors reported that
tidal effects lead to an increase in the $M/L$ ratio in most cases, in
contrast to the findings of the present study. This discrepancy is
possibly due to the fact that the stellar distribution in the
\citet{Penarrubia_etal08} dwarfs followed the King profile which is
characterized by a density core. As a result, the stars were loosely
bound within the potential of the dwarf galaxy and the stellar
component was much more heavily stripped compared to what is typically
found here. In our experiments, the formation of tidally-induced bars,
which is obviously missing from the \citet{Penarrubia_etal08} models,
is crucial. This is because bar formation enhances the resilience of
the dwarf galaxies to mass loss and tidal stripping by increasing the
stellar density and, correspondingly, the depth of the potential well.

Given the existence of a number of qualitatively different scenarios
for the formation and evolution of dSphs, it is critical to be able to
test these scenarios and discriminate among them. The growing
kinematic data sets for dSph galaxies can already facilitate detailed
comparisons with theoretical models \citep[e.g.,][]{Walker_etal09}.
For example, the level of residual stellar rotation in the remnants
can be used to constrain the competing models. Indeed, according to
the results presented in Figure~\ref{fig11} and detailed analysis of
similar simulations \citep[see][]{Lokas_etal10b}, if the present-day
dSph galaxies originated from disky dwarfs they should, at least in
some cases, show signatures of intrinsic rotation.  Interestingly, in
addition to the cases of the isolated dSphs Leo I
\citep{Sohn_etal07,Mateo_etal08}, Cetus \citep{Lewis_etal07}, and
Tucana \citep{Fraternali_etal09} mentioned already, detection of
rotation at different levels of significance has also been claimed for
Ursa Minor \citep{Hargreaves_etal94, Armandroff_etal95} and Sculptor
\citep{Battaglia_etal08}. It is important to stress, however, that
intrinsic rotation may be difficult to distinguish from the velocity
gradients induced either by the presence of tidal tails or, for nearby
systems, caused by transverse motions (the so-called ``perspective
rotation''; see \citealt{Walker_etal08}). Due to such complications,
Leo I seems to be the most promising candidate for detection of
intrinsic rotation (see discussion in \citealt{Lokas_etal08}). In
addition, as discussed above, our simulations predict a positive
correlation between the magnitude of stellar rotation in dSphs and
their distances from the center of the host galaxy (or equivalently
the time of infall of the progenitor dwarf onto the primary). Such a
correlation would be difficult to establish, for example, in models
that propose the {\it ab initio} formation of isolated dSphs from
cosmological initial conditions
\citep[e.g.,][]{Ricotti_Gnedin05,Sawala_etal10}. Indeed, in this case
the amount of rotation in the stellar components of dSphs should
depend on the specifics of the formation process and the intrinsic
properties of these systems (e.g., the initial spin parameter of the
DM halo of the dwarf) and thus be uncorrelated with the distance from
the center of the host galaxy.

In the context of specific observational signatures of the tidal
stirring model, bars play a prominent role.  According to our results,
bar-like structures should be relatively common in dSphs as the bar
phase is one of the longest stages in the transformation process. In
fact, some of the less evolved dSphs in the LG may still be in the bar
stage and show signs of bar-like distortions. One irrefutable example
of a LG dwarf that contains a bar is the LMC.  Orbital evolution
models using three-dimensional velocities constrained by recent proper
motion measurements \citep{Kallivayalil_etal06} suggest that the LMC
may currently be on the first passage having just crossed the
pericenter of its orbit around the MW \citep{Besla_etal07}. Confirming
that the LMC is on a such extended orbit with a pericentric distance
of $r_{\rm peri} \sim 50$~kpc would be in line with our findings, as
we expect its stellar component to be less evolved and thus in some
early, transitory stage of the transformation between a disk and a
spheroid.

Except for the LMC, however, the overall number of detections of
bar-like structures among the dSphs in the LG is relatively
low\footnote{We note that there are a number of reasons as to why bars
  in dwarf galaxies may escape detection in the LG.  First, bars are
  oriented randomly with respect to the observer, so some bar-like
  structures may appear as only slightly flattened.  Second, the
  smoothing procedures usually applied when measuring the surface
  density distributions of stars in dwarf galaxies can decrease the
  detectability of bars in these systems (for a thorough discussion
  pertaining to the difficulties in identifying bars in dSphs, see
  \citealt{Klimentowski_etal09a}).}. Indeed, markedly elongated
isophotes that could be attributed to a residual bar-like component
are observed in only a few cases, including Ursa Minor
\citep{Irwin_Hatzidimitriou95} and the newly discovered Hercules dSph
\citep{Coleman_etal07}.  Recently, the strongly non-spherical shape of
the core of the disrupting nearby Sagittarius dSph has also been
ascribed to a pre-existing bar \citep{Lokas_etal10a}.  Interestingly,
the recently discovered ultra-faint MW satellites also exhibit
substantial degrees of flattening \citep[e.g.,][]{Martin_etal08}. Of
course, the high elongation of the stellar component in some of the
dSphs can also be caused by other effects, including the triaxiality
of the surrounding DM halos \citep{Kazantzidis_etal10} or tidal
deformation in the gravitational field of the MW. Regarding the
latter, \citet{Martin_Jin10} have recently proposed that Hercules dSph
is in fact a stellar stream in formation, thus suggesting tidal
disruption as the most valid scenario for the extreme shape of this
system.  Deeper observations to track evidence of such tidal
interactions would be able to settle these issues.

The findings of the present study also indicate that the formation of
tidally-induced bars is strongly linked with the transformation of
rotationally-supported dwarfs to dSphs (for similar conclusions, see
\citealt{Mayer_etal01a} and \citealt{Klimentowski_etal09a}). Indeed,
as shown in Table~\ref{table:summary}, in only one of the $14$
simulations that produced dSphs in the end, specifically experiment
R16, a bar did not form at some point during the evolution of the
progenitor disky dwarf inside the host galaxy. The following sequence
of events is typically observed in our simulations. First, the strong
tidal forces at the initial pericentric approach trigger a bar
instability in the disk of the dwarf. The tidally-induced bar
transports angular momentum outwards to the outer regions of the disk
and to the DM halo.  As tidal stripping removes the outer parts of the
dwarf, the entire angular momentum content gradually decreases and the
ability of the dwarf galaxy to be supported by rotation progressively
diminishes.  Second, subsequent tidal shocks destroy the centrophilic
stellar orbits which support the bar and increase the stellar velocity
dispersion. As a result, the bar continuously loses its elongation and
is tidally heated into a more isotropic, diffuse spheroid.  The
ultimate outcome of these two physical processes is the formation of
pressure-supported stellar systems with values of $V_{\rm
  rot}/\sigma_{\ast} \lesssim 1$ that are appropriate for dSphs.

We stress that the above discussion describes only one channel for the
transformation of a disky dwarf into a dSph via tidal stirring.  An
alternative mechanism involves the buckling of the bar due to the
amplification of vertical $m=2$ (bending) modes
\citep{Mayer_etal01a,Mayer_etal01b}.  While the loss of angular
momentum occurs in exactly the same way as previously described, the
buckling of the bar now becomes the main process that leads to the
significant vertical heating and the increase of the velocity
dispersion of the system (see, e.g., \citealt{Debattista_etal06} for a
detailed description of these processes). In this picture, the
spheroidal shape of the stellar distribution constitutes the end
result of the instability and is not driven by tidal heating as in the
simulations of the present study.  Because the buckling instability
requires a fairly strong bar to develop, this transformation mechanism
is relevant to relatively massive systems with high surface densities.
Indeed, none of the bars in our low surface density dwarfs showed any
signs of buckling.

To summarize, it seems that there are at least two channels for the
formation of the spheroidal component in the tidal stirring scenario:
prolonged impulsive tidal heating and the bar buckling instability.
The former is favored in low mass, low surface density disky dwarfs
for which heating is particularly efficient. This mechanism likely
applies to the progenitors of the faintest classic dSphs such as
Draco, Sculptor or Leo I. On the other hand, bar buckling requires
strongly self-gravitating systems with higher surface densities. This
channel plausibly operates on the progenitors of the brightest dSphs
such as Fornax and Sagittarius, and even more so on those of the
bright dwarf elliptical satellites of M31, such as NGC185 and NGC167.
This mechanism was also found to be applicable to the transformation
of relatively massive and bright spiral galaxies in galaxy clusters
\citep{Mastropietro_etal05}.

Our simulations also suggest that the amount of mass loss that the
dwarf galaxies experience can be considerable.  Indeed, in the
reference experiment R1, the dwarf lost $\sim 90\%$ of its initial
mass within $r_{\rm max}$ and still survived as a bound entity (the
amount of mass loss for the most heavily stripped dwarfs in
simulations R2 and R4 reached $\sim 99\%$).  Typically, the maximum
circular velocities, $V_{\rm max}$, decreased by a factor of $\sim 2$
during the orbital evolution in the cases where dSphs were produced
(this factor increased to $\sim 3$ in experiments R2 and R4).  Such
substantial evolution in $V_{\rm max}$ occurred despite the presence
of the baryons which tend to moderate the effect of tidal shocks by
making the potential well deeper, especially after bar formation.
These findings are in agreement with those of
\citet{Klimentowski_etal09a} as well as with results of other studies.
For example, \citet{Kravtsov_etal04}, using fully cosmological DM-only
simulations, reported that the $V_{\rm max}$ of their most evolved
subhalos decreased by a factor of $1.5-2$ on average during $10$~Gyr
of tidal evolution. More recently, \citet{Diemand_etal07} and
\citet{Madau_etal08} found that satellites in the Via Lactea
simulation lost between $\sim 30$ and $\sim 99\%$ of their pre-infall
mass, and that $V_{\rm max}$ typically decreased by a factor of $\sim
2-3$.

The previous discussion suggests that for subhalos which have been
accreted by their hosts at early cosmic epochs and have completed
several orbits with fairly small pericenters, $V_{\rm max}$ is
expected to have evolved significantly. Therefore, the DM halos of
present-day dSphs may have had considerably larger masses and circular
velocities when they entered the halo of the MW (see, e.g.,
\citealt{Kravtsov_etal04} and \citealt{Madau_etal08}). This has
important implications for galaxy formation models. Indeed, the
substantial observational work during the past few years suggests that
the present-day classic dSphs of the MW and M31 have central stellar
velocity dispersions in the range $\sigma_{\ast} \sim 7-13 \kms$ (see,
e.g., \citealt{Walker_etal09} and references therein).  Assuming that
$V_{\rm max} \sim \sqrt{3} \,\sigma_{\ast}$, in accordance with the
results of \S~\ref{subsec:Vmax_sigma}, implies current values of
$V_{\rm max}$ in the range $V_{\rm max} \sim 12-22 \kms$ and initial
$V_{\rm max}$ values, namely before infall and tidal mass loss, in the
range $V_{\rm max} \sim 24-44 \kms$. These numbers are quite
important.  Indeed, with such relatively high values of initial halo
$V_{\rm max}$, photoevaporation of the gas by the cosmic ultraviolet
background after reionization as well as supernovae feedback could not
be the factors that shaped the baryonic content and nature of these
dSphs. In fact, according to the radiative transfer simulations of
\citet{Susa_Umemura04}, photoevaporation would be effective and remove
most of the gas only for $V_{\rm max} \lesssim 20 \kms$. This is
consistent with the lack of a clear signature of the reionization
epoch in the star formation histories of dSphs
\citep{Grebel_Gallagher04,Orban_etal08}.

In this paper, we have also demonstrated how the orbital parameters
and initial structures of the progenitor late-type disky dwarfs
determine the final properties of the dSphs. Therefore, the fact that
Fornax and Draco have roughly similar masses at present, as inferred
from the $V_{\rm max}$ of their halos
\citep[e.g.,][]{Kazantzidis_etal04a}, but differ by about an order of
magnitude in luminosity and $M/L$ ratio, can be explained in two ways.
One possibility is that the progenitors of these two dwarfs began with
very different relative amounts of DM and baryons, for reasons related
to their formation history and not to the environment.  Alternatively,
Fornax and Draco originated from systems that had comparable DM and
baryonic masses, but experienced dissimilar tidal evolutions because
they entered the primary galaxy at different epochs and/or on
different orbits.

Regarding the latter, according to the hydrodynamical simulations of
\citet{Mayer_etal07}, gas-rich disky satellites that were accreted by
their hosts when the intensity of the cosmic ultraviolet background
was much higher than today ($z \sim 2$), can be completely stripped of
their gas by ram pressure in one or two pericentric passsages. In this
case, the final systems would correspond to dSphs with truncated star
formation histories such as Draco \citep[e.g.,][]{Orban_etal08}. While
the baryonic content of the progenitor dwarfs decreased significantly
as a result of gas stripping, the initial DM mass in the central
regions around the surviving baryonic core is largely preserved.  This
is because DM is affected only by tides and not by ram pressure.  As a
result, exceptional $M/L$ values, of the order of $100$, similar to
those inferred for Draco and Ursa Minor, can arise.  On the other
hand, disky dwarfs that were accreted by their hosts at $z \lesssim
1$, when the intensity of the cosmic ultraviolet radiation dropped by
more than an order of magnitude compared to $z \sim 2$, were able to
retain some gas because tides and ram pressure could not strip it
completely.  Under these conditions, the infalling dwarf galaxies can
undergo tidally-triggered bursts of star formation associated with
bar-driven gas inflows at pericentric approaches.  Such a model would
be applicable to dSphs with extended star formation histories such as
Fornax, Carina, and Leo I. This mechanism can produce dSphs that are
brighter for a given halo mass (or central stellar velocity
dispersion) compared to the ones that were accreted earlier. While
this generic scenario seem to explain the differences in the
properties of Fornax and Draco despite their similar masses, it would
be important to revisit it in future work with models capable of
capturing the multi-phase structure of the ISM in dwarf galaxies.

Lastly, it has long been debated whether the inferred high $M/L$
ratios of dSphs indeed signify exceptional DM content or are simply a
reflection of strong tidal effects and of the fact that these systems
are in reality on the verge of disruption.  The findings of the
present study have intriguing implications for earlier attempts to
model dSph galaxies as unbound systems without DM
\citep[e.g.,][]{Kroupa97,Klessen_Kroupa98}. Indeed, the tidal stirring
model demonstrates how mass loss and the formation of tidal tails can
be consistent with the presence of a bound stellar component embedded
in a relatively massive CDM halo, even after several Gyr of tidal
evolution inside the host. Our results indicate that substantial mass
loss and the existence of a gravitationally bound dSph galaxy with a
relatively high $M/L$ ratio are not mutually exclusive, confirming
earlier claims based on a smaller set of lower resolution simulations
\citep{Mayer_etal02}. The findings of the present study also suggest
that the claimed detection of extra-tidal stars in a number of dSphs,
including Ursa Minor \citep{Martinez-Delgado_etal01}, Fornax
\citep{Coleman_etal05}, Carina \citep{Munoz_etal06}, and Leo I
\citep{Sohn_etal07} is consistent with satellite accretion in CDM
models. We note in passing that the number of such detections may be
low due to the intrinsic difficulties associated with separating the
tidal tails from the bound core of the dwarfs (see, e.g.,
\citealt{Klimentowski_etal09b}). Furthermore, in the context of tidal
stirring, dSphs embedded in CDM halos exhibit stellar distributions
that are adequately fit by exponential or King profiles in agreement
with observed dSphs (see review by \citealt{Mayer10}).  Stellar
profiles of this type are difficult to accommodate within models that
represent dSphs as systems devoid of DM. Indeed, such models predict
nearly flat profiles as expected for objects close to complete
disruption \citep[e.g.,][]{Kroupa97}.

\subsection{Caveats and Future Directions}
\label{subsec:caveats_directions}

Certainly the approach presented in this paper is not without caveats.
A first limitation is related to the fact that we have adopted a
single primary galaxy with the present-day structural properties of
the MW.  In general, at the time when the dwarfs are accreted by the
primary galaxy at high redshift, the DM and baryonic masses of the
host would be different compared to those of the present time Our
methodology thus neglects the cosmological evolution of the host
galaxy structure via mergers and smooth accretion during the
interactions with the dwarfs. However, this simplification may be
justified to a certain degree by recent {\LCDM} galaxy formation
simulations, which have shown that MW-sized disk galaxies assemble
most of the mass in their inner regions between $\sim 8-10$~Gyr ago
\citep[e.g.,][]{Governato_etal07,Governato_etal09}. This is indeed the
timescale that we follow in our simulations. Nonetheless, a more
complete investigation would have to include the ongoing formation of
the host galaxy. In addition, we have assumed a host DM halo that is
spherical (except in the very inner regions that are dominated by the
potential of the disk), instead of triaxial as postulated by CDM
models \citep[e.g.,][]{Frenk_etal88}.  Halo triaxiality and the
complexity of halo formation in a realistic cosmological context, with
continuous mergers, accretion, and rapidly changing potential wells
can have an impact on the orbital evolution of infalling satellites
\citep[e.g.,][]{Kravtsov_etal04}, with consequences for the efficiency
of their transformation. It will be important to explore these issues
in future investigations of tidal stirring.

A second shortcoming of our work is related to the fact that we have
focused on experiments where the alignments between the internal
angular momenta of the dwarfs, those of the primary disks and the
orbital angular momenta were all prograde ($45\degrees$ in most cases;
see Table~\ref{table:simulations}). Prograde alignments between the
orbital angular momenta of the dwarfs and the spins of the primary
disks, or between the angular momenta of the two disks, are expected
to be important mainly for orbits with pericentic distances smaller
than a characteristic radius containing a non-negligible fraction of
the mass of the primary disks. In this case, the tidal effects of the
host disks would be enhanced and the orbits of the dwarfs would
quickly decay due to the additional dynamical friction provided by the
disks of the primary galaxies \citep[e.g.,][]{Quinn_Goodman86}. As our
typical pericentric distances are $r_{\rm peri} \gtrsim 15$~kpc, we
expect the efficiency of the transformation reported here to be weakly
affected by such effects.

On the other hand, the orientation between the orbital angular momenta
of the dwarfs and the internal angular momenta of their disks is
particularly relevant for our experiments.  For retrograde alignments
between these two angular momenta, tidal stripping is considerably
reduced \citep[e.g.,][]{Read_etal06b} and both bar formation and the
efficiency of transformation via the tidal stirring mechanism are also
suppressed \citep{Mayer_etal01a}.  These facts highlight the
importance of coupling between orbital and internal motions.
Determining the statistics of alignments between internal and orbital
angular momenta would require a series of fully cosmological studies,
focusing on the accretion of satellites within halos of disk galaxies,
and is therefore beyond the scope of this paper. We stress, however,
that the orbital parameters adopted in the present study correspond to
only moderate alignments, and therefore our results should not be
biased by any strong coupling of angular momenta. Unless there is some
yet unknown cosmological bias {\it against} the mildly prograde
alignments that we have adopted here, our findings should be able to
illuminate at least some of the details of the typical transformation
experienced by infalling disky dwarfs in the LG and similar
environments.

Lastly, the most evident limitation of this study is that we have
addressed the efficiency of the tidal stirring mechanism only in the
collisionless regime. Our results should therefore be viewed as
preliminary. A more complete treatment including hydrodynamics is
required to illuminate one of the most distinct properties of dSphs,
namely their low gas content (see, e.g., \citealt{Grcevich_Putman09}
and references therein) and fully refine the conclusions presented
here. The present-day structure of dwarf galaxies originates from a
complex interplay of effects and a full explanation requires detailed
knowledge of their star formation histories and chemical evolution,
amongst others. Adding star formation as a further ingredient will
offer the possibility to determine the magnitude of starbursts induced
in the dwarfs at pericentric passages, while gaining a deeper
understanding of the wide diversity in their star formation histories
\citep[e.g.,][]{Grebel00,Orban_etal08}.  Furthermore, specific
predictions for the metallicity of the remnants formed by tidal
stirring, let alone comparisons with the luminosity-metallicity
relation for nearby dwarf galaxies \citep{Tolstoy_etal09}, are not
currently available.  We plan to extend the present investigation in
these directions in due course.

\citet{Mayer_etal07} have also demonstrated how the efficiency of
tidal stirring is affected by the presence of a dissipative component,
and how it varies depending on the balance between heating and cooling
in the gas.  This is especially relevant for disk-like progenitors
that are able to overcome ram pressure and retain their gas for a
longer period of time. Satellites infalling at $z \lesssim 1$ when the
intensity of the cosmic ultraviolet background is weaker and the gas
can settle in a colder and denser configuration within the potential
of the dwarf would fall in this category. In this case, the tidal
heating of the bar into a spheroid becomes less efficient. This is
because the bar instability causes the gas to flow towards the central
region of the dwarf, increasing its central density and causing a more
adiabatic response of the system to the external tidal perturbation.

Despite the aforementioned limitations, several facts do suggest that
our results for the efficiency of tidal stirring should be regarded as
conservative. First, as we discussed above, some of the progenitors of
present-day dSphs might have suffered significant tidal perturbations
before being accreted by their hosts \citep[e.g.,][]{Kravtsov_etal04}.
Therefore, it is plausible that these systems might have entered their
primary galaxies already partially transformed. Such a condition would
facilitate their complete transformation inside the tidal field of the
primary galaxy.  Second, if the progenitor disky dwarfs had either
very low halo concentration parameters or core-like density profiles,
as it is indeed suggested by both the modeling of rotation curves of
present-day LSB and dIrr galaxies \citep[e.g.,][]{deBlok10} and recent
cosmological simulations of dwarf galaxy formation
\citep{Governato_etal10}, our conclusions regarding the effectiveness
of the transformation would be reinforced.  Indeed, in such
circumstances the response of the dwarfs to the tidal shocks would be
much more impulsive compared to the cases of steep NFW-like profiles
like the ones that we used in our experiments.  This would give rise
to augmented tidal mass loss \citep[e.g.,][]{Kazantzidis_etal04b} and
to a more effective transformation into a dSph. Lastly, none of the
disky dwarfs in our experiments passed through the disk of the host
galaxy. This is important as it has been recently shown that disk
shocking, namely tidal shocks induced by passages through the disk,
can affect significantly the evolution of satellites having masses
$\lesssim 10^9 M_{\odot}$ and pericentic distances $\lesssim 30$~kpc,
and even cause the disruption of a fraction of them
\citep{D'Onghia_etal10}.

As a final remark, we reiterate that accumulating observational and
theoretical evidence suggests that dwarf galaxies are not formed as
thin disks, but rather are born as thick, puffy systems
\citep[e.g.,][]{Dalcanton_etal04,Kaufmann_etal07}.  The effect of
thermal support, as opposed to rotation, ought to be thoroughly
investigated in forthcoming studies of tidal stirring. Indeed, a
thicker, more diffuse stellar component suggests a stronger effect of
direct tidal heating \citep{Spitzer58}, but the bar instability, an
essential element of tidal stirring, is associated with thin stellar
configurations. Moreover, gas stripping by ram pressure should be
enhanced in an initially thicker, more diffuse stellar system owing to
the reduced gravitational restoring force of the gas.  The interplay
between all these aspects of the modeling will be assessed with future
work where realistic gas-rich dIrrs will be evolved inside MW-sized
hosts with recipes of radiative cooling, star formation, and
supernovae feedback (Kazantzidis et al., 2010b, in preparation).
These dwarfs have been formed self-consistently in cosmological
hydrodynamical simulations \citep{Governato_etal10} and are
characterized by a turbulent and multi-phase realistic ISM.

\section{Summary}
\label{sec:summary}

Using a suite of collisionless $N$-body simulations we have
investigated the efficiency of the tidal stirring mechanism for the
origin of dSphs. Specifically, we have examined the degree to which
the sizes, masses, shapes, and kinematics of late-type,
rotationally-supported dwarfs are affected by the gravitational field
of MW-sized host galaxies for a range of dwarf orbital and structural
parameters. Unlike previous work on the subject, we have employed
equilibrium numerical models of dwarf galaxies constructed from
composite DFs and consisting of exponential stellar disks embedded in
cosmologically-motivated DM halos. The self-consistency of the adopted
models is crucial for confirming the complex transformation process of
a disky dwarf into a dSph. This aspect of the modeling constitutes the
major improvement that we introduce in the present study.
Furthermore, we have extended earlier contributions on the subject by
conducting a simulation campaign which is carefully designed to allow
an investigation of a much larger parameter space than before. Lastly,
the fairly high numerical resolution of our experiments combined with
the growing observational data sets for dSph galaxies provide unique
opportunities for systematic and quantitative comparisons between the
theoretical models and the data, and we undertake such a task in a
companion paper ({\L}okas et al. 2010, in preparation).

Our main results and conclusions can be summarized as follows.

\begin{itemize}
  
\item[1.] Tidal interactions between rotationally-supported dwarf
  galaxies and MW-sized hosts can lead to the formation of
  pressure-supported, spheroidal stellar systems with kinematic and
  structural properties akin to those of the classic dSphs in the LG
  and similar environments. Our exploration of a wide variety of
  initial conditions for the progenitor disky dwarfs suggests that
  such tidal transformations are fairly efficient and should thus be
  common occurrences within the currently favored cosmological
  paradigm. Due to the fact that satellite accretion is a generic
  feature of hierarchical models of structure formation, the
  transformation process described in this study should be applicable
  to at least some of the dSph galaxies in the universe.
  
\item[2.] The transformation mechanism is complex and involves a
  combination of tidally-induced bar instabilities in stellar disks
  and impulsive tidal heating of the stellar distribution. Given the
  self-consistency of our dwarf galaxy models, we can safely conclude
  that the formation of dSphs can be entirely attributed to tidal
  perturbations, rather than being a consequence of the initial
  conditions. In the context of the tidal stirring model, bar
  formation is intimately linked to the formation of dSphs (see
  Table~\ref{table:summary}). Loss of angular momentum caused by the
  bar instability and simultaneous increase of the stellar velocity
  dispersion due to tidal heating lead to low values of $V_{\rm
    rot}/\sigma_{\ast}$ in the simulated remnants comparable to those
  of observed dSphs ($V_{\rm rot}/\sigma_{\ast} \lesssim 1$). Heating
  via tidal shocks at pericentric passages decreases continuously the
  elongation of the bar and causes the initially disky stellar
  distributions to transform into spheroids with projected axis ratios
  of $c/a \gtrsim 0.5$.
  
\item[3.] Bar formation constitutes a sufficient but not necessary
  condition for the formation of dSphs via the tidal stirring of
  rotationally-supported dwarfs. In cases where bar instabilities are
  not triggered by the tidal interactions with the host galaxies,
  spheroidal stellar systems with negligible amounts of rotation can
  still be produced solely via the action of impulsive tidal heating
  (R16; see \S~\ref{subsec:halo_mass}).
  
\item[4.] The effectiveness of the transformation into a dSph depends
  crucially on the orbital parameters of the progenitor disky dwarfs
  (\S~\ref{sec:orbital_parameters}).  For a fixed eccentricity,
  $r_{\rm apo}/r_{\rm peri}$, tighter orbits which are characterized
  by shorter orbital times, $T_{\rm orb}$, and smaller apocentric,
  $r_{\rm apo}$, and pericentric distances, $r_{\rm peri}$, lead to
  more rapid and complete transformations. For a fixed apocentric
  distance, orbits with higher eccentricities also induce stronger
  transformations.  Under the right combination of orbital parameters,
  tidal stirring can yield spherically-symmetric ($b/a \approx c/a
  \gtrsim 0.95$) and isotropic ($\beta \approx 0$) stellar systems
  with negligible amounts of rotation ($V_{\rm rot}/\sigma_{\ast}
  \lesssim 0.1$) (R2 and R4; see Table~\ref{table:summary}).
  
\item[5.] The degree of the tidal transformation depends on both the
  number and the strength of the tidal shocks, which are determined by
  the orbital time, $T_{\rm orb}$, and the pericentric distance,
  $r_{\rm peri}$, respectively. Small pericentric distances and short
  orbital times, corresponding to orbits associated with a large
  number of strong tidal shocks, produce the strongest and most
  complete transformations. However, in order to simply transform a
  disky dwarf galaxy into a dSph without necessarily inducing the
  strongest transformation, $r_{\rm peri}$ is the more salient orbital
  parameter. The last conclusion holds provided that $T_{\rm orb}$ is
  short enough to allow the dwarfs to complete at least two
  pericentric passages inside their hosts. Such a condition is likely
  satisfied for a significant fraction of dwarf galaxies in the LG
  (see \S~\ref{subsec:Torb_rperi}).
  
\item[6.]  The efficiency of the transformation via tidal stirring is
  notably affected by the structure of the progenitor disky dwarfs.
  Specifically, it is enhanced considerably for those with less
  massive and more extended disks, as well as for dwarfs embedded in
  halos of lower concentration (see \S~\ref{sec:struct_properties}).
  These properties are akin to those of LSB, gas-rich dIrr galaxies
  which reside in the outskirts of the LG, a fact that highlights
  tidal stirring as a plausible causal mechanism for the origin of the
  morphology-density relation.
  
\item[7.] The robustness of dwarf galaxies to tides and mass loss is
  increased significantly for those with more massive and more compact
  disks, as well as for dwarfs embedded in halos of higher
  concentration (see \S~\ref{sec:struct_properties}). This enhanced
  resilience to tidal effects has important consequences for the
  missing satellites problem as well as for determining the radial
  distribution of satellites inside host halos
  \citep[e.g.,][]{Diemand_etal04}.
  
\item[8.] The products of tidal stirring satisfy the relation $V_{\rm
    max} = \sqrt{3} \,\sigma_{\ast}$, where $\sigma_{\ast}$ is the
  one-dimensional, central stellar velocity dispersion and $V_{\rm
    max}$ is the maximum halo circular velocity (see
  \S~\ref{subsec:Vmax_sigma}).  Such a small conversion factor between
  $\sigma_{\ast}$ and $V_{\rm max}$, formally valid for a tracer
  stellar population with an isotropic velocity dispersion tensor, is
  in agreement with those originally adopted to formulate the missing
  satellites problem \citep{Moore_etal99,Klypin_etal99}.
  
\item[9.] The mass-to-light ratios, $M/L$, of the orbiting disky
  dwarfs galaxies decrease monotonically with time as the extended DM
  halos are preferentially tidally stripped. In some cases, however,
  the $M/L$ ratios start to increase later in the evolution when
  stellar mass loss becomes more effective (see \S~\ref{subsec:M/L}).
  These cases are associated with either enhanced tidal mass loss (R2
  and R4) or more extended initial stellar distributions (R13) and
  demand that the dwarfs are stripped down to the scales where the
  alignment between the angular momenta of the stars and the orbital
  angular momenta of the dwarfs has an important effect on stellar
  stripping. This mechanism causes only a moderate increase of the
  $M/L$ ratio and it may thus not be able to account for the extreme
  DM content in some of the classic dSphs (e.g., Draco or Ursa Minor).
  Producing such systems likely requires hydrodynamical processes.
  
\item[10.] Distant dSphs in the LG, such as Leo I, Tucana, and Cetus,
  which are likely moving on very wide orbits, should have only been
  partially stirred by their hosts, assuming that their properties
  originate from tidal stirring. As a result, these remote dwarfs
  should exhibit higher values of $V_{\rm rot}/\sigma_{\ast}$ compared
  to those of dSphs located closer to the primary galaxies. Future
  conclusive measurements of rotation in these systems will serve to
  validate (or falsify) this prediction.

\end{itemize}

\vspace{-0.5cm}

\acknowledgments

The authors acknowledge useful discussions with James Bullock, Mandeep
Gill, Andrey Kravtsov, Andrea Macci\`{o}, Chiara Mastropietro, Mario
Mateo, Chris Orban, Michael Stamatikos, Justin Read, and David
Weinberg. S.K. would like to thank Victor Debattista for valuable
conversations on bar instabilities and related issues which
significantly informed this work. S.K. also acknowledges the
hospitality of the Nicolaus Copernicus Astronomical Center during a
visit when the final stages of this work were completed. S.K. is
funded by the Center for Cosmology and Astro-Particle Physics (CCAPP)
at The Ohio State University.  E.L.{\L}. is grateful for the
hospitality of CCAPP during her visit.  This research was partially
supported by the Polish Ministry of Science and Higher Education under
grant NN203025333. The work of L.A.M. was carried out at the Jet
Propulsion Laboratory (JPL), California Institute of Technology, under
a contract with NASA.  L.A.M.  acknowledges support from the NASA ATFP
program.  The numerical simulations were performed on the Cosmos
cluster at JPL.  Cosmos was provided by funding from the JPL Office of
the Chief Information Officer. This work was also supported by an
allocation of computing time from the Ohio Supercomputer Center
(http://www.osc.edu).

\bibliography{dwarfs}

\end{document}